\documentclass[twocolumn,tighten]{aastex631}
\usepackage{xcolor,amsmath}
\usepackage[flushmargin]{footmisc}

\newcommand{\fesc}{$f_{esc}^{\rm LyC}$}

\newcommand{\mathfesc}{f_{esc}^{\rm LyC}}
\newcommand{\fg}{$f_{\rm g}$}
\newcommand{\fst}{$f_{\rm s}$}

\newcommand{\fesclya}{$f_{esc}^{{\rm Ly}\alpha}$}
\newcommand{\nhi}{$N_{\rm H I}$}
\newcommand{\orat}{$\rm O_{32}$}
\newcommand{\sigsfr}{$\Sigma_{\rm SFR}$}
\newcommand{\ssfr}{${\rm sSFR}$}
\newcommand{\kms}{${\rm km~s^{-1}}$}
\newcommand{\logoh}{$\rm \log(\frac O H)+12$}
\newcommand{\ebv}{$\rm E({B-V})$}
\newcommand{\xiion}{$\xi_{ion}$}
\newcommand{\emech}{$E_{mech}$}
\newcommand{\fyng}{$f_{\star}(t<3)$}
\newcommand{\fwr}{$f_{\star}(3<t<6)$}
\newcommand{\fold}{$f_{\star}(t>8)$}
\newcommand{\muv}{$\rm M_{UV}$}
\newcommand{\rfl}{$R_{f,\rm LIS}$}
\newcommand{\rfh}{$R_{f,\rm H~I}$}
\newcommand{\cfl}{$C_{f,\rm LIS}$}
\newcommand{\cfh}{$C_{f,\rm H~I}$}
\newcommand{\wfl}{$W_{\lambda,\rm LIS}$}
\newcommand{\wfh}{$W_{\lambda,\rm H~I}$}

\newcommand{\pfesc}{$P(<f_{\rm s}|f_{\rm g})$}

\graphicspath{{./}{./plots/}}

\begin{document}

\title{The Low-Redshift Lyman Continuum Survey:\\The Roles of Stellar Feedback and ISM Geometry in LyC Escape}

\shorttitle{LzLCS: Stellar Populations, Gas Geometry, and LyC Escape}
\shortauthors{Flury et al}
\submitjournal{ApJ}
\received{}
\revised{}
\accepted{}

\correspondingauthor{Sophia Flury}
\email{sflury@umass.edu}

\author[0000-0002-0159-2613]{Sophia R. Flury}
\affiliation{Department of Astronomy, University of Massachusetts Amherst, Amherst, MA 01002, United States}
\affiliation{Institute for Astronomy, University of Edinburgh, Royal Observatory, Edinburgh, EH9 3HJ, UK}

\author[0000-0002-6790-5125]{Anne E. Jaskot}
\affiliation{Department of Astronomy, Williams College, Williamstown, MA 01267, United States}

\author[0000-0001-8419-3062]{Alberto Saldana-Lopez}
\affiliation{Department of Astronomy, Oskar Klein Centre, Stockholm University, 106 91 Stockholm, Sweden}

\author[0000-0002-5808-1320]{M. S. Oey}
\affiliation{University of Michigan, Department of Astronomy, 323 West Hall,
1085 S. University Ave, Ann Arbor, MI 48109, United States}

\author[0000-0002-0302-2577]{John Chisholm}
\affiliation{Department of Astronomy, The University of Texas at Austin, 2515
Speedway, Stop C1400, Austin, TX 78712-1205, United States}

\author[0000-0001-5758-1000]{Ricardo Amor\'in}
\affiliation{ARAID Foundation, Centro de Estudios de Física del Cosmos de
Aragón (CEFCA), Unidad Asociada al CSIC, Plaza San Juan 1,
44001 Teruel, Spain}
\affiliation{Departamento de Astronomía, Universidad de La Serena, Avda.
Juan Cisternas 1200, La Serena, Chile}

\author[0000-0003-2722-8841]{Omkar Bait}
\affiliation{SKA Observatory, Jodrell Bank, Lower Withington, Macclesfield, SK11 9FT, UK}

\author[0000-0002-2724-8298]{Sanchayeeta Borthakur}
\affiliation{School of Earth and Space Exploration, Arizona State University, 781 Terrace Mall, Tempe, AZ 85287, United States}

\author[0000-0003-4166-2855]{Cody Carr}
\affiliation{Center for Cosmology and Computational Astrophysics, Institute
for Advanced Study in Physics, Zhejiang University, Hangzhou
310058, PR China}
\affiliation{Institute of Astronomy, School of Physics, Zhejiang University,
Hangzhou 310058, PR China}

\author[0000-0001-7113-2738]{Henry C. Ferguson}
\affiliation{Space Telescope Science Institute, 3700 San Martin Drive, Baltimore, MD 21218, United States}

\author[0000-0002-7831-8751]{Mauro Giavalisco}
\affiliation{Department of Astronomy, University of Massachusetts Amherst, Amherst, MA 01002, United States}

\author[0000-0001-8587-218X]{Matthew Hayes}
\affiliation{Department of Astronomy, Oskar Klein Centre, Stockholm University, 106 91 Stockholm, Sweden}

\author[0000-0001-6670-6370]{Timothy Heckman}
\affiliation{Center for Astrophysical Sciences, Department of Physics \&
Astronomy, Johns Hopkins University, Baltimore, MD 21218, United States}

\author[0000-0002-6586-4446]{Alaina Henry}
\affiliation{Center for Astrophysical Sciences, Department of Physics \&
Astronomy, Johns Hopkins University, Baltimore, MD 21218, USA}
\affiliation{Space Telescope Science Institute, 3700 San Martin Drive, Baltimore, MD 21218, United States}

\author[0000-0001-7673-2257]{Zhiyuan Ji}
\affiliation{Steward Observatory, University of Arizona, 933 N. Cherry Avenue,
Tucson, AZ 85721, United States}

\author[0000-0002-5235-7971]{Lena Komarova}
\affiliation{University of Michigan, Department of Astronomy, 323 West Hall,
1085 S. University Ave, Ann Arbor, MI 48109, United States}

\author[0000-0002-6085-5073]{Floriane Leclercq}
\affiliation{Department of Astronomy, The University of Texas at Austin, 2515
Speedway, Stop C1400, Austin, TX 78712-1205, United States}

\author[0000-0003-1767-6421]{Alexandra Le Reste}
\affiliation{Minnesota Institute for Astrophysics, University of Minnesota, Minneapolis, MN 55455, United States}

\author[0000-0003-0503-4667]{Stephan McCandliss}
\affiliation{Center for Astrophysical Sciences, Department of Physics \&
Astronomy, Johns Hopkins University, Baltimore, MD 21218, USA}

\author[0000-0001-8442-1846]{Rui Marques-Chaves}
\affiliation{Department of Astronomy, University of Geneva, 51 Chemin Pegasi,
1290 Versoix, Switzerland}

\author[0000-0002-3005-1349]{G{\"o}ran {\"O}stlin}
\affiliation{Department of Astronomy, Oskar Klein Centre, Stockholm University, 106 91 Stockholm, Sweden}

\author[0000-0001-8940-6768]{Laura Pentericci}
\affiliation{INAF – Osservatorio Astronomico di Roma, Via di Frascati 33,
00078 Monte Porzio Catone, Italy}

\author[0000-0002-5269-6527]{Swara Ravindranath}
\affiliation{Astrophysics Science Division, NASA Goddard Space Flight Center, 8800 Greenbelt Road, Greenbelt, MD 20771, USA}
\affiliation{Center for Research and Exploration in Space Science and Technology II, Department of Physics, Catholic University of America, 620 Michigan Ave N.E., Washington DC 20064, USA}

\author[0000-0002-6085-5073]{Michael Rutkowski}
\affiliation{Minnesota State University-Mankato, Telescope Science Institute, TN141, Mankato, MN 56001, United States}

\author[0000-0002-9136-8876]{Claudia Scarlata}
\affiliation{Minnesota Institute for Astrophysics, University of Minnesota, Minneapolis, MN 55455, United States}

\author[0000-0001-7144-7182]{Daniel Schaerer}
\affiliation{Department of Astronomy, University of Geneva, 51 Chemin Pegasi,
1290 Versoix, Switzerland}

\author[0000-0001-5331-2030]{Trinh Thuan}
\affiliation{Astronomy Department, University of Virginia, PO Box 400325,
Charlottesville, VA 22904-4325, United States}

\author[0000-0002-6849-5375]{Maxime Trebitsch}
\affiliation{Kapteyn Astronomical Institute, University of Groningen, PO Box
800, 9700 AV Groningen, The Netherlands}

\author[0000-0001-5228-9326]{Eros Vanzella}
\affiliation{INAF, Osservatorio Astronomico di Bologna, via Gobetti 93/3 I-40129 Bologna, Italy}

\author[0000-0002-2201-1865]{Anne Verhamme}
\affiliation{Department of Astronomy, University of Geneva, 51 Chemin Pegasi,
1290 Versoix, Switzerland}

\author[0000-0001-9269-5046]{Bingjie Wang}
\affiliation{Department of Astronomy \& Astrophysics, The Pennsylvania State
University, University Park, PA 16802, United States}
\affiliation{Institute for Computational \& Data Sciences, The Pennsylvania
State University, University Park, PA 16802, United States}
\affiliation{Institute for Gravitation and the Cosmos, The Pennsylvania State
University, University Park, PA 16802, United States}

\author[0000-0003-0960-3580]{G{\'a}bor Worseck}
\affiliation{VDI/VDE Innovation+Technik, Berlin, Germany}

\author[0000-0002-9217-7051]{Xinfeng Xu}
\affiliation{Department of Physics and Astronomy, Northwestern University,
2145 Sheridan Road, Evanston, IL, 60208, USA.}
\affiliation{Center for Interdisciplinary Exploration and Research in
Astrophysics (CIERA), Northwestern University, 1800 Sherman Avenue,
Evanston, IL, 60201, USA.}

\begin{abstract}
One of the fundamental questions of cosmology is the origin and mechanism(s) responsible for the reionization of the Universe beyond $z\sim6$. To address this question, many studies over the past decade have focused on local ($z\sim0.3$) galaxies which leak ionizing radiation (Lyman continuum or LyC). However, line-of-sight effects and data quality have prohibited deeper insight into the nature of LyC escape. To circumvent these limitations, we analyze stacks of a consolidated sample of \emph{HST}/COS observations of the LyC in 89 galaxies at $z\sim0.3$. From fitting of the continuum, we obtain information about the underlying stellar populations and neutral ISM geometry.
We find that most LyC non-detections are not leaking appreciable LyC (\fesc$<1$\%) but also that exceptional cases point to spatial variations in the LyC escape fraction \fesc. Stellar populations younger than 3 Myr lead to an increase in ionizing feedback, which in turn increases the isotropy of LyC escape. Moreover, mechanical feedback from supernovae in 8-10 Myr stellar populations is important for anisotropic gas distributions needed for LyC escape. While mechanical feedback is necessary for any LyC escape, high \fesc\ ($>5$\%) also requires a confluence of young stars and ionizing feedback.
A two-stage burst of star formation is critical to producing this optimal LyC escape scenario and should be considered fundamental to identifying LyC emitters at the epoch of reionization.
\end{abstract}

\section{Introduction\label{sec:intro}}

Identifying the galaxies responsible for cosmic reionization is fundamental to understanding galaxies and the intergalactic medium (IGM) in the early Universe. Over the last decade, the number of galaxies with detected Lyman continuum (LyC) has increased dramatically from a handful of candidates to dozens of confirmed LyC emitters (LCEs) at both $z\lesssim0.5$ \citep[e.g.,][]{2016MNRAS.461.3683I,2018MNRAS.478.4851I,2022ApJS..260....1F} and $z\sim3$ \citep[e.g.,][]{2019ApJ...878...87F,2022MNRAS.511..120S}. While these exciting results provide the foundation necessary for identifying star-forming LCEs at the epoch of reionization \citep[e.g.,][]{2021ApJ...916....3W,2022ApJ...930..126F,2022A&A...663A..59S,2022MNRAS.517.5104C,2023A&A...672A.155M,2024arXiv240610171J,2024arXiv240610179J}, the mechanisms which promote the escape of ionizing photons are not well understood.

Recently, \citet{2024arXiv240104278A} established a connection between outflows and the fraction of LyC photons which escape from a galaxy (the escape fraction, or \fesc), indicating that stellar feedback is a likely culprit in LyC escape \citep[as anticipated in previous studies, e.g.,][]{2001ApJ...558...56H,2002MNRAS.337.1299C,2011ApJ...730....5H,2020MNRAS.494.3541H,2022ApJ...940..160M}. Concentrated star formation produces more extreme feedback \citep{2019ApJ...886...74M,2020ApJ...889L..22C,2023A&A...676A..53L,2024MNRAS.527.7871C}, which may explain the relationship between star formation rate surface density \sigsfr\ and \fesc\ \citep{2022ApJ...930..126F}. Simulations predict that this feedback can fully account for the escape of LyC photons from star-forming regions and out into the intergalactic medium \citep[e.g.,][]{2010ApJ...710.1239R,2017MNRAS.470..224T,2020ApJ...902L..39B,2020MNRAS.498.2001M}. 
However, the exact mechanisms for feedback, and the associated galaxy properties, are neither well understood nor constrained observationally.

Competing hypotheses for the relationship between feedback and LyC escape suggest different stellar populations and feedback mechanisms. Simulations by \citet{2017MNRAS.466.4826K} suggest ionizing feedback from very young ($<3$ Myr) stellar populations is critical, which may be consistent with results pointing to young stellar populations as key drivers of LyC escape \citep[e.g.,][]{2019ApJ...885...96J,2021ApJ...908...30K}, particularly since ionizing radiation may be sufficient to drive superwinds \citep{2021ApJ...920L..46K}. Other studies point to stellar populations with ages of 3-5 Myr \citep[e.g.,][]{2013ApJ...779...76Z,2020MNRAS.498.2001M,2023arXiv230408526C}, invoking stellar winds and initial supernovae (SNe) as sources of feedback responsible for clearing out channels optically thin to the LyC. Older (8-10 Myr) populations may also be significant, with LyC escape requiring a time delay between star formation and the effects of feedback-driven outflows \citep[e.g.,][]{2017MNRAS.470..224T,2020ApJ...902L..39B}. Still others suggest two-stage bursts, the first to generate feedback via winds and SNe and the second to produce LyC photons, are necessary for LyC escape \citep[e.g.,][]{2017ApJ...845..165M,2023A&A...672A..11E} with binary star populations sustaining a strong LyC budget for long periods of time \citep[e.g.,][]{2018MNRAS.479..994R}. However, substantial mechanical feedback may disrupt subsequent star formation, thereby inhibiting multi-burst LyC escape \citep[e.g.,][]{2020ApJ...902L..39B}.

In the context of LyC escape, feedback may play a key role in reshaping the distribution of gas in the interstellar medium (ISM). Studies often consider two simplified descriptions of the ISM geometry in LCEs: ubiquitously thin gas that allows isotropic LyC escape \citep[the so-called ``density-bounded'' scenario, e.g.,][]{2013ApJ...777...39Z,2014MNRAS.442..900N} and patchy, anisotropic gas distributions \citep[the so-called ``picket fence'' scenario, e.g.,][]{2001ApJ...558...56H,2017ApJ...836...78Z}. Feedback can produce or affect these geometries in various ways. SNe and winds can redistribute material from remnant birth clouds, generating picket fence morphology close to the ionizing cluster, promoting LyC escape \citep[e.g.,][]{2019ApJ...885...96J}, but can also blow out chimneys in dense gas through which the LyC can escape \citep[e.g.,][]{2001ApJ...558...56H,2011ApJ...730....5H}. However, catastropic cooling can mitigate the effects of SNe and winds, particulalry at low metallicities, thereby maintaining large opening angles of low-density gas \citep{2021ApJ...921...91D}. Moreover, delays in SNe onset by 5 Myr or more due to metallicity effects can drastically reduce early cumulative mechanical feedback, further mitigating its effects on the surrounding ISM \citep{2023ApJ...958..149J}. And if very young stellar populations are at play, then ionizing LyC feedback can also play a significant role in the geometry by driving winds \citep{2021ApJ...920L..46K} or by ionizing the gas to produce a density-bounded scenario \citep[e.g.,][]{2020A&A...639A..85G}.

Previous efforts to address the roles of feedback and ISM geometry in LyC escape have relied on rest-frame FUV spectroscopy across a range of redshifts. Geometry measurements suggest LyC escape increases as optically thin gas becomes increasingly isotropic \citep[e.g.,][]{2018A&A...616A..30C,2020A&A...639A..85G,2022A&A...663A..59S}. Due to the sensitivity of both \fesc\ and absorption lines to line-of-sight, these results are biased toward chance (mis)alignments of LCEs with the observer and may not fully capture the geometry of LyC escape. Averaging over multiple sightlines is key to painting a complete picture of the ISM conditions. Investigations of stellar populations consistently indicate ages $<10$ Myr are most closely associated with LyC escape \citep[e.g.,][]{2022MNRAS.517.5104C,2023A&A...672A..11E,2023MNRAS.522.6295S,2023ApJ...955L..17K}. More detailed assessments, including distinguishing contributions from stellar populations before and after the onset of core-collapse supernovae (ages $<3$ and $>6$ Myr, respectively) in observed spectra, are necessary to fully understand the relationships between star formation, feedback, and LyC escape. Such detail requires good detection of diagnostic features such as the \ion{O}{6}$\lambda\lambda1032,38$ and \ion{N}{5}$\lambda\lambda1238,40$ P Cygni lines and the \ion{S}{4}$\lambda1060$ and \ion{C}{3}$\lambda1175$ photospheric lines \citep{2019ApJ...882..182C}.

To obtain constraints on the stellar populations, feedback, and average geometry as they pertain to \fesc, we perform a detailed analysis of stacked \emph{HST}/COS spectra from the Low-redshift Lyman Continuum Survey \citep[LzLCS,][]{2022ApJS..260....1F} and other LyC programs. In \S\ref{sec:lzlcs}, we review the combined LzLCS and published COS surveys of LCE candidates at $z\sim0.3$ and, in the subsequent \S\ref{sec:stacks}, outline our stacking procedure. We analyze the stacked spectra to determine stellar populations (\S\ref{sec:specfit}), \fesc (\S\ref{sec:nondets}), and ISM properties (\S\ref{sec:absn}). With these results, we demonstrate relationships between stellar populations, feedback, and geometry in \S\ref{sec:discuss}. Finally, in \S\ref{sec:lycescape}, we use these results to assemble a framework for interpreting LyC escape both locally and at $z>6$.

For this study, we assume $H_0=70\rm~km~s^{-1}~Mpc^{-1}$, $\Omega_m=0.3$, and $\Omega_\Lambda=1$. Unless specified, all atomic data are taken from \citet{2003ApJS..149..205M}.

\section{Sample Selection}\label{sec:lzlcs}

The Low-redshift Lyman Continuum Survey \citep[LzLCS][]{2022ApJS..260....1F} is a large \emph{HST}/COS observing program (GO 15626, PI Jaskot) which includes observations of the LyC window and FUV continuum for 66 galaxies at $z\sim0.3$ that cover a wide range of properties. The LzLCS provides the first statistically robust test of indirect diagnostics for identifying LCEs \citep[e.g.,][]{2022ApJ...930..126F,2022A&A...663A..59S,2021ApJ...916....3W}. One of the major objectives of the LzLCS is to provide an empirical foundation for selecting LCE candidates from \emph{JWST} observations of galaxies in the epoch of reionization at $z\gtrsim6$ \citep{2024arXiv240610171J,2024arXiv240610179J}. However, the LzLCS also offers a unique opportunity to investigate the properties and underlying mechanisms of LCEs.

With 23 additional observations of $z\lesssim0.5$ LCE candidates in the literature \citep{2016Natur.529..178I,2016MNRAS.461.3683I,2018MNRAS.474.4514I,2018MNRAS.478.4851I,2021MNRAS.503.1734I,2019ApJ...885...57W}, the combined LzLCS, henceforth LzLCS+, comprises COS G140L spectra uniformly processed using { FaintCOS} \citep{2021ApJ...912...38M} for 89 objects spanning a wide range of UV luminosities, starburst ages, stellar masses, and ionization parameters. \citet{2022A&A...663A..59S} fit these spectra with stellar population SEDs to determine properties like the intrinsic LyC, burst age, and \fesc. Accompanying the UV observations are rest-frame optical spectra from the Sloan Digital Sky Survey \citep[SDSS,][]{2000AJ....120.1579Y}. As outlined in \citet{2022ApJS..260....1F}, the LzLCS+ includes ancillary information obtained from the nebular emission lines in the SDSS spectra such as flux ratios, direct-method gas phase chemical abundances, and star formation rates.

\section{Stacking}\label{sec:stacks}

\subsection{Motivation for Stacking}

Our motivation for stacking is threefold: (i) to boost the S/N of the non-ionizing FUV spectra to obtain strong constraints on the stellar population ages and absorption line properties; (ii) to average over line-of-sight variations to account for orientation and pencil-beam effects, and (iii) to push below the \fesc\ upper limit threshold imposed by the exposure times and detector sensitivity in the individual observations.

The characteristic S/N of the non-ionizing spectra for the LzLCS galaxies is quite low, with individual pixels detecting continuum at $\approx1.5\sigma$ ($\sigma$ per resolution element) at rest-frame 1100 \AA\ where the starlight continuum is bright and featureless. The continuum fluctuates by $\approx20$\% on average at the rest-frame 1100 \AA\ for pixels binned to match the COS G140L spectral resolution. As indicated by the large uncertainties reported in \citet{2022A&A...663A..59S}, the low S/N reduces constraining power of important spectroscopic features, including age diagnostics in the starlight continuum (e.g., \ion{O}{6} $\lambda$ 1037, \ion{S}{4} $\lambda\lambda1062,73$, \ion{P}{5} $\lambda\lambda1118,28$, \ion{C}{3} $\lambda1175,1247$, \ion{N}{5} $\lambda1240$) and ISM absorption lines (e.g., \ion{Si}{2} $\lambda1193,1260$, \ion{Si}{2}+\ion{O}{1} $\lambda1302$, \ion{C}{2} $\lambda1334$), in some cases to the extent of preventing detection altogether. Stacking will increase the S/N of the spectra, clarifying many of these key stellar population and ISM signatures.

One of the principal issues with assessing the geometry of the ISM for LyC escape is line-of-sight effects \citep[e.g.,][]{2024ApJ...967..117K}. Depending on the orientation of the \ion{H}{2} region(s) and the host galaxy with respect to the observer, \fesc\ can take on a broad range of values \citep[e.g.,][cf. results in \citealt{2022MNRAS.517.5104C,2022A&A...663A..59S}]{2015ApJ...801L..25C}, even to the extent that no LyC flux is detected from a genuine LCE due to misalignment of the source with the observer \citep[e.g.,][]{2019ApJ...878...87F,2024ApJ...967..117K}. Corresponding absorption lines are also sensitive to the ``pencil beam'' effect, tracing only the gas geometry and column density along sight lines illuminated by the background stellar population \citep[e.g.][]{2018A&A...616A..30C,2018ApJ...869..123S,2020A&A...639A..85G,2022A&A...663A..59S}. These line-of-sight effects due to orientation and pencil-beaming lead to substantial scatter in relationships between \fesc\ and most properties \citep{2022ApJ...930..126F,2022A&A...663A..59S}. Stacking averages over different sight lines, thereby allowing us to investigate the characteristic ISM geometry and \fesc\ of galaxies without the substantial uncertainty or strong biases of line-of-sight effects inherent to individual galaxies.

{ Finally, detection of the LyC flux is sensitive to several factors \citep[cf.][]{2022ApJS..260....1F}, including where the LyC falls on the COS FUV chip, and may not always reflect whether an object is cosmologically irrelevant. A target's redshift determines where its LyC falls on the COS FUVA detector. The FUVA sensitivity declines precipitously below 1100 \AA\ in the observed frame. Moreover, geocoronal Ly$\alpha$ prevents signal detection between observed-frame 1190 and 1235 \AA\ (at $z=0.35$, 900 \AA\ rest-frame corresponds roughly to the peak of telluric Ly$\alpha$).}
Exposure times for the LzLCS were originally designed for detection of the LyC at \fesc$=5$\%. While the LzLCS+ was immensely successful in detecting the LyC at even lower \fesc\ than anticipated (29 of the 50 LyC detections had \fesc$<0.05$), 39 targets still yielded non-detections. \citet{2022ApJS..260....1F} demonstrate a LyC flux detection threshold of $3\times10^{-18}\rm~erg~s^{-1}~cm^{-2}$ over the majority of the redshift range, although some fainter sources have upper limits of \fesc$\sim0.05$. { Particularly faint LCEs might appear as non-LCEs simply by having emergent LyC flux below this limit. Stacking will push down below the COS G140L FUVA sensitivity threshold to determine if non-detections are cosmologically insignificant LyC leakers or instead are faint LCEs lurking beneath the detection limit.}

\subsection{Method for Stacking \emph{HST}/COS Spectra\label{sec:UVStacks}}

Below, we outline the generalized procedure used to stack arbitrary sets of spectra from the combined LzLCS+ sample from \citet{2022ApJS..260....1F}, which includes uniformly processed \emph{HST}/COS G140L spectra from the LzLCS (GO 15626) and from previous GO programs 13744 \citep{2016MNRAS.461.3683I}, 14635 \citep{2018MNRAS.478.4851I}, 15341 \citep{2019ApJ...885...57W}, and 15639 \citep{2021MNRAS.503.1734I}. Geocoronal features, notably Ly$\alpha$ and \ion{O}{1}$\lambda1304$, are masked. When available, we replace the region affected by telluric \ion{O}{1}$\lambda1304$ with orbital night flux instead of masking it, as the shadow data are not contaminated by this feature. Absorption lines due to the Galactic interstellar medium (ISM) are also masked to further limit any contamination of the stacked flux.

To prevent biasing the stack towards the brightest objects, we normalize each spectrum by its flux summed over a 1 \AA\ bin centered on 1100 \AA\ in the rest frame. We choose this location for the window because it is sufficiently far from any telluric, ISM (Galactic or otherwise), or stellar features. Averaging over a 1 \AA\ width reduces any potential contamination from fluctuations in the noise. After normalizing the spectrum, we sum up its flux in 0.5 \AA\ rest-frame wavelength bins over the rest-frame interval of 800 to 1450 \AA, corresponding to 1125-1850 \AA\ in the observed frame, imposed by the sensitivity of the COS detector. The 0.5 \AA\ rest-frame wavelength binning is chosen to approximate Nyquist sampling of the COS G140L spectra at a characteristic redshift of 0.3.

Once all the spectra have been normalized and binned in the rest frame, we compute the median flux of all the spectra in each wavelength bin to obtain the stacked flux. While the number of objects used to build the total stack does not change, differences in redshift will change the rest-frame wavelength range for each object as well as the location of telluric lines, which can affect how many objects can contribute to each wavelength bin in the stack. To determine the uncertainty in the stacked flux, we bootstrap fluxes from all the spectra contributing to the bin and take the median $10^4$ times, essentially obtaining a standard error of the mean for each bin's median flux within the stack. This uncertainty assessment is robust because it accounts for both the range of fluxes contributing to the median flux within the bin and the number of spectra contributing to the wavelength bin. After visual inspection, we find that wavelength bins containing at least four objects will have sufficiently small uncertainties in the flux to constrain relevant features, which we attribute to the factor of 2 reduction in standard error of the mean relative to that of the flux measured for the individual galaxies. In our subsequent analysis, we mask out any wavelength bins with fewer than four contributing objects. As consequence of this requirement, stacks with four or fewer galaxies are unusable.

\begin{figure*}
    \centering
    \includegraphics[width=\linewidth,trim={0cm 1cm 0cm 0cm},clip]{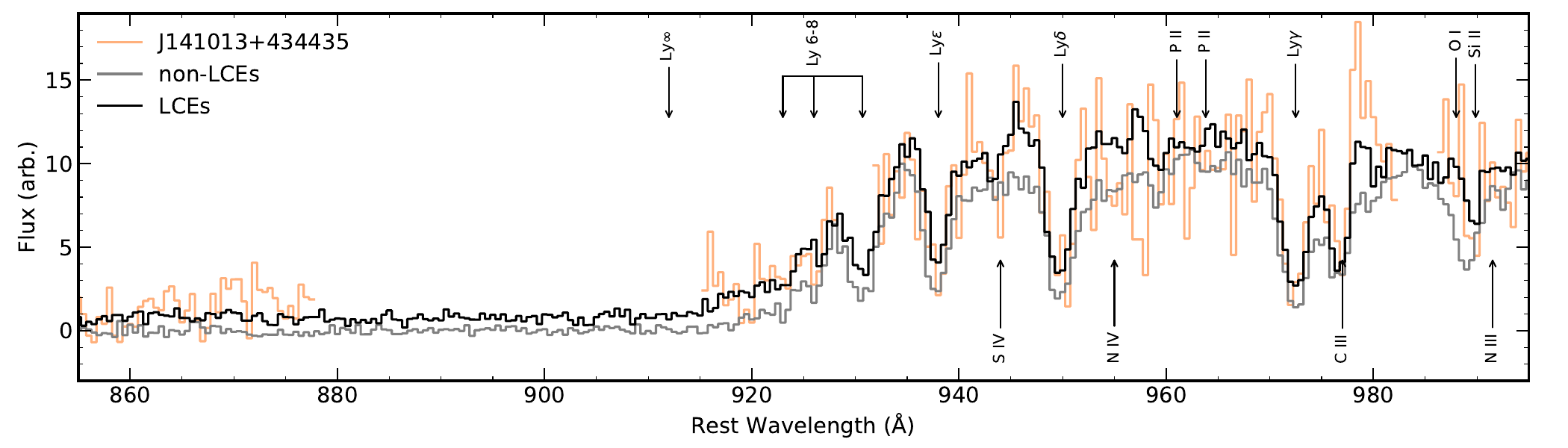}
    \includegraphics[width=\linewidth,trim={0cm 1cm 0cm 0cm},clip]{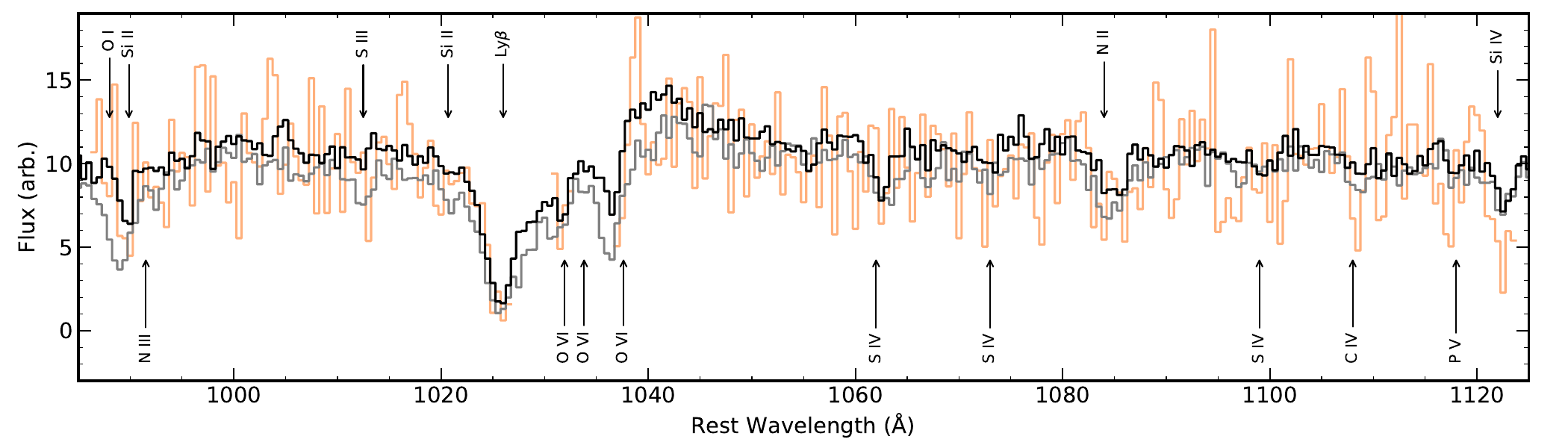}
    \includegraphics[width=\linewidth,trim={0cm 1cm 0cm 0cm},clip]{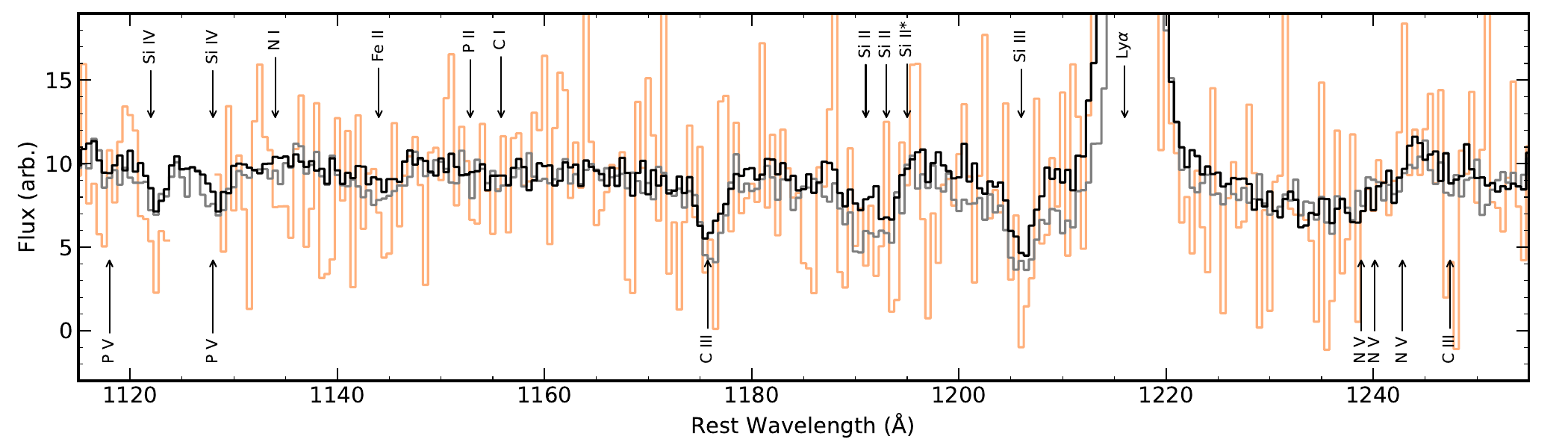}
    \includegraphics[width=\linewidth,trim={0cm 0cm 0cm 0cm},clip]{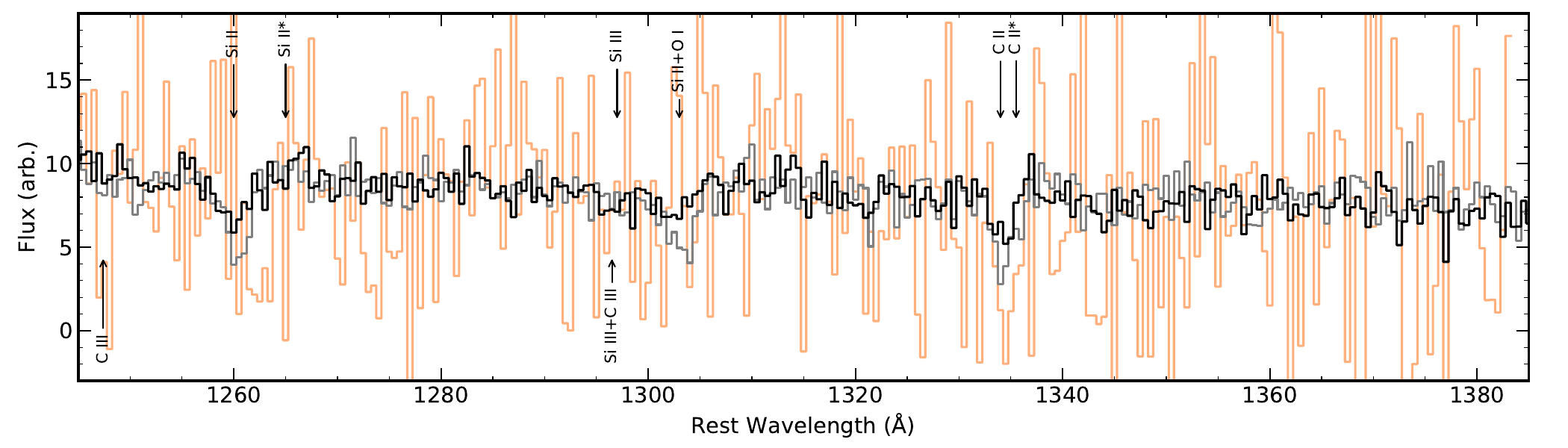}
    \caption{Full stacks of LCEs (black) and non-detections (grey). To illustrate the improvement in signal via stacking,
 we include the spectrum for strong LCE J141013+434435 (yellow), the galaxy with the median 1100 $\rm\AA$\ flux in the LzLCS. ISM features are indicated from above, stellar wind and photospheric lines indicated from below.}
    \label{fig:stack_features}
\end{figure*}

We show two stacks, one of LCEs (LyC detected with $>97.725$\% confidence) and one of non-LCEs (LyC not detected), in Figure \ref{fig:stack_features}. Myriad ISM and stellar features are identifiable in the stacks but are largely undetected in the individual spectra, as highlighted in Figure \ref{fig:stack_features}. Moreover, the emergent LyC flux in the LCE stack is quite clear whereas the non-LCE stack exhibits no LyC whatsoever, indicating that, on average, non-LCEs do not exhibit undetected LyC.

To explore differences between and among LCEs and non-detections, we also produce stacked spectra for subsamples of LCE and non-LCEs.
We select subsamples based on thresholds associated with high fractions of LCEs in the literature \citep{2022MNRAS.517.5104C,2022ApJ...930..126F}, including properties like burst age (EW H$\beta$), ionization parameter (\orat=[\ion{O}{3}]$\lambda5007$/[\ion{O}{2}]$\lambda3726,29$
), and dust attenuation ($\beta_{1550}$). Divisions among subsets are chosen to optimally select for LCEs based on apparent thresholds in \citet{2022ApJ...930..126F}. For all galaxies which meet a given criterion, we stack the entire subset, the LCEs, the non-LCEs, the strong LyC leakers (\fesc$>0.05$), the weak LyC leakers (\fesc$\in[0.01,0.05]$), and the non-leakers (\fesc$<0.01$). Thus, for each given criterion, we produce a total of six different stacks.

We list the full set of stacking criteria and the number of galaxies matching those criteria in Table \ref{tab:stack_props}. Accounting for duplicates or criteria which include no objects, we build a total of 127 stacks. 124 of these stacks include at least 5 objects, the threshold below which stacking does not reliably boost the signal due to a combination of poor sampling of each wavelength bin and discontinuous wavelength coverage as a result of geocoronal features and sensitivity dropoff towards the edges of the COS FUVA detector.

\begin{longrotatetable}
\startlongtable
\begin{deluxetable*}{crcccccccccccc}
\tablecaption{Stellar population properties of stacks of \emph{HST}/COS G140L rest-UV spectra with direct LyC measurements. Columns indicate (i) the criterion/a for inclusion of a target in the stack, (ii) the number of targets N included in the stack, (iii) the LyC escape fraction \fesc, (iv) probability $P(<\mathfesc)$ that the stack \fesc\ is representative of its constituents' \fesc, (v) UV slope $\beta$, (vi) FUV magnitude \muv, (vii-viii) fraction of intrinsic SED light contributed by stellar populations of the given age ranges, (ix) ionizing photon production efficiency \xiion\ of the stellar populations, and (x) cumulative mechanical energy injected by feedback from stellar populations.\label{tab:stack_props}}
\tablewidth{\textwidth}
\renewcommand{\arraystretch}{0.75}
\tablehead{
\colhead{criterion} & \colhead{N} & \colhead{\fesc} & \colhead{$P(<\mathfesc)$} &
    \colhead{$\beta_{1550}$} & \colhead{\muv} & \colhead{\fyng} & \colhead{\fold} &
    \colhead{\xiion} & \colhead{\emech} \\
     & & & & & & & &
        \colhead{$[\rm10^{25}~phot~Hz~erg^{-1}]$} & \colhead{$[\rm10^{54}~erg]$} }
\startdata
\\[-8pt]
\multicolumn{14}{c}{\bf total} \\[2pt]
\hline
$\beta_{1550}>-2$                            & 51 & 0.007 $\pm$ 0.002 & 0.522 & -1.214 $\pm$ 0.024 & -22.096 $\pm$ 0.017 & 0.007 $\pm$ 0.003 & 0.157 $\pm$ 0.065 & 1.064 $\pm$ 0.064 & 54.220 $\pm$ 0.226 \\
$\beta_{1550}<-2$                            & 38 & 0.082 $\pm$ 0.008 & 0.546 & -1.716 $\pm$ 0.027 & -21.515 $\pm$ 0.013 & 0.097 $\pm$ 0.015 & 0.455 $\pm$ 0.085 & 1.315 $\pm$ 0.061 & 54.375 $\pm$ 0.091 \\
EW H$\beta$ $>100$ \AA                       & 31 & 0.150 $\pm$ 0.016 & 0.726 & -1.499 $\pm$ 0.030 & -20.903 $\pm$ 0.013 & 0.342 $\pm$ 0.043 & 0.497 $\pm$ 0.073 & 1.669 $\pm$ 0.049 & 54.387 $\pm$ 0.076 \\
EW H$\beta$ $<100$ \AA                       & 58 & 0.009 $\pm$ 0.001 & 0.470 & -1.319 $\pm$ 0.021 & -22.243 $\pm$ 0.015 & 0.000             & 0.154 $\pm$ 0.059 & 0.927 $\pm$ 0.059 & 54.213 $\pm$ 0.201 \\
EW Ly$\alpha$ $>50$ \AA                      & 44 & 0.112 $\pm$ 0.013 & 0.723 & -1.492 $\pm$ 0.029 & -21.051 $\pm$ 0.017 & 0.142 $\pm$ 0.037 & 0.363 $\pm$ 0.116 & 1.396 $\pm$ 0.064 & 54.388 $\pm$ 0.182 \\
EW Ly$\alpha$ $<50$ \AA                      & 45 & 0.004 $\pm$ 0.001 & 0.363 & -1.192 $\pm$ 0.022 & -22.353 $\pm$ 0.016 & 0.000             & 0.142 $\pm$ 0.062 & 0.983 $\pm$ 0.061 & 54.211 $\pm$ 0.226 \\
\fesclya$>0.2$                               & 42 & 0.068 $\pm$ 0.008 & 0.666 & -1.563 $\pm$ 0.024 & -21.457 $\pm$ 0.014 & 0.084 $\pm$ 0.019 & 0.391 $\pm$ 0.101 & 1.218 $\pm$ 0.064 & 54.348 $\pm$ 0.143 \\
\fesclya$<0.2$                               & 47 & 0.006 $\pm$ 0.001 & 0.413 & -1.072 $\pm$ 0.022 & -22.200 $\pm$ 0.011 & 0.019 $\pm$ 0.004 & 0.201 $\pm$ 0.052 & 1.067 $\pm$ 0.057 & 54.164 $\pm$ 0.133 \\
\muv$>-19.5$                                 & 31 & 0.120 $\pm$ 0.018 & 0.747 & -1.498 $\pm$ 0.036 & -20.791 $\pm$ 0.019 & 0.222 $\pm$ 0.051 & 0.359 $\pm$ 0.094 & 1.584 $\pm$ 0.083 & 54.385 $\pm$ 0.135 \\
\muv$<-19.5$                                 & 58 & 0.011 $\pm$ 0.001 & 0.409 & -1.315 $\pm$ 0.018 & -22.311 $\pm$ 0.013 & 0.000             & 0.280 $\pm$ 0.098 & 1.164 $\pm$ 0.053 & 54.304 $\pm$ 0.194 \\
$\rm M_\star>10^9~M_\odot$                   & 36 & 0.004 $\pm$ 0.001 & 0.347 & -1.166 $\pm$ 0.027 & -22.678 $\pm$ 0.015 & 0.000             & 0.018 $\pm$ 0.009 & 0.975 $\pm$ 0.071 & 54.160 $\pm$ 0.132 \\
$\rm M_\star<10^9~M_\odot$                   & 52 & 0.061 $\pm$ 0.006 & 0.734 & -1.583 $\pm$ 0.025 & -21.384 $\pm$ 0.015 & 0.123 $\pm$ 0.030 & 0.362 $\pm$ 0.122 & 1.181 $\pm$ 0.062 & 54.336 $\pm$ 0.179 \\
\orat$>5$                                    & 29 & 0.216 $\pm$ 0.030 & 0.766 & -1.644 $\pm$ 0.034 & -20.761 $\pm$ 0.018 & 0.195 $\pm$ 0.061 & 0.277 $\pm$ 0.119 & 1.512 $\pm$ 0.073 & 54.319 $\pm$ 0.232 \\
\orat$<5$                                    & 60 & 0.009 $\pm$ 0.001 & 0.506 & -1.295 $\pm$ 0.021 & -22.212 $\pm$ 0.016 & 0.012 $\pm$ 0.004 & 0.292 $\pm$ 0.106 & 0.954 $\pm$ 0.057 & 54.272 $\pm$ 0.201 \\
\logoh$>8$                                   & 62 & 0.013 $\pm$ 0.001 & 0.474 & -1.310 $\pm$ 0.022 & -22.080 $\pm$ 0.015 & 0.018 $\pm$ 0.005 & 0.106 $\pm$ 0.043 & 0.896 $\pm$ 0.058 & 54.209 $\pm$ 0.191 \\
\logoh$<8$                                   & 27 & 0.139 $\pm$ 0.021 & 0.743 & -1.675 $\pm$ 0.032 & -20.987 $\pm$ 0.015 & 0.186 $\pm$ 0.047 & 0.481 $\pm$ 0.154 & 1.329 $\pm$ 0.065 & 54.386 $\pm$ 0.173 \\
$r_{50}>0.6~\rm kpc$                         & 41 & 0.006 $\pm$ 0.002 & 0.512 & -1.310 $\pm$ 0.026 & -21.959 $\pm$ 0.017 & 0.000             & 0.297 $\pm$ 0.117 & 0.961 $\pm$ 0.067 & 54.282 $\pm$ 0.220 \\
$r_{50}<0.6~\rm kpc$                         & 48 & 0.049 $\pm$ 0.005 & 0.543 & -1.404 $\pm$ 0.020 & -21.602 $\pm$ 0.010 & 0.168 $\pm$ 0.029 & 0.417 $\pm$ 0.098 & 1.340 $\pm$ 0.059 & 54.371 $\pm$ 0.115 \\
\sigsfr$\rm >10~\frac{M_{\sun}}{yr~kpc^{2}}$ & 32 & 0.018 $\pm$ 0.002 & 0.264 & -1.306 $\pm$ 0.027 & -22.519 $\pm$ 0.014 & 0.098 $\pm$ 0.021 & 0.300 $\pm$ 0.079 & 1.416 $\pm$ 0.071 & 54.355 $\pm$ 0.135 \\
\sigsfr$\rm <10~\frac{M_{\sun}}{yr~kpc^{2}}$ & 56 & 0.019 $\pm$ 0.004 & 0.613 & -1.427 $\pm$ 0.024 & -21.668 $\pm$ 0.013 & 0.025 $\pm$ 0.007 & 0.223 $\pm$ 0.076 & 0.895 $\pm$ 0.065 & 54.246 $\pm$ 0.177 \\
\hline
\\[-8pt]
\multicolumn{14}{c}{\bf LCEs} \\[2pt]
\hline
$\beta_{1550}>-2$                            & 22 & 0.023 $\pm$ 0.004 & 0.564 & -1.265 $\pm$ 0.030 & -22.174 $\pm$ 0.011 & 0.000             & 0.537 $\pm$ 0.148 & 0.899 $\pm$ 0.092 & 54.352 $\pm$ 0.142 \\
$\beta_{1550}<-2$                            & 28 & 0.107 $\pm$ 0.012 & 0.458 & -1.738 $\pm$ 0.029 & -21.642 $\pm$ 0.015 & 0.128 $\pm$ 0.029 & 0.382 $\pm$ 0.121 & 1.271 $\pm$ 0.066 & 54.386 $\pm$ 0.166 \\
EW H$\beta$ $>100$ \AA                       & 20 & 0.270 $\pm$ 0.029 & 0.693 & -1.673 $\pm$ 0.030 & -20.883 $\pm$ 0.015 & 0.221 $\pm$ 0.046 & 0.446 $\pm$ 0.113 & 1.528 $\pm$ 0.057 & 54.368 $\pm$ 0.134 \\
EW H$\beta$ $<100$ \AA                       & 30 & 0.017 $\pm$ 0.003 & 0.248 & -1.397 $\pm$ 0.029 & -22.529 $\pm$ 0.010 & 0.025 $\pm$ 0.006 & 0.550 $\pm$ 0.165 & 0.853 $\pm$ 0.077 & 54.356 $\pm$ 0.153 \\
EW Ly$\alpha$ $>50$ \AA                      & 31 & 0.132 $\pm$ 0.011 & 0.647 & -1.540 $\pm$ 0.029 & -21.348 $\pm$ 0.014 & 0.179 $\pm$ 0.029 & 0.467 $\pm$ 0.090 & 1.442 $\pm$ 0.063 & 54.359 $\pm$ 0.098 \\
EW Ly$\alpha$ $<50$ \AA                      & 19 & 0.014 $\pm$ 0.002 & 0.329 & -1.454 $\pm$ 0.030 & -22.655 $\pm$ 0.011 & 0.000             & 0.511 $\pm$ 0.174 & 0.840 $\pm$ 0.084 & 54.356 $\pm$ 0.179 \\
\fesclya$>0.2$                               & 29 & 0.097 $\pm$ 0.011 & 0.554 & -1.573 $\pm$ 0.030 & -21.535 $\pm$ 0.014 & 0.080 $\pm$ 0.016 & 0.477 $\pm$ 0.114 & 1.143 $\pm$ 0.066 & 54.370 $\pm$ 0.122 \\
\fesclya$<0.2$                               & 21 & 0.011 $\pm$ 0.002 & 0.190 & -1.323 $\pm$ 0.032 & -22.716 $\pm$ 0.018 & 0.008 $\pm$ 0.003 & 0.205 $\pm$ 0.103 & 1.195 $\pm$ 0.084 & 54.185 $\pm$ 0.260 \\
\muv$>-19.5$                                 & 16 & 0.215 $\pm$ 0.029 & 0.632 & -1.505 $\pm$ 0.041 & -20.849 $\pm$ 0.021 & 0.282 $\pm$ 0.061 & 0.437 $\pm$ 0.116 & 1.609 $\pm$ 0.073 & 54.387 $\pm$ 0.145 \\
\muv$<-19.5$                                 & 34 & 0.022 $\pm$ 0.003 & 0.301 & -1.403 $\pm$ 0.025 & -22.394 $\pm$ 0.010 & 0.115 $\pm$ 0.022 & 0.520 $\pm$ 0.141 & 0.997 $\pm$ 0.063 & 54.367 $\pm$ 0.135 \\
$\rm M_\star>10^9~M_\odot$                   & 18 & 0.018 $\pm$ 0.003 & 0.392 & -1.367 $\pm$ 0.031 & -22.728 $\pm$ 0.012 & 0.000 $\pm$ 0.000 & 0.430 $\pm$ 0.079 & 1.133 $\pm$ 0.077 & 54.353 $\pm$ 0.093 \\
$\rm M_\star<10^9~M_\odot$                   & 31 & 0.101 $\pm$ 0.011 & 0.596 & -1.690 $\pm$ 0.031 & -21.430 $\pm$ 0.015 & 0.170 $\pm$ 0.038 & 0.426 $\pm$ 0.118 & 1.297 $\pm$ 0.064 & 54.362 $\pm$ 0.146 \\
\orat$>5$                                    & 21 & 0.374 $\pm$ 0.043 & 0.748 & -1.719 $\pm$ 0.027 & -20.693 $\pm$ 0.011 & 0.171 $\pm$ 0.044 & 0.246 $\pm$ 0.087 & 1.542 $\pm$ 0.059 & 54.313 $\pm$ 0.192 \\
\orat$<5$                                    & 29 & 0.017 $\pm$ 0.002 & 0.313 & -1.416 $\pm$ 0.028 & -22.499 $\pm$ 0.016 & 0.021 $\pm$ 0.007 & 0.320 $\pm$ 0.143 & 0.904 $\pm$ 0.082 & 54.327 $\pm$ 0.234 \\
\logoh$>8$                                   & 30 & 0.019 $\pm$ 0.003 & 0.284 & -1.381 $\pm$ 0.025 & -22.416 $\pm$ 0.017 & 0.006 $\pm$ 0.002 & 0.214 $\pm$ 0.098 & 1.049 $\pm$ 0.083 & 54.301 $\pm$ 0.234 \\
\logoh$<8$                                   & 20 & 0.186 $\pm$ 0.026 & 0.693 & -1.659 $\pm$ 0.031 & -21.053 $\pm$ 0.017 & 0.133 $\pm$ 0.040 & 0.344 $\pm$ 0.136 & 1.441 $\pm$ 0.073 & 54.357 $\pm$ 0.226 \\
$r_{50}>0.6~\rm kpc$                         & 18 & 0.029 $\pm$ 0.006 & 0.725 & -1.681 $\pm$ 0.033 & -21.995 $\pm$ 0.017 & 0.000             & 0.223 $\pm$ 0.085 & 0.560 $\pm$ 0.064 & 54.236 $\pm$ 0.193 \\
$r_{50}<0.6~\rm kpc$                         & 32 & 0.070 $\pm$ 0.006 & 0.501 & -1.460 $\pm$ 0.030 & -21.837 $\pm$ 0.013 & 0.205 $\pm$ 0.038 & 0.429 $\pm$ 0.090 & 1.465 $\pm$ 0.061 & 54.362 $\pm$ 0.114 \\
\sigsfr$\rm >10~\frac{M_{\sun}}{yr~kpc^{2}}$ & 27 & 0.024 $\pm$ 0.003 & 0.216 & -1.367 $\pm$ 0.027 & -22.523 $\pm$ 0.012 & 0.074 $\pm$ 0.015 & 0.373 $\pm$ 0.105 & 1.338 $\pm$ 0.083 & 54.364 $\pm$ 0.141 \\
\sigsfr$\rm <10~\frac{M_{\sun}}{yr~kpc^{2}}$ & 23 & 0.082 $\pm$ 0.010 & 0.712 & -1.595 $\pm$ 0.031 & -21.532 $\pm$ 0.015 & 0.000             & 0.478 $\pm$ 0.100 & 1.112 $\pm$ 0.071 & 54.358 $\pm$ 0.112 \\
all                                          & 50 & 0.052 $\pm$ 0.005 & 0.585 & -1.499 $\pm$ 0.024 & -21.880 $\pm$ 0.011 & 0.107 $\pm$ 0.018 & 0.484 $\pm$ 0.116 & 1.152 $\pm$ 0.060 & 54.366 $\pm$ 0.120 \\
\hline\\[-8pt]
\multicolumn{14}{c}{\bf non-detections} \\[2pt]
\hline
$\beta_{1550}>-2$                            & 29 & 0.002 $\pm$ 0.002 & 0.000 & -1.131 $\pm$ 0.031 & -22.022 $\pm$ 0.016 & 0.032 $\pm$ 0.009 & 0.017 $\pm$ 0.009 & 1.015 $\pm$ 0.077 & 54.142 $\pm$ 0.166 \\
$\beta_{1550}<-2$                            & 10 & 0.022 $\pm$ 0.007 & 0.000 & -1.427 $\pm$ 0.054 & -21.128 $\pm$ 0.025 & 0.064 $\pm$ 0.024 & 0.045 $\pm$ 0.022 & 1.790 $\pm$ 0.095 & 54.217 $\pm$ 0.284 \\
EW H$\beta$ $>100$ \AA                       & 11 & 0.011 $\pm$ 0.007 & 0.000 & -1.114 $\pm$ 0.074 & -21.019 $\pm$ 0.028 & 0.118 $\pm$ 0.058 & 0.020 $\pm$ 0.020 & 1.482 $\pm$ 0.117 & 54.166 $\pm$ 0.288 \\
EW H$\beta$ $<100$ \AA                       & 28 & 0.001 $\pm$ 0.001 & 0.000 & -1.209 $\pm$ 0.032 & -22.069 $\pm$ 0.017 & 0.039 $\pm$ 0.009 & 0.052 $\pm$ 0.015 & 0.987 $\pm$ 0.084 & 54.186 $\pm$ 0.138 \\
EW Ly$\alpha$ $>50$ \AA                      & 13 & 0.030 $\pm$ 0.010 & 0.000 & -1.337 $\pm$ 0.057 & -20.896 $\pm$ 0.027 & 0.238 $\pm$ 0.061 & 0.038 $\pm$ 0.007 & 1.734 $\pm$ 0.092 & 54.195 $\pm$ 0.154 \\
EW Ly$\alpha$ $<50$ \AA                      & 26 & 0.001 $\pm$ 0.001 & 0.000 & -1.163 $\pm$ 0.031 & -22.145 $\pm$ 0.016 & 0.029 $\pm$ 0.007 & 0.019 $\pm$ 0.003 & 1.000 $\pm$ 0.080 & 54.145 $\pm$ 0.120 \\
\fesclya$>0.2$                               & 13 & 0.037 $\pm$ 0.019 & 0.000 & -1.802 $\pm$ 0.051 & -21.190 $\pm$ 0.024 & 0.000             & 0.479 $\pm$ 0.206 & 0.398 $\pm$ 0.098 & 54.355 $\pm$ 0.223 \\
\fesclya$<0.2$                               & 26 & 0.001 $\pm$ 0.001 & 0.000 & -0.982 $\pm$ 0.029 & -22.052 $\pm$ 0.011 & 0.010 $\pm$ 0.003 & 0.014 $\pm$ 0.002 & 0.997 $\pm$ 0.067 & 54.136 $\pm$ 0.077 \\
\muv$>-19.5$                                 & 15 & 0.010 $\pm$ 0.010 & 0.000 & -1.204 $\pm$ 0.054 & -20.877 $\pm$ 0.027 & 0.110 $\pm$ 0.033 & 0.020 $\pm$ 0.005 & 1.296 $\pm$ 0.109 & 54.160 $\pm$ 0.206 \\
\muv$<-19.5$                                 & 24 & 0.001 $\pm$ 0.001 & 0.000 & -1.146 $\pm$ 0.026 & -22.309 $\pm$ 0.013 & 0.000             & 0.021 $\pm$ 0.012 & 1.147 $\pm$ 0.071 & 54.141 $\pm$ 0.181 \\
$\rm M_\star>10^9~M_\odot$                   & 18 & 0.001 $\pm$ 0.001 & 0.000 & -1.118 $\pm$ 0.039 & -22.325 $\pm$ 0.017 & 0.176 $\pm$ 0.020 & 0.679 $\pm$ 0.101 & 1.014 $\pm$ 0.059 & 54.465 $\pm$ 0.072 \\
$\rm M_\star<10^9~M_\odot$                   & 21 & 0.010 $\pm$ 0.006 & 0.000 & -1.411 $\pm$ 0.036 & -21.087 $\pm$ 0.019 & 0.127 $\pm$ 0.035 & 0.030 $\pm$ 0.012 & 1.350 $\pm$ 0.082 & 54.176 $\pm$ 0.197 \\
\orat$>5$                                    &  8 & 0.021 $\pm$ 0.021 & 0.000 & -1.377 $\pm$ 0.118 & -20.612 $\pm$ 0.039 & 0.287 $\pm$ 0.079 & 0.234 $\pm$ 0.053 & 1.962 $\pm$ 0.114 & 54.426 $\pm$ 0.110 \\
\orat$<5$                                    & 31 & 0.002 $\pm$ 0.002 & 0.000 & -1.204 $\pm$ 0.028 & -22.009 $\pm$ 0.014 & 0.010 $\pm$ 0.003 & 0.022 $\pm$ 0.013 & 1.044 $\pm$ 0.070 & 54.144 $\pm$ 0.187 \\
\logoh$>8$                                   & 32 & 0.002 $\pm$ 0.002 & 0.000 & -1.244 $\pm$ 0.030 & -21.901 $\pm$ 0.016 & 0.000             & 0.044 $\pm$ 0.021 & 0.830 $\pm$ 0.073 & 54.175 $\pm$ 0.183 \\
\logoh$<8$                                   &  7 & 0.015 $\pm$ 0.015 & 0.000 & -1.010 $\pm$ 0.029 & -20.955 $\pm$ 0.010 & 0.291 $\pm$ 0.071 & 0.412 $\pm$ 0.106 & 1.970 $\pm$ 0.056 & 54.399 $\pm$ 0.138 \\
$r_{50}>0.6~\rm kpc$                         & 23 & 0.002 $\pm$ 0.002 & 0.000 & -1.036 $\pm$ 0.036 & -22.015 $\pm$ 0.022 & 0.061 $\pm$ 0.020 & 0.330 $\pm$ 0.131 & 1.336 $\pm$ 0.091 & 54.390 $\pm$ 0.230 \\
$r_{50}<0.6~\rm kpc$                         & 16 & 0.007 $\pm$ 0.004 & 0.000 & -1.361 $\pm$ 0.039 & -21.349 $\pm$ 0.017 & 0.034 $\pm$ 0.011 & 0.027 $\pm$ 0.006 & 1.316 $\pm$ 0.112 & 54.160 $\pm$ 0.181 \\
\sigsfr$\rm >10~\frac{M_{\sun}}{yr~kpc^{2}}$ &  5 & 0.000             & 0.000 & -0.242 $\pm$ 0.052 & -21.833 $\pm$ 0.022 & 0.309 $\pm$ 0.111 & 0.233 $\pm$ 0.073 & 2.420 $\pm$ 0.101 & 54.393 $\pm$ 0.150 \\
\sigsfr$\rm <10~\frac{M_{\sun}}{yr~kpc^{2}}$ & 33 & 0.002 $\pm$ 0.002 & 0.000 & -1.298 $\pm$ 0.029 & -21.788 $\pm$ 0.015 & 0.000             & 0.023 $\pm$ 0.003 & 0.853 $\pm$ 0.069 & 54.146 $\pm$ 0.102 \\
all                                          & 39 & 0.002 $\pm$ 0.002 & 0.000 & -1.226 $\pm$ 0.025 & -21.829 $\pm$ 0.011 & 0.029 $\pm$ 0.007 & 0.020 $\pm$ 0.003 & 0.959 $\pm$ 0.067 & 54.145 $\pm$ 0.096 \\
\hline\\[-8pt]
\multicolumn{14}{c}{\bf non-LCEs: \fesc$<0.01$} \\[2pt]
\hline
$\beta_{1550}>-2$                            & 33 & 0.001 $\pm$       & 0.538 & -1.198 $\pm$ 0.030 & -22.049 $\pm$ 0.014 & 0.014 $\pm$ 0.004 & 0.020 $\pm$ 0.003 & 0.965 $\pm$ 0.078 & 54.143 $\pm$ 0.113 \\
$\beta_{1550}<-2$                            & 10 & 0.022 $\pm$ 0.007 & 0.000 & -1.425 $\pm$ 0.057 & -21.133 $\pm$ 0.026 & 0.073 $\pm$ 0.027 & 0.042 $\pm$ 0.017 & 1.765 $\pm$ 0.105 & 54.209 $\pm$ 0.235 \\
EW H$\beta$ $>100$ \AA                       & 12 & 0.025 $\pm$ 0.011 & 1.000 & -1.476 $\pm$ 0.048 & -20.935 $\pm$ 0.023 & 0.119 $\pm$ 0.025 & 0.038 $\pm$ 0.006 & 1.309 $\pm$ 0.109 & 54.176 $\pm$ 0.121 \\
EW H$\beta$ $<100$ \AA                       & 31 & 0.001 $\pm$ 0.001 & 0.558 & -1.228 $\pm$ 0.029 & -22.037 $\pm$ 0.013 & 0.000             & 0.019 $\pm$ 0.007 & 0.973 $\pm$ 0.082 & 54.143 $\pm$ 0.129 \\
EW Ly$\alpha$ $>50$ \AA                      & 14 & 0.034 $\pm$ 0.012 & 1.000 & -1.369 $\pm$ 0.064 & -20.947 $\pm$ 0.027 & 0.082 $\pm$ 0.040 & 0.250 $\pm$ 0.138 & 1.515 $\pm$ 0.120 & 54.365 $\pm$ 0.319 \\
EW Ly$\alpha$ $<50$ \AA                      & 29 & 0.001 $\pm$       & 0.564 & -1.199 $\pm$ 0.026 & -22.159 $\pm$ 0.014 & 0.000             & 0.020 $\pm$ 0.003 & 0.880 $\pm$ 0.072 & 54.148 $\pm$ 0.109 \\
\fesclya$>0.2$                               & 14 & 0.027 $\pm$ 0.014 & 1.000 & -1.737 $\pm$ 0.051 & -21.207 $\pm$ 0.030 & 0.000             & 0.177 $\pm$ 0.114 & 0.494 $\pm$ 0.098 & 54.280 $\pm$ 0.327 \\
\fesclya$<0.2$                               & 29 & 0.001 $\pm$       & 0.595 & -0.952 $\pm$ 0.030 & -22.082 $\pm$ 0.011 & 0.040 $\pm$ 0.008 & 0.014 $\pm$ 0.001 & 1.029 $\pm$ 0.059 & 54.139 $\pm$ 0.064 \\
\muv$>-19.5$                                 & 15 & 0.010 $\pm$       & 0.000 & -1.189 $\pm$ 0.051 & -20.881 $\pm$ 0.026 & 0.122 $\pm$ 0.035 & 0.021 $\pm$ 0.005 & 1.317 $\pm$ 0.102 & 54.161 $\pm$ 0.197 \\
\muv$<-19.5$                                 & 28 & 0.002 $\pm$ 0.001 & 0.527 & -1.225 $\pm$ 0.026 & -22.238 $\pm$ 0.014 & 0.000             & 0.038 $\pm$ 0.022 & 0.997 $\pm$ 0.067 & 54.152 $\pm$ 0.213 \\
$\rm M_\star>10^9~M_\odot$                   & 21 & 0.001 $\pm$ 0.001 & 0.595 & -1.005 $\pm$ 0.034 & -22.400 $\pm$ 0.022 & 0.047 $\pm$ 0.015 & 0.306 $\pm$ 0.138 & 0.847 $\pm$ 0.091 & 54.327 $\pm$ 0.240 \\
$\rm M_\star<10^9~M_\odot$                   & 22 & 0.009 $\pm$ 0.005 & 1.000 & -1.406 $\pm$ 0.037 & -21.289 $\pm$ 0.018 & 0.065 $\pm$ 0.018 & 0.029 $\pm$ 0.006 & 1.269 $\pm$ 0.091 & 54.166 $\pm$ 0.171 \\
\orat$>5$                                    &  8 & 0.030 $\pm$       & 0.000 & -1.665 $\pm$ 0.093 & -20.506 $\pm$ 0.029 & 0.299 $\pm$ 0.052 & 0.494 $\pm$ 0.101 & 1.743 $\pm$ 0.095 & 54.473 $\pm$ 0.106 \\
\orat$<5$                                    & 35 & 0.001 $\pm$       & 0.536 & -1.222 $\pm$ 0.026 & -22.037 $\pm$ 0.013 & 0.000             & 0.020 $\pm$ 0.005 & 0.895 $\pm$ 0.070 & 54.143 $\pm$ 0.107 \\
\logoh$>8$                                   & 36 & 0.002 $\pm$ 0.001 & 0.521 & -1.215 $\pm$ 0.026 & -21.993 $\pm$ 0.013 & 0.000             & 0.018 $\pm$ 0.005 & 0.818 $\pm$ 0.072 & 54.151 $\pm$ 0.103 \\
\logoh$<8$                                   &  7 & 0.015 $\pm$       & 0.000 & -1.013 $\pm$ 0.029 & -20.954 $\pm$ 0.010 & 0.295 $\pm$ 0.070 & 0.418 $\pm$ 0.103 & 1.967 $\pm$ 0.056 & 54.399 $\pm$ 0.129 \\
$r_{50}>0.6~\rm kpc$                         & 27 & 0.002 $\pm$ 0.001 & 0.620 & -1.109 $\pm$ 0.032 & -22.020 $\pm$ 0.021 & 0.023 $\pm$ 0.007 & 0.017 $\pm$ 0.011 & 1.124 $\pm$ 0.093 & 54.164 $\pm$ 0.219 \\
$r_{50}<0.6~\rm kpc$                         & 16 & 0.007 $\pm$ 0.004 & 0.000 & -1.334 $\pm$ 0.037 & -21.356 $\pm$ 0.017 & 0.032 $\pm$ 0.011 & 0.026 $\pm$ 0.006 & 1.340 $\pm$ 0.113 & 54.159 $\pm$ 0.184 \\
\sigsfr$\rm >10~\frac{M_{\sun}}{yr~kpc^{2}}$ &  4 & 0.005 $\pm$ 0.002 & 0.524 & -1.353 $\pm$ 0.039 & -21.882 $\pm$ 0.014 & 0.000             & 0.206 $\pm$ 0.058 & 0.880 $\pm$ 0.153 & 54.383 $\pm$ 0.128 \\
\sigsfr$\rm <10~\frac{M_{\sun}}{yr~kpc^{2}}$ & 39 & 0.002 $\pm$       & 0.766 & -1.348 $\pm$ 0.029 & -21.757 $\pm$ 0.015 & 0.000             & 0.023 $\pm$ 0.006 & 0.726 $\pm$ 0.061 & 54.149 $\pm$ 0.110 \\
detections                                   &  7 & 0.000             & 0.438 & -1.519 $\pm$ 0.085 & -22.715 $\pm$ 0.030 & 0.000             & 0.292 $\pm$ 0.201 & 1.106 $\pm$ 0.185 & 54.358 $\pm$ 0.411 \\
non-detections                               & 35 & 0.002 $\pm$       & 0.000 & -1.208 $\pm$ 0.025 & -21.834 $\pm$ 0.011 & 0.035 $\pm$ 0.008 & 0.020 $\pm$ 0.003 & 1.003 $\pm$ 0.066 & 54.145 $\pm$ 0.098 \\
all                                          & 43 & 0.002 $\pm$ 0.002 & 0.436 & -1.248 $\pm$ 0.024 & -21.818 $\pm$ 0.010 & 0.000             & 0.021 $\pm$ 0.003 & 0.893 $\pm$ 0.062 & 54.143 $\pm$ 0.088 \\
\hline\\[-8pt]
\multicolumn{14}{c}{\bf weak LCEs: \fesc$\in[0.01,0.05]$} \\[2pt]
\hline
$\beta_{1550}>-2$                            & 16 & 0.038 $\pm$ 0.006 & 0.909 & -1.301 $\pm$ 0.041 & -21.860 $\pm$ 0.015 & 0.000             & 0.458 $\pm$ 0.180 & 0.948 $\pm$ 0.103 & 54.353 $\pm$ 0.204 \\
$\beta_{1550}<-2$                            &  9 & 0.010 $\pm$ 0.003 & 0.080 & -1.635 $\pm$ 0.028 & -22.327 $\pm$ 0.016 & 0.000             & 0.062 $\pm$ 0.024 & 0.607 $\pm$ 0.033 & 54.188 $\pm$ 0.127 \\
EW H$\beta$ $>100$ \AA                       &  6 & 0.083 $\pm$ 0.023 & 1.000 & -1.282 $\pm$ 0.041 & -21.088 $\pm$ 0.016 & 0.169 $\pm$ 0.043 & 0.251 $\pm$ 0.065 & 1.953 $\pm$ 0.090 & 54.365 $\pm$ 0.122 \\
EW H$\beta$ $<100$ \AA                       & 19 & 0.014 $\pm$ 0.004 & 0.167 & -1.375 $\pm$ 0.030 & -22.380 $\pm$ 0.011 & 0.000             & 0.440 $\pm$ 0.120 & 1.207 $\pm$ 0.076 & 54.352 $\pm$ 0.142 \\
EW Ly$\alpha$ $>50$ \AA                      & 14 & 0.039 $\pm$ 0.007 & 0.741 & -1.253 $\pm$ 0.038 & -21.491 $\pm$ 0.019 & 0.014 $\pm$ 0.008 & 0.028 $\pm$ 0.029 & 1.306 $\pm$ 0.117 & 54.064 $\pm$ 0.394 \\
EW Ly$\alpha$ $<50$ \AA                      & 11 & 0.012 $\pm$ 0.002 & 0.081 & -1.328 $\pm$ 0.039 & -22.499 $\pm$ 0.017 & 0.000             & 0.382 $\pm$ 0.197 & 0.708 $\pm$ 0.119 & 54.327 $\pm$ 0.282 \\
\fesclya$>0.2$                               & 13 & 0.035 $\pm$ 0.008 & 0.654 & -1.484 $\pm$ 0.033 & -21.756 $\pm$ 0.014 & 0.003 $\pm$ 0.001 & 0.365 $\pm$ 0.092 & 1.100 $\pm$ 0.100 & 54.313 $\pm$ 0.138 \\
\fesclya$<0.2$                               & 12 & 0.010 $\pm$ 0.003 & 0.069 & -1.282 $\pm$ 0.039 & -22.646 $\pm$ 0.020 & 0.000             & 0.032 $\pm$ 0.009 & 1.556 $\pm$ 0.078 & 53.981 $\pm$ 0.190 \\
\muv$>-19.5$                                 &  6 & 0.073 $\pm$ 0.023 & 1.000 & -1.082 $\pm$ 0.101 & -21.113 $\pm$ 0.037 & 0.134 $\pm$ 0.037 & 0.066 $\pm$ 0.015 & 2.117 $\pm$ 0.102 & 53.907 $\pm$ 0.129 \\
\muv$<-19.5$                                 & 19 & 0.018 $\pm$ 0.003 & 0.275 & -1.376 $\pm$ 0.029 & -22.381 $\pm$ 0.011 & 0.005 $\pm$ 0.002 & 0.449 $\pm$ 0.135 & 1.050 $\pm$ 0.075 & 54.354 $\pm$ 0.162 \\
$\rm M_\star>10^9~M_\odot$                   & 10 & 0.019 $\pm$ 0.005 & 0.430 & -1.353 $\pm$ 0.040 & -22.558 $\pm$ 0.016 & 0.059 $\pm$ 0.014 & 0.374 $\pm$ 0.112 & 1.386 $\pm$ 0.077 & 54.391 $\pm$ 0.154 \\
$\rm M_\star<10^9~M_\odot$                   & 15 & 0.042 $\pm$ 0.009 & 0.800 & -1.593 $\pm$ 0.035 & -21.601 $\pm$ 0.021 & 0.071 $\pm$ 0.022 & 0.048 $\pm$ 0.027 & 1.493 $\pm$ 0.083 & 54.196 $\pm$ 0.242 \\
\orat$>5$                                    &  9 & 0.051 $\pm$ 0.014 & 1.000 & -1.089 $\pm$ 0.026 & -21.183 $\pm$ 0.011 & 0.156 $\pm$ 0.027 & 0.202 $\pm$ 0.043 & 2.025 $\pm$ 0.094 & 54.259 $\pm$ 0.102 \\
\orat$<5$                                    & 16 & 0.012 $\pm$ 0.002 & 0.059 & -1.277 $\pm$ 0.034 & -22.485 $\pm$ 0.013 & 0.016 $\pm$ 0.005 & 0.427 $\pm$ 0.145 & 1.106 $\pm$ 0.079 & 54.352 $\pm$ 0.180 \\
\logoh$>8$                                   & 15 & 0.021 $\pm$ 0.004 & 0.308 & -1.598 $\pm$ 0.029 & -22.351 $\pm$ 0.016 & 0.000             & 0.318 $\pm$ 0.092 & 0.827 $\pm$ 0.095 & 54.343 $\pm$ 0.145 \\
\logoh$<8$                                   & 10 & 0.068 $\pm$ 0.014 & 1.000 & -1.451 $\pm$ 0.039 & -21.268 $\pm$ 0.017 & 0.000             & 0.233 $\pm$ 0.108 & 1.263 $\pm$ 0.109 & 54.287 $\pm$ 0.261 \\
$r_{50}>0.6~\rm kpc$                         & 12 & 0.041 $\pm$ 0.008 & 0.987 & -1.667 $\pm$ 0.038 & -21.984 $\pm$ 0.015 & 0.010 $\pm$ 0.003 & 0.616 $\pm$ 0.184 & 0.659 $\pm$ 0.077 & 54.362 $\pm$ 0.149 \\
$r_{50}<0.6~\rm kpc$                         & 13 & 0.026 $\pm$ 0.005 & 0.505 & -1.404 $\pm$ 0.026 & -21.948 $\pm$ 0.011 & 0.000             & 0.141 $\pm$ 0.050 & 1.467 $\pm$ 0.148 & 54.302 $\pm$ 0.169 \\
\sigsfr$\rm >10~\frac{M_{\sun}}{yr~kpc^{2}}$ & 11 & 0.008 $\pm$ 0.003 & 0.056 & -1.173 $\pm$ 0.040 & -22.738 $\pm$ 0.018 & 0.067 $\pm$ 0.018 & 0.249 $\pm$ 0.073 & 1.593 $\pm$ 0.100 & 54.323 $\pm$ 0.148 \\
\sigsfr$\rm <10~\frac{M_{\sun}}{yr~kpc^{2}}$ & 14 & 0.030 $\pm$ 0.006 & 0.705 & -1.306 $\pm$ 0.035 & -21.715 $\pm$ 0.014 & 0.000             & 0.424 $\pm$ 0.082 & 1.251 $\pm$ 0.091 & 54.350 $\pm$ 0.103 \\
all                                          & 25 & 0.026 $\pm$ 0.004 & 0.470 & -1.468 $\pm$ 0.026 & -22.055 $\pm$ 0.013 & 0.000             & 0.239 $\pm$ 0.068 & 1.162 $\pm$ 0.082 & 54.261 $\pm$ 0.156 \\
\hline\\[-8pt]
\multicolumn{14}{c}{\bf strong LCEs: \fesc$>0.05$} \\[2pt]
\hline
$\beta_{1550}>-2$                            &  2 &        --         &  --   &         --         &          --         &        --         &        --         &        --         &         --         \\
$\beta_{1550}<-2$                            & 19 & 0.206 $\pm$ 0.022 & 0.586 & -1.672 $\pm$ 0.036 & -21.304 $\pm$ 0.018 & 0.268 $\pm$ 0.031 & 0.448 $\pm$ 0.057 & 1.552 $\pm$ 0.065 & 54.404 $\pm$ 0.067 \\
EW H$\beta$ $>100$ \AA                       & 13 & 0.410 $\pm$ 0.053 & 0.656 & -1.661 $\pm$ 0.026 & -20.824 $\pm$ 0.010 & 0.378 $\pm$ 0.026 & 0.550 $\pm$ 0.042 & 1.693 $\pm$ 0.060 & 54.397 $\pm$ 0.036 \\
EW H$\beta$ $<100$ \AA                       &  8 & 0.030 $\pm$ 0.004 & 0.067 & -1.313 $\pm$ 0.046 & -22.561 $\pm$ 0.018 & 0.011 $\pm$ 0.003 & 0.365 $\pm$ 0.099 & 1.347 $\pm$ 0.077 & 54.414 $\pm$ 0.134 \\
EW Ly$\alpha$ $>50$ \AA                      & 16 & 0.306 $\pm$ 0.036 & 0.623 & -1.701 $\pm$ 0.034 & -21.093 $\pm$ 0.016 & 0.299 $\pm$ 0.036 & 0.493 $\pm$ 0.064 & 1.587 $\pm$ 0.067 & 54.404 $\pm$ 0.071 \\
EW Ly$\alpha$ $<50$ \AA                      &  5 & 0.004 $\pm$ 0.001 & 0.019 & -1.190 $\pm$ 0.031 & -22.836 $\pm$ 0.017 & 0.000             & 0.103 $\pm$ 0.054 & 0.884 $\pm$ 0.067 & 54.243 $\pm$ 0.201 \\
\fesclya$>0.2$                               & 15 & 0.215 $\pm$ 0.026 & 0.484 & -1.693 $\pm$ 0.036 & -21.419 $\pm$ 0.015 & 0.234 $\pm$ 0.032 & 0.587 $\pm$ 0.100 & 1.288 $\pm$ 0.065 & 54.382 $\pm$ 0.088 \\
\fesclya$<0.2$                               &  6 & 0.018 $\pm$ 0.004 & 0.055 & -1.177 $\pm$ 0.061 & -22.425 $\pm$ 0.027 & 0.000             & 0.362 $\pm$ 0.176 & 1.119 $\pm$ 0.098 & 54.312 $\pm$ 0.249 \\
\muv$>-19.5$                                 & 10 & 0.287 $\pm$ 0.043 & 0.532 & -1.266 $\pm$ 0.065 & -20.735 $\pm$ 0.025 & 0.199 $\pm$ 0.070 & 0.267 $\pm$ 0.085 & 1.803 $\pm$ 0.108 & 54.376 $\pm$ 0.151 \\
\muv$<-19.5$                                 & 11 & 0.040 $\pm$ 0.007 & 0.067 & -1.400 $\pm$ 0.047 & -22.460 $\pm$ 0.021 & 0.212 $\pm$ 0.034 & 0.501 $\pm$ 0.088 & 1.366 $\pm$ 0.081 & 54.381 $\pm$ 0.088 \\
$\rm M_\star>10^9~M_\odot$                   &  5 & 0.000             & 0.017 & -0.991 $\pm$ 0.026 & -23.043 $\pm$ 0.013 & 0.000             & 0.023 $\pm$ 0.004 & 1.701 $\pm$ 0.045 & 53.763 $\pm$ 0.116 \\
$\rm M_\star<10^9~M_\odot$                   & 15 & 0.221 $\pm$ 0.025 & 0.559 & -1.752 $\pm$ 0.040 & -21.336 $\pm$ 0.019 & 0.233 $\pm$ 0.028 & 0.612 $\pm$ 0.084 & 1.276 $\pm$ 0.075 & 54.387 $\pm$ 0.069 \\
\orat$>5$                                    & 12 & 0.520 $\pm$ 0.077 & 0.773 & -1.698 $\pm$ 0.037 & -20.644 $\pm$ 0.019 & 0.272 $\pm$ 0.040 & 0.468 $\pm$ 0.077 & 1.559 $\pm$ 0.074 & 54.402 $\pm$ 0.086 \\
\orat$<5$                                    &  9 & 0.044 $\pm$ 0.006 & 0.090 & -1.542 $\pm$ 0.032 & -22.480 $\pm$ 0.014 & 0.202 $\pm$ 0.027 & 0.649 $\pm$ 0.079 & 1.185 $\pm$ 0.038 & 54.459 $\pm$ 0.055 \\
\logoh$>8$                                   & 11 & 0.041 $\pm$ 0.005 & 0.068 & -1.378 $\pm$ 0.049 & -22.433 $\pm$ 0.024 & 0.088 $\pm$ 0.029 & 0.427 $\pm$ 0.207 & 1.074 $\pm$ 0.099 & 54.349 $\pm$ 0.253 \\
\logoh$<8$                                   & 10 & 0.291 $\pm$ 0.052 & 0.523 & -1.612 $\pm$ 0.046 & -21.036 $\pm$ 0.018 & 0.223 $\pm$ 0.041 & 0.066 $\pm$ 0.009 & 1.658 $\pm$ 0.084 & 54.216 $\pm$ 0.099 \\
$r_{50}>0.6~\rm kpc$                         &  2 &        --         &  --   &         --         &          --         &        --         &        --         &        --         &         --         \\
$r_{50}<0.6~\rm kpc$                         & 19 & 0.160 $\pm$ 0.019 & 0.459 & -1.583 $\pm$ 0.042 & -21.622 $\pm$ 0.017 & 0.121 $\pm$ 0.021 & 0.432 $\pm$ 0.098 & 1.269 $\pm$ 0.085 & 54.409 $\pm$ 0.129 \\
\sigsfr$\rm >10~\frac{M_{\sun}}{yr~kpc^{2}}$ & 14 & 0.077 $\pm$ 0.010 & 0.234 & -1.520 $\pm$ 0.041 & -22.251 $\pm$ 0.019 & 0.187 $\pm$ 0.026 & 0.488 $\pm$ 0.073 & 1.264 $\pm$ 0.080 & 54.387 $\pm$ 0.073 \\
\sigsfr$\rm <10~\frac{M_{\sun}}{yr~kpc^{2}}$ &  7 & 0.322 $\pm$ 0.055 & 0.618 & -1.938 $\pm$ 0.050 & -21.272 $\pm$ 0.014 & 0.156 $\pm$ 0.038 & 0.659 $\pm$ 0.095 & 1.075 $\pm$ 0.121 & 54.388 $\pm$ 0.057 \\
all                                          & 21 & 0.134 $\pm$ 0.013 & 0.427 & -1.529 $\pm$ 0.038 & -21.640 $\pm$ 0.017 & 0.285 $\pm$ 0.040 & 0.521 $\pm$ 0.076 & 1.491 $\pm$ 0.063 & 54.387 $\pm$ 0.074
\enddata
\end{deluxetable*}
\end{longrotatetable}

\subsection{Method for Stacking SDSS Nebular Line Fluxes\label{sec:OptStacks}}

To complement the UV spectra stacks, we also stack the rest-optical emission line fluxes using the same UV flux normalizations as used for the \emph{HST}/COS spectra stacks. Following the methodology of \citet{2022ApJS..260....1F}, we subsequently deredden the spectra and compute flux ratios and direct-method oxygen abundances \logoh.

\subsection{A Note on Correlation Coefficients}

In our subsequent analysis of these stacks, we compute correlation coefficients and multivariate loading factors to determine how various properties may relate to one another. Above, we outlined the construction of 124 stacks of 89 galaxies with cuts along various criteria. Properties of these stacks are inherently correlated with one another since each galaxy can contribute to many different stacks. Thus, our reported trends and various correlation quantities ought to be considered with this caveat in mind. However, we note that uncorrelated stacks, specifically those built from different \fesc\ subsets with no other criteria, exhibit the same trends as the total set of stacks. We highlight these three uncorrelated stacks in figures below to emphasize that these trends are genuine despite the fact that many stacks correlate with one another.

\section{Properties from UV SEDs}\label{sec:specfit}

\begin{figure}
    \centering
    \includegraphics[width=\columnwidth,clip=true,trim={0.05in 0in 0.15in 0in}]{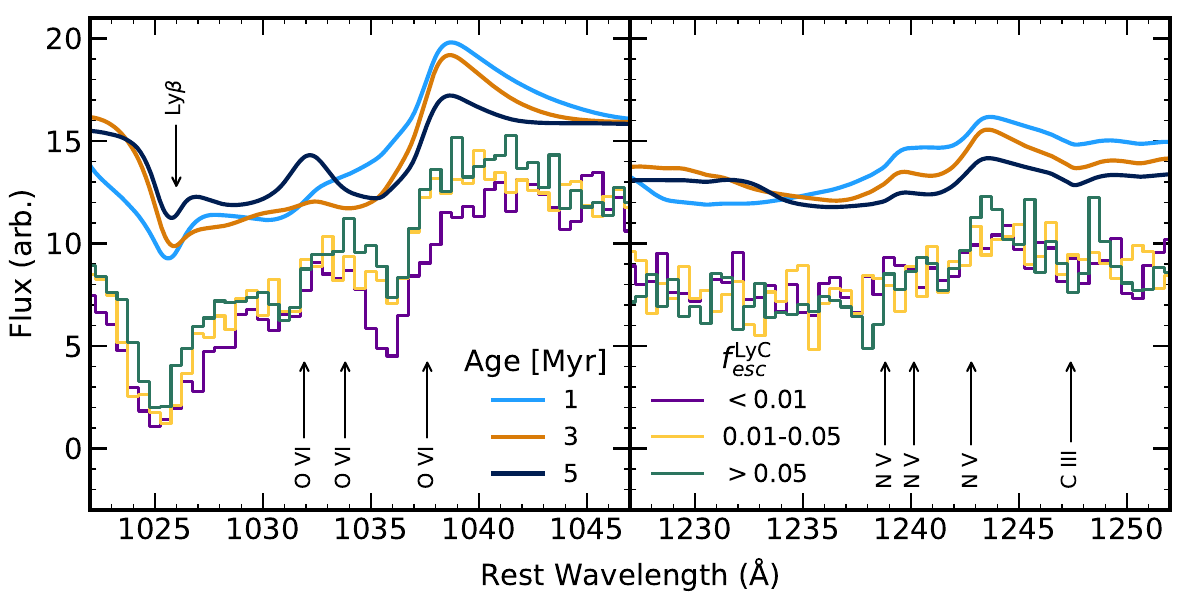}
    \caption{P Cygni profiles for \ion{O}{6} ({\it left}) and \ion{N}{5} ({\it right}) for stacks of the LzLCS+ on \fesc\ (steps) and for {\tt Starburst99} models at 10\% solar metallicity for different ages (lines). Models are attenuated by the \citet{2016ApJ...828..107R} curve assuming E$_{\rm B-V}$=0.05, convolved with the COS line-spread function (see Appendix \ref{apx:conv}), and arbitrarily offset for visualization.}
    \label{fig:pcyg}
\end{figure}

\begin{figure}
    \centering
    \includegraphics[width=\columnwidth,clip=true,trim={0.05in 0in 0.15in 0in}]{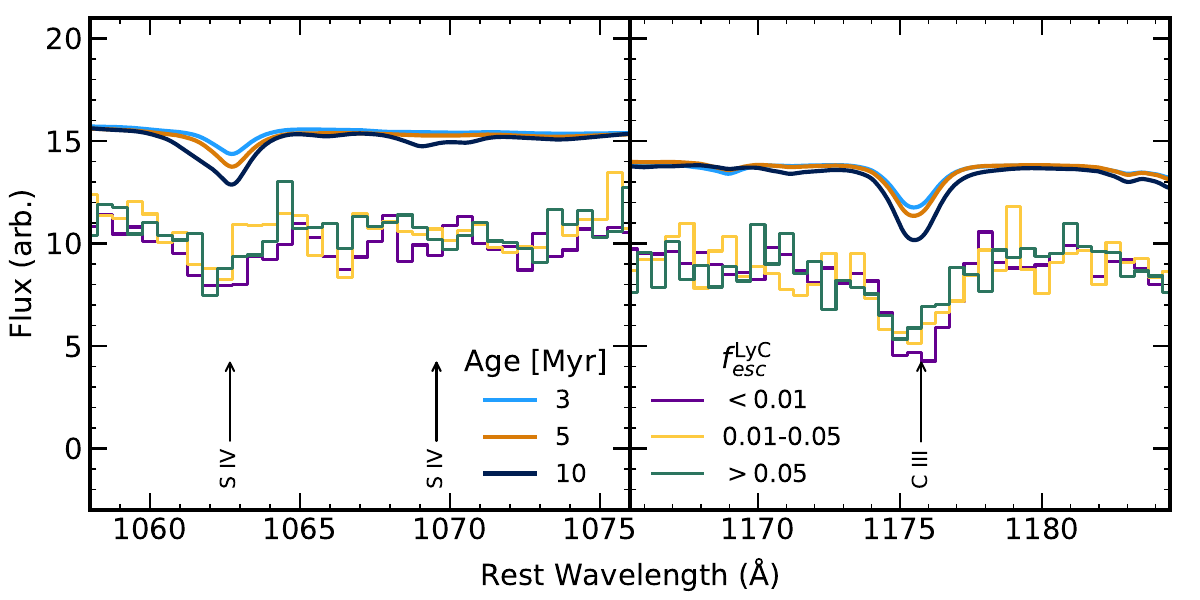}
    \caption{ Same as Figure \ref{fig:pcyg} but for photospheric lines \ion{S}{4} (\emph{left}) and \ion{C}{3} (\emph{right}) tracing slightly older populations.}
    \label{fig:phot}
\end{figure}

With stacking, we have increased signal in stellar features, which are key to uncovering the underlying stellar populations of LCEs and their counterparts. Critical features for constraining young stellar population ages and metallicities include the prominent \ion{O}{6} $\lambda\lambda1033,36,38$ and \ion{N}{5} $\lambda\lambda1239,40,43$ P Cygni wind features as well as the fainter \ion{P}{5} $\lambda\lambda1118,28$ resonance doublet \citep[cf.][]{2002ApJS..143..159P,2002ApJS..141..443W}. To illustrate the constraining power offered by the P Cygni features in our stacks, we compare stacks on different \fesc\ subsamples with {\tt Starburst99} templates convolved with the COS line spread function (LSF; see below and Appendix \ref{apx:conv}) in Figure \ref{fig:pcyg}. The prominence of the P Cygni features traces burst age well in the templates and is a clear signature in the stacks themselves. We illustrate these wind lines as age diagnostics in Figure \ref{fig:pcyg}, where the P Cygni profiles change both in model templates and in stacks of different \fesc\ subsets as the underlying population age changes (in this particular case, suggesting younger populations may be more prevalent in galaxies with higher \fesc). Other constraining features, particularly for slightly older stellar populations, include the \ion{S}{4} $\lambda1062$ and \ion{C}{3} $\lambda1175$ photospheric lines owing to their slightly lower ionization potentials (see Figure \ref{fig:phot}). Below, we elaborate on the use of the stacked UV SEDs to infer quantitative information about dust attenuation, stellar populations, and \fesc.

\subsection{SED Fitting with {\tt FiCUS}}

\begin{figure*}
    \centering
    \includegraphics[width=\linewidth,trim={0in 0.125in 0in 0.5in},clip=True]{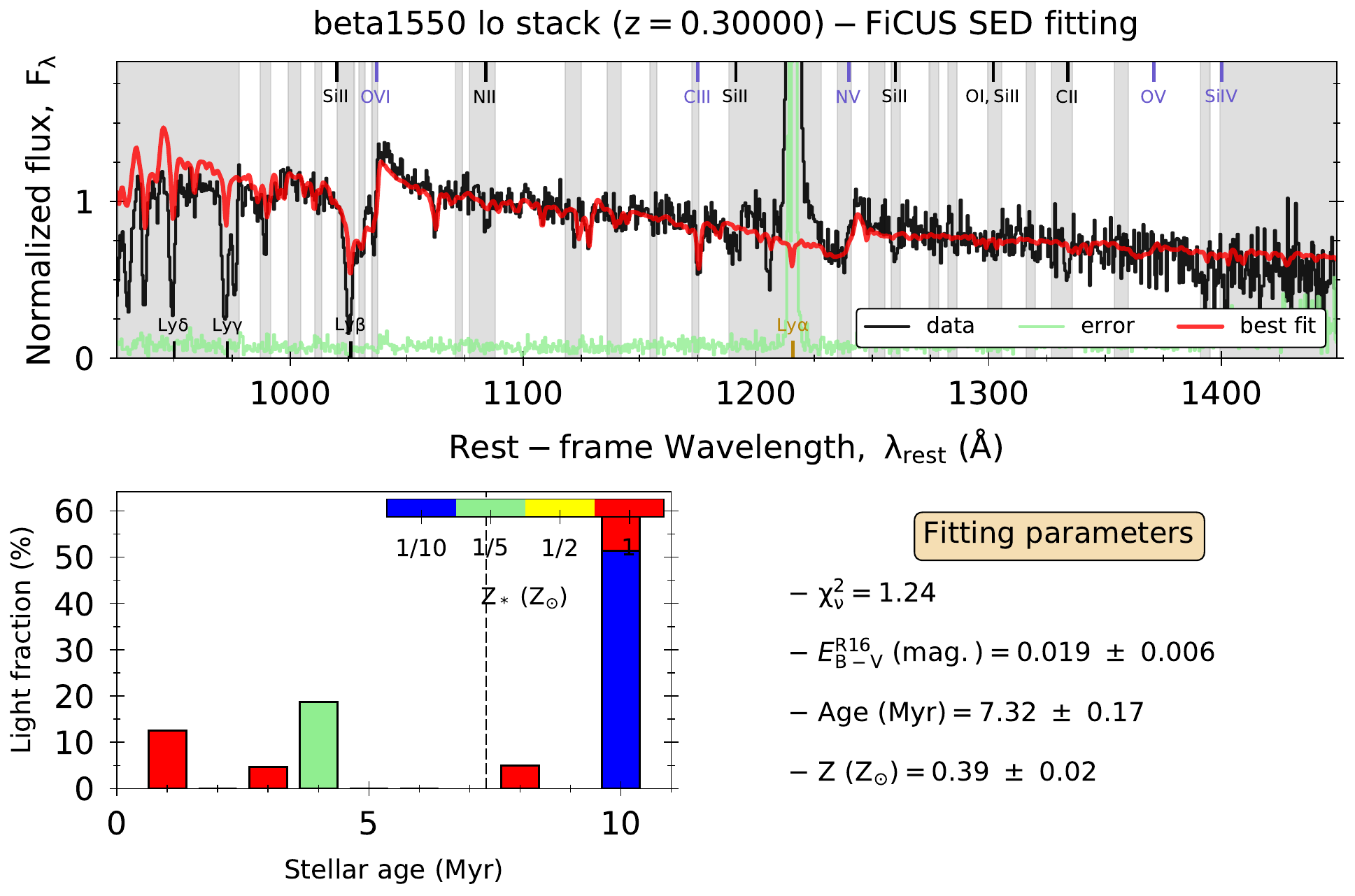}
    \caption{{\tt FiCUS} fit of {\tt Starburst99}+{\tt Cloudy} templates to the stack of \emph{HST}/COS spectra for all galaxies in the LzLCS+ sample with $\beta_{1550}<-2$. \emph{Top}: Stack (black) and accompanying model SED fit (red) with grey shaded regions indicated regions masked due to ISM contamination or poor COS sensitivity. \emph{Bottom left}: Light fractions $w$ for the best-fit templates sorted according to age and colored by stellar metallicity. Dashed line indicates the light-weighted ``age'' of the stellar populations.  \emph{Bottom right}: Light-weighted properties and dust attenuation.}
    \label{fig:lce_fit}
\end{figure*}

Following previous work \citep{2019ApJ...882..182C,2022A&A...663A..59S,2023MNRAS.522.6295S}, we fit template stellar spectra to the stacks using the {\tt FiCUS} code\footnote{\url{https://github.com/asalda/FiCUS}} assuming that the observed flux $\hat{\boldsymbol{F}}_\lambda$ is a positive linear combination of model fluxes $\boldsymbol{F}_{\lambda}^{\prime}=[{F}_{\lambda,1},{F}_{\lambda,2},...,{F}_{\lambda,n}]$ attenuated by dust such that
\begin{equation}
\begin{split}
    \hat{F}_\lambda &= 10^{-0.4A_\lambda}\sum\limits_{i}^{N} w_i F_{\lambda,i} \\
    &= 10^{-0.4A_\lambda}\left(\boldsymbol{\rm W}^\prime\cdot\boldsymbol{F}_{\lambda}\right)
\end{split}
\end{equation}
for wavelength-independent light fractions $\boldsymbol{\rm W}^\prime=[w_1,w_2,...,w_n]$ and extinction $A_\lambda=k_\lambda E_{\rm B-V}$ assuming the coefficients from \citet{2016ApJ...828..107R}. This template-fitting approach constitutes a non-parametric star-formation history. A continuous star formation history \citep[e.g.,][]{2018ApJ...869..123S}, while an attainable solution within our method, is unlikely due to the stochasticity of cluster formation over short ($<$10 Myr) timescales in low mass ($<10^9$ M$_\odot$) starburst galaxies \citep[cf.][for analogs of the LzLCS+]{2017ApJ...845..165M,2022MNRAS.510.4819S}.

Templates consist of a range of young burst ages (1, 2, 3, 4, 5, 6, 8, 10 Myr) and stellar metallicities (0.1, 0.2, 0.5, 1 $Z_\sun$) for a total of $N=32$. Templates are limited to $\leq$ 10 Myr due to poor isochrone sampling in the high resolution {\tt Starburst99} models \citep{2022MNRAS.517.5104C}. { While the effects of binary stars may be important for reionization \citep{2016MNRAS.459.3614M,2018MNRAS.479..994R},} we do not consider the prevalent {\tt BPASS} models \citep{2018MNRAS.479...75S} because the predicted spectra do not accurately reproduce the critical age diagnostic wind lines like \ion{O}{6}, particularly in the emission part of the P-Cygni profile \citep{2022MNRAS.517.5104C}. { However, neglecting binaries will not likely affect our results since (i) binary affects on \fesc\ are limited to older stellar populations \citep[e.g.,][]{2022ApJS..260....1F} and (ii) comparisons with {\tt Starburst99} indicate relatively little difference in the predicted feedback.}

Each starlight template is supplemented with {\tt Cloudy} model predictions for the nebular continuum, which can contribute up to 6-7\% of the flux redward of Ly$\alpha$ via two-photon emission \citep[e.g.,][]{2016ApJ...826..159S,2019ApJ...882..182C}. To predict the continuum, we assume $\log U = -2.5$ \citep{2019ApJ...882..182C}, $n_{\rm H}=100~\rm cm^{-3}$ \citep[consistent with many of the LzLCS targets,][]{2022ApJS..260....1F}, isobaric structure with $\log P/k = 6$, and abundance patterns from the Galactic Concordance \citep{2017MNRAS.466.4403N} matched to the metallicity of the input stellar template. 
Finally, we convolve each model template with the wavelength-dependent COS G140L line spread function in the observed frame assuming a typical redshift of $z=0.3$ to match the stacked spectra (see Appendix \ref{apx:conv}). We show an example fit with {\tt FiCUS} in Figure \ref{fig:lce_fit} to illustrate the non-parametric fit and constraining power of higher signal-to-noise in the stacks. { From the best-fit templates, {\tt FiCUS} also provides properties such as \xiion, $\beta_{1550}$, and E(B-V).}

{ One surprising result from these fits, particularly in the case of very young stellar populations, is that {\tt FiCUS} finds best-fit templates with uncharacteristically high metallicities (see the 1 and 3 Myr populations in the lower left panel in Figure \ref{sec:specfit}). Closer inspection suggests that {\tt FiCUS} chooses these high metallicity templates to better describe the prominent \ion{O}{6} features in these stacks. As a test of this interpretation, we refit stacks with (i) masking of the \ion{O}{6} P Cygni line and (ii) fixed metallicity of the model starlight templates. For (i), {\tt FiCUS} selects templates in the 10-50\% $Z_\odot$ range, indicating that {\tt FiCUS} invokes high metallicity templates simply to account for the \ion{O}{6} feature. In (ii), only the solar metallicity templates can describe the \ion{O}{6} feature but, in doing so, over-predict the depths of stellar absorption features such as \ion{N}{5}, \ion{O}{5}, and \ion{C}{3}. Therefore, we cannot interpret the best-fit template metallicities as being representative of the stellar populations contributing to the stacked SED. Instead, a ``high'' metallicity indicates very young, very low metallicity stellar populations with enough high mass stars to provide the radiation necessary to produce \ion{O}{6}. {Given the $\rm O^{+4}$ ionization potential of 113.8 eV, a prominent \ion{O}{6} feature would thus require substantial soft X-rays, a component which is not empirically well constrained, particularly at low metallicity. In most cases, the prescription for X-rays in model atmospheres is heuristic at best (as with {\tt Starburst99}) and sometimes absent altogether (as with {\tt BPASS}, which is why those templates were not used in this analysis).}

We note that the mechanical luminosities predicted by {\tt Starburst99} are dominated by SNe, which depend very weakly on metallicity, and that, as a result, \emech\ is not affected by the best-fit metallicity. While the ionizing continuum is more sensitive to metallicity, we find that \xiion\ changes insignificantly in our test cases.}

\subsection{Light Fractions}

{ As an assessment of the underyling stellar population ages, we calculate the fraction of light in given age ranges corresponding to different feedback regimes: ionization ($<3$ Myr), stellar winds (4-6 Myr), and SNe ($>8$ Myr). We consider age bins rather than individual template ages as in many cases the stacks do not provide sufficient signal to constrain the individual template weights.} We calculate the light fraction $f_\star$ for the different age bins such that
\begin{equation}
    f_{\star j}(t<3{\rm~Myr}) = \left(\sum\limits_{i} w_{ij}\right)^{-1}\sum\limits_{i}^{t_i<3} w_{ij}
\end{equation}
for the youngest stellar populations,
\begin{equation}
    f_{\star j}(3<t<6{\rm~Myr}) = \left(\sum\limits_{i} w_{ij}\right)^{-1}\sum\limits_{i}^{3<t_i<6} w_{ij}
\end{equation}
for the Wolf-Rayet stage stellar populations, and
\begin{equation}
    f_{\star j}(t>8{\rm~Myr}) = \left(\sum\limits_{i} w_{ij}\right)^{-1}\sum\limits_{i}^{8<t_i<10} w_{ij}
\end{equation}
for the SNe-onset stellar populations. These metrics provide an advantage over other age measurements in that they directly indicate the fraction of starlight contributed by stellar populations of certain ages. The oft-employed light-weighted or mass-weighted age can be difficult to interpret, particularly owing to the fact that the weighted mean does not necessarily represent the true distribution of population ages, as illustrated in the bottom left panel of Figure \ref{fig:lce_fit}. { Typically, the light fractions of our stacks have uncertainties of $\sim$25\% and are calculated from the best-fit SED via Monte Carlo sampling as in \citet{2022A&A...663A..59S}.} 

\subsection{\fesc}

From the stacked spectra, we measure the emergent LyC flux $F_{\rm LyC}$ from 880 to 900 \AA. We infer the intrinsic LyC flux $F_{\rm LyC,0}$ from the best-fit stellar populations. The ratio of emergent to intrinsic flux gives the line-of-sight LyC escape fraction such that
\begin{equation}
    f_{esc}^{\rm LyC} = \frac{F_{\rm LyC}}{F_{\rm LyC,0}}
\end{equation}
as in \citet{2022A&A...663A..59S}. Measured values of \fesc\ are listed in Table \ref{tab:stack_props}. Most \fesc\ results for the stacks reflect the distribution of \fesc\ values of their constituent galaxies. Cases where they do not, while, seldom, are worth noting. We discuss these further below.

\begin{figure}
    \centering
    \includegraphics[width=\columnwidth]{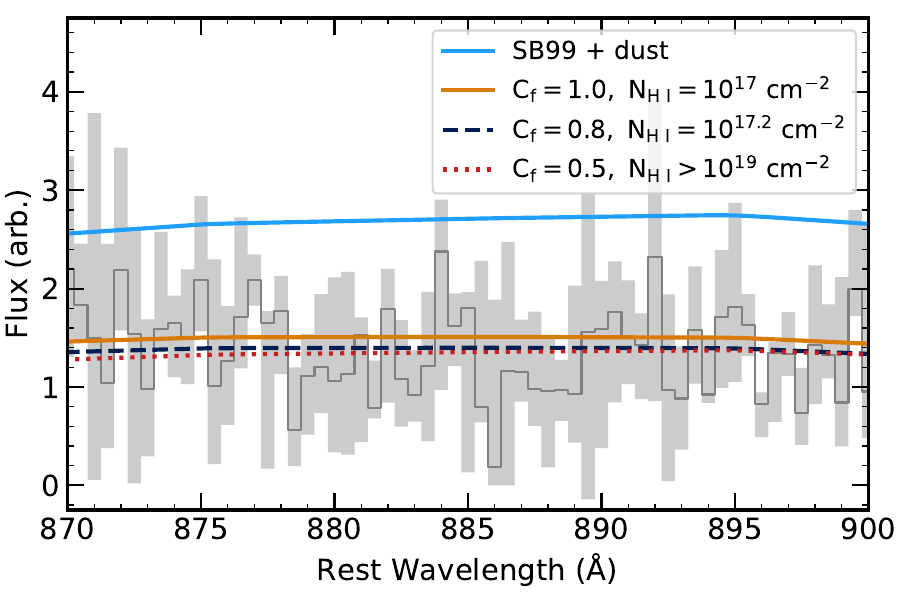}
    \caption{Comparison of Lyman continuum in LCEs with \orat$>5$ (grey line) with uncertainties (light grey shaded region) with the best-fit {\tt Starburst99} templates accounting for dust-attenuation (blue) and additionally for photionization of \ion{H}{1} (orange).}
    \label{fig:lycbump}
\end{figure}

\subsection{Nebular Contributions to the LyC}

Several studies have proposed possible nebular contributions to the {emergent} LyC, producing the so-called LyC ``bump'', a nebular continuum which increases dramatically with wavelength between 800 and 912 \AA\ \citep[e.g.,][]{2010MNRAS.401.1325I,2024MNRAS.530.2133S}. We do not consider that possibility here for several reasons. First, the LyC bump is most prominent in the 900-912 \AA\ range, which is subject to contamination from non-ionizing starlight due to the COS LSF \citep[e.g.][]{2022ApJS..260....1F}. Second, the LzLCS+ does not always sample the optimal LyC bump well because redshifts of 0.32-0.36 (25\% of the combined sample) place the LyC bump within the broad geocoronal Ly$\alpha$, which severely contaminates or entirely swamps the signal. Third, the most likely ``bump'' emitters, high ionization (\orat$>5$)
LCEs that may have near-isotropic LyC escape \citep{2022ApJ...930..126F}, exhibit LyC shapes which \emph{decrease} with wavelength (see Fig \ref{fig:lycbump}), a trend opposite to that expected from nebular emission and instead consistent with \ion{H}{1} absorption via photoionization.

To test our visual inspection of the LyC shape for evidence of nebular continuum, we consider a photoionization-only scenario for the stack of high ionization LCEs (\orat$>5$). We predict the {\it emergent} LyC shape from the best-fit starlight continuum using the following expression
\begin{equation}
    F_{\rm LyC} = F_{\star}\left[C_f \exp\left(-a_\lambda N_{\rm~H~I}\right)+1-C_f\right]
\end{equation}
from radiative transfer for pure \ion{H}{1} photoionization \citep[cf.][]{2006agna.book.....O} where $C_f$ is the gas covering fraction, $N_{\rm H I}$ is the \ion{H}{1} column density, and $a_\lambda$ is the wavelength-dependent photoionization cross section taken from atomic data in \citet{1996ApJ...465..487V} and $C_f\in[0,1]$ is the covering fraction representing the geometric percentage of the source area obstructed by \ion{H}{1} along the line of sight. In Figure \ref{fig:lycbump}, we compare the predicted emergent LyC continuum with that of the stack for three different combinations of $C_f$ and $N_{\rm H~I}$ chosen to match the LyC flux. For all three cases, photoionization of \ion{H}{1} can fully account for the observed shape of the LyC, indicating no nebular contribution to the emergent LyC or the measured \fesc.

{ This result not only rules out nebular contributions to the \emph{emergent} LyC but also hints at a ``picket fence'' ISM geometry in which some gas is entirely opaque to the LyC while the remaining gas is almost entirely transparent \citep[e.g.,][]{2016ApJ...828..108R,2018A&A...616A..29G,2020A&A...639A..85G}. According to \citet{2024MNRAS.527.6139S}, nebular continuum should maximally affect the LyC shape over $\lambda=800$-900 \AA\ for column densities of \nhi$10^{17}$-$10^{18}\rm~cm^{-2}$. Given the column density - covering fraction pairs in Figure \ref{fig:lycbump}, the only case we have considered which is consistent with unobserved nebular LyC is that of low covering fraction ($C_f=0.5$) with high column density (\nhi=$10^{19}\rm~cm^{-2}$). In such a scenario, any significant production of nebular continuum occurs in optically thick regions with no emergent LyC, meaning \fesc=0 and the LyC bump signature is lost to photoionization of \ion{H}{1}. Meanwhile, optically thin channels lack sufficient \ion{H}{1} to produce significant amounts of nebular continuum to affect the shape of the emergent LyC, with \fesc\ approaching unity.}

\section{Non-Detections vs Non-Emitters}\label{sec:nondets}


As discussed in \S\ref{sec:stacks}, one of our objectives is to explore \fesc\ below the detection limits of COS
. How low is \fesc\ for non-detections, and are these galaxies cosmically insignificant LCEs? Under the picket-fence framework, a sightline has either very high \fesc\ or no LyC escape whatsoever \citep[e.g., $z\sim3$ results in][]{2019ApJ...878...87F}, although see \citet{2020A&A...639A..85G} for a scenario where extreme ionization leads to isotropic escape with spatial variations in \fesc\ along different sightlines. Thus, by determining \fesc\ for stacks of non-detections, we can assess the (an)isotropy of LyC escape. If non-emitters do have ubiquitously obscured sightlines, we anticipate gas covering fractions which approach unity, which lead to \fesc$< 0.01$.


\subsection{LyC Below the COS LyC Detection Limit}

As shown in Table \ref{tab:stack_props}, most, but not all, of the non-detection stacks exhibit \fesc$<0.01$, indicating no LyC escape down to very low \fesc. Indeed, the stack of all non-detections has \fesc$=0.15$\%. { Because the stacks of non-detections are selected on optical thickness to the LyC, this result indicates only that the non-detections do not emit cosmologically relevant quantities of LyC along the sightlines observed.} Whether they exhibit LyC escape in other directions cannot be demonstrated from these stacks alone.

Some stacks of non-detections do contain measurable LyC with \fesc$\sim0.01-0.03$. These stacks include non-detections with concentrated star formation (\sigsfr$>10$), strong Ly$\alpha$ (EW Ly$\alpha>50$ \AA\ or \fesclya$>0.2$), high ionization (\orat$>5$), low dust content ($\beta_{1550}<-2$), or low metallicity (\logoh$<8$). Given that stacks of LCEs with similar properties can exhibit \fesc$\sim0.03-0.06$, the non-detections selected for these particular stacks could arise from near-isotropic LyC escape with spatially varying \fesc, as suggested by \citet[][see their Fig 16b]{2020A&A...639A..85G}. Whether this
scenario can be substantiated remains to be seen. But certainly, these galaxies with properties similar to more prodigious LCEs are leaking LyC at considerably lower fractions. 

\subsection{Stack vs Galaxy \fesc}

\begin{figure}
    \centering
    \includegraphics[width=\columnwidth]{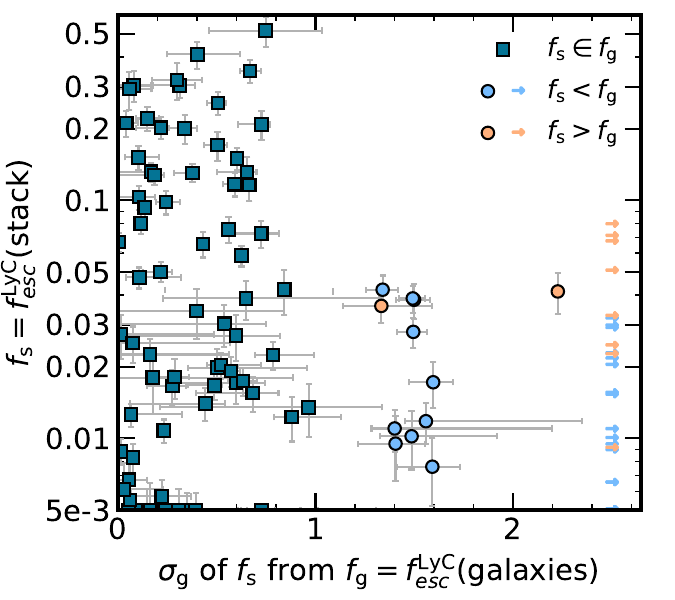}
    \caption{The \fesc\ for stacks (\fst) compared to how confidently we can reject the null hypothesis that \fst\ represents the escape fraction \fg\ of galaxies used to build the stack. Uncertainties in $\sigma$ correspond to projections of the \fst\ uncertainty onto the Kaplan-Meier survival curve. Dark teal squares indicate cases where \fst\ is consistent with \fg\ within $1\sigma$ ($P\in[0.1587,0.8413]$, null hypothesis accepted), circles \fst\ within 2-3$\sigma$ of \fg\ ($P\in[0.025,0.975]$, null hypothesis tentative), and arrows with no agreement between \fst\ and \fg\ ($P\notin[0.01,0.99]$, null hypothesis rejected at $>3\sigma$ confidence). Cases where \fst\ is significantly lower (higher) than its constituent \fg\ are shown in light blue (orange).}
    \label{fig:kmresults}
\end{figure}

For a more robust assessment of whether non-detections are characteristically insignificant LyC leakers, we compare \fst, the \fesc\ measured for each stack, to the distribution of \fg, the \fesc\ values of the galaxies contributing to that stack. For this comparison, we want to know the probability \pfesc\ (confidence $\sigma_0$) with which we can reject the null hypothesis that \fst\ is representative of \fg. We calculate \pfesc\ using the Kaplan-Meier estimator \citep[][for details, see Appendix \ref{apx:kaplanmeier}]{KaplanMeier1958}. We list \pfesc\ in Table \ref{tab:stack_props} and compare $\sigma_0$ with \fst\ in Fig \ref{fig:kmresults}.

As evidenced in Figure \ref{fig:kmresults}, the majority of stacks, including all stacks with \fesc$>0.1$, have \pfesc\ within 0.1587-0.8413 interval (i.e., $-1<\sigma_0<1$, thus \fst\ is within $1\sigma$ of the median of \fg), accepting the null hypothesis that these stacks are representative in \fesc\ of the galaxies used to build them. Moreover, the fact that the stacks with \fesc$>0.1$ are all consistent with their constituent galaxies, suggests LyC escape increases in isotropy with increasing \fesc.

There are few cases where \fst\ is outside this interval: either non-detection stacks or stacks with both low \fesc\ and strong LCE properties as a selection criterion (e.g., \fesc$<0.01$ and high \orat). In the former scenario, the \fg\ are predominantly upper limits, meaning that \fst\ is lower than the measured upper limits on \fesc\ and suggesting \fst\ is consistent with both the measurements of the galaxies and the interpretation that these galaxies are non-leakers.
For other cases, it is entirely possible that the stack is not sufficiently representative of the comprised galaxies due to noise in and under-sampling of certain key age diagnostic features such as \ion{O}{6}, \ion{C}{3}, and \ion{N}{5}. In such instances, a small number of galaxies may be over-represented in certain features in the stack, which would adversely affect the inferred intrinsic LyC.

Alternatively, some galaxies which lack well-measured features may also have inaccurate intrinsic LyC estimates owing to poor constraints on the underlying stellar populations in the individual spectrum. We test this possibility using a Monte Carlo simulation of $10^4$ trials sampling distributions of each \fesc$\in$\fg. This test finds substantial variance in the computed \pfesc\ when \fst$<5$\% and \pfesc$\in(0,1)$, suggesting that the tenuous \fst\ outliers (\pfesc\ corresponding to 1-2$\sigma$) are in fact in agreement with their constituent \fg. At the extrema, where \pfesc$=0$ or \pfesc$=1$, uncertainties in \fg\ \fesc\ values cannot account for the discrepancy. The overwhelming majority of these extrema have \pfesc$=0$ and correspond to the stacks of non-detections, meaning the apparent discrepancy solely arises from the stacks' ability to push below the COS detection limit. In the remaining 8 cases, the stacks with $P(<$\fst$|$\fg$)=1$ are built from subsets of weak LCEs (\fesc$\in[0.01,0.05]$) or non-LCEs (\fesc$<0.01$) with properties similar to strong LCEs (M$_\star<10^{9}\rm~M_\odot$, \logoh$<8$, \orat$>5$, H$\beta$ EW $>100\rm~\AA$, and \muv$>-19.5$). These \pfesc$=1$ extrema point to the scenario of isotropic LyC escape with spatially varying \fesc.

As a whole, the \fesc\ of our stacks are typically in agreement with the \fesc\ of the galaxies used to build them. When the stack \fesc\ is significantly lower than the galaxies (\pfesc$<0.05$), we have achieved our objective of pushing below the COS detection threshold by measuring \fesc\ for galaxies with undetected LyC which are genuinely not leaking LyC photons along the line of sight. In cases where \fesc\ is significantly higher in the stack than in the galaxies (\pfesc$>0.95$), the stack criteria match properties of strong LCEs but are built on subsets of weak or low \fesc, suggesting that the strongest LCEs undergo isotropic escape with spatial variations in \fesc\ along different sightlines. This comparison between stacks and their constituent galaxies indicates that individual non-detections are consistent with insignificant LyC escape along the line of sight. Moreover, the few cases where \fst\ and \fg\ are discrepant are cases where \fg\ is not well-constrained; the stacks in these instances may indicate stronger LyC escape than measured directly from the individual COS spectra.

\section{ISM Absorption Lines}\label{sec:absn}

\subsection{Measuring Absorption Line Properties}

As the stacks lack sufficient S/N and resolution to perform detailed analysis via modeling of the observed ISM absorption line profiles \citep[e.g.,][]{2015ApJ...801...43S,2016MNRAS.463..541C,2017A&A...605A..67C,2018A&A...616A..29G,2020A&A...639A..85G,xu2022}, we take a more simplistic approach to deriving relevant information from the absorption lines observed in the stacked spectra. Following \citet{2022A&A...663A..59S}, we measure the residual flux and equivalent widths of the absorption features by normalizing the stacked flux $F$ by the estimated underlying continuum $F_0$. We measure the equivalent width of each line by
\begin{equation}
    W_\lambda = \int\limits_{\lambda_0-\delta\lambda}^{\lambda_0+\delta\lambda} 1-\frac{F}{F_0}{\rm d}\lambda
\end{equation}
where $\delta\lambda$ corresponds to 1 250 km s$^{-1}$ to include the entire line profile \citep[see ][]{2022A&A...663A..59S} and the integral is approximated using the trapezoid method. We also measure the residual flux
\begin{equation}
    R_f = \left\langle\frac{F}{F_0}\right\rangle
\end{equation}
where the average is calculated over the inner 300 km s$^{-1}$ of the line profile. $R_f$ measures whether gas producing the absorption feature covers the entire line-of-sight, with $R_f=0$ indicating total coverage and $R_f=1$ indicating no coverage \citep[see][]{2022A&A...663A..59S}. Uncertainties in residual flux and equivalent width are estimated by Monte Carlo sampling of the stacked fluxes and recalculating both values $10^4$ times. To reduce the total uncertainty in our results, we take the variance-weighted average of the measured $W_\lambda$ and $R_f$ for different groups of ion species as listed in Table \ref{tab:lines} following Eqns 6-7 in \citet{2022A&A...663A..59S}.

{ For estimates of $F_0$, we compare the best-fit stellar SED to an approximation of the local continuum.} The former is discussed in \S\ref{sec:specfit}. The latter is obtained individually for each line using a variance-weighted linear least squares fit to continuum within 1 500 km s$^{-1}$ of the minimum line flux. Then, we correct for stellar absorption effects by first measuring $W_{\lambda\star}$ and $R_{f\star}$ in each of the stellar templates, taking the weighted average using the best-fit SED weights for each stack, and finally subtracting $W_{\lambda\star}$ from $W_\lambda$ and scaling $R_f$ by $R_{f\star}$. We find similar values for $R_f$ and $W_\lambda$ regardless of the continuum fit method in most cases, suggesting that the choice of $F_0$ introduces no systematic effects. Visual inspection indicates that the linear fit better approximates the continuum immediately adjacent to the absorption features, particularly in the case of the Lyman series. In the five cases where the linear fit does not well-approximate the continuum near the low ionization species (LIS) lines, we assume the SED fit but note that the LIS lines in these cases are not significantly detected. In all other cases, we adopt $W_\lambda$ and $R_f$ values measured assuming the linear continuum fit for $F_0$.

Both the COS instrument and stacking may affect these measured properties. To test this possibility, we simulate COS G140L observations of model absorption lines, accounting for the non-Gaussian, wavelength-dependent line spread function (LSF). We subsequently stack mock observations and measure the weighted-average of the LIS and \ion{H}{1} absorption line properties. $R_f$ is not fully recoverable at very low values due to artificial line infilling by the COS optics, resulting in a limiting $R_f=0.27$ for the case of 100\% covering fraction. The equivalent widths are consistently more recoverable because the $W_\lambda$ is integrated over the entire line profile and the LSF convolution kernel integrates to unity. Details of these simulations are provided in Appendix \ref{apx:abs}.

\begin{deluxetable}{l l l}
\tablecaption{Absorption lines used to calculate variance-weighted average $W_\lambda$ and $R_f$ for different species groups. Groups are defined by the ionization potential energy $E_{ion}$ of the corresponding species. Thresholds selected correspond to \ion{H}{1} ($E_{ion}=13.6$ eV) and \ion{He}{1} ($E_{ion}=24.6$ eV).\label{tab:lines}}
\tablewidth{\textwidth}
\tablehead{
\colhead{Group} & $E_{ion}$ (eV) & \colhead{Lines Included} }
\startdata
\ion{H}{1} & 13.6 eV & Ly$\gamma$, Ly$\delta$, Ly$\epsilon$, Ly$\zeta$\\
LIS & $<13.6$ & \ion{Si}{2}$\lambda1190,93$, \ion{Si}{2}$\lambda1260$, \\
    &  & \ion{O}{1}$+$\ion{Si}{2}$\lambda1303$, \ion{C}{2}$\lambda1334$ \\
MIS & $13.6$-$24.6$ & \ion{S}{3}$\lambda1012$, \ion{N}{2}$\lambda1084$, \ion{Si}{3}$\lambda1206$ \\
HIS & $>24.6$ & \ion{Si}{4}$\lambda1393$, \ion{Si}{4}$\lambda1402$\\
\enddata
\end{deluxetable}


\begin{figure}
    \centering
    \includegraphics[width=\columnwidth]{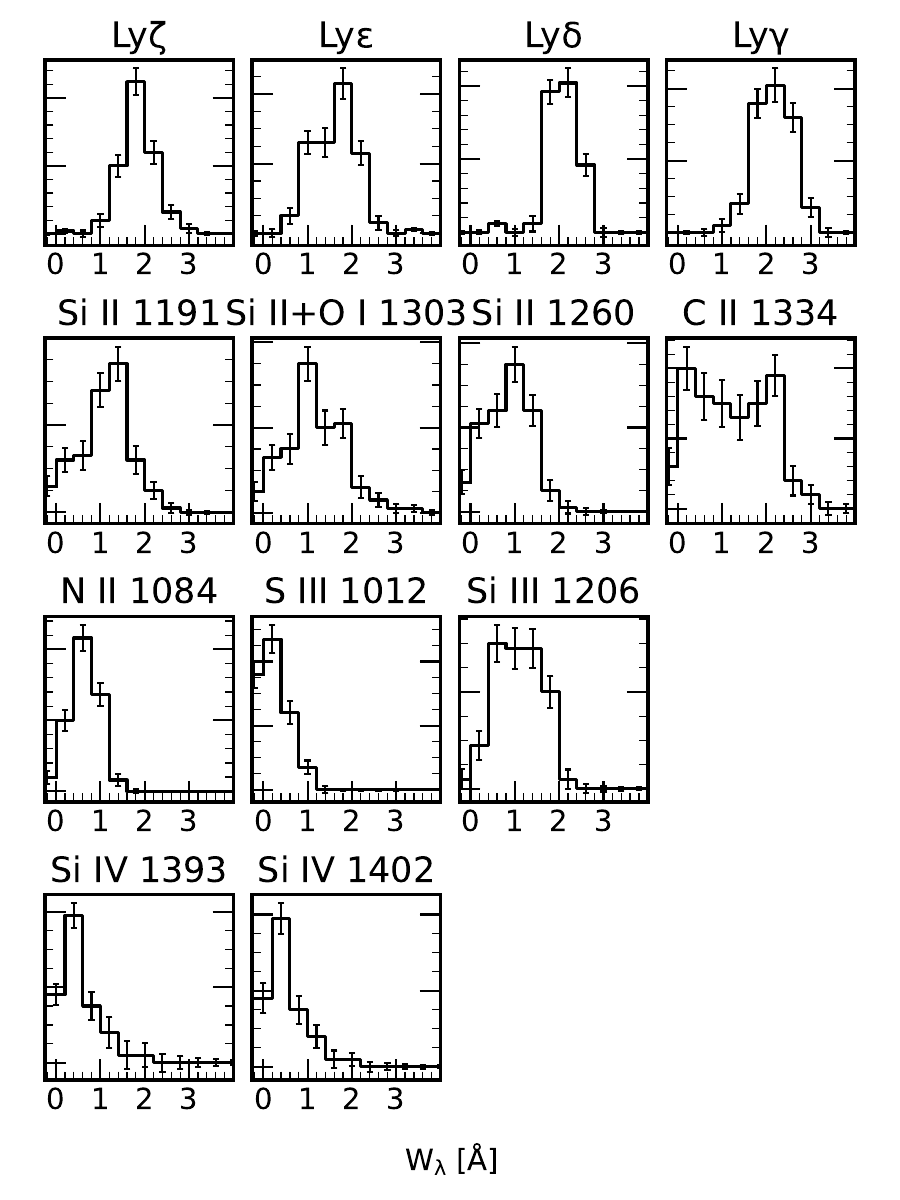}
    \caption{Distribution of equivalent widths $W_\lambda$ measured from absorption lines in the stacks. From top to bottom, we show lines included in the \ion{H}{1}, LIS, MIS, and HIS groups. Error bars indicate the Poisson binomial uncertainties calculated from the uncertainties in each $W_\lambda$.}
    \label{fig:Wl_hists}
\end{figure}

\begin{figure}
    \centering
    \includegraphics[width=\columnwidth]{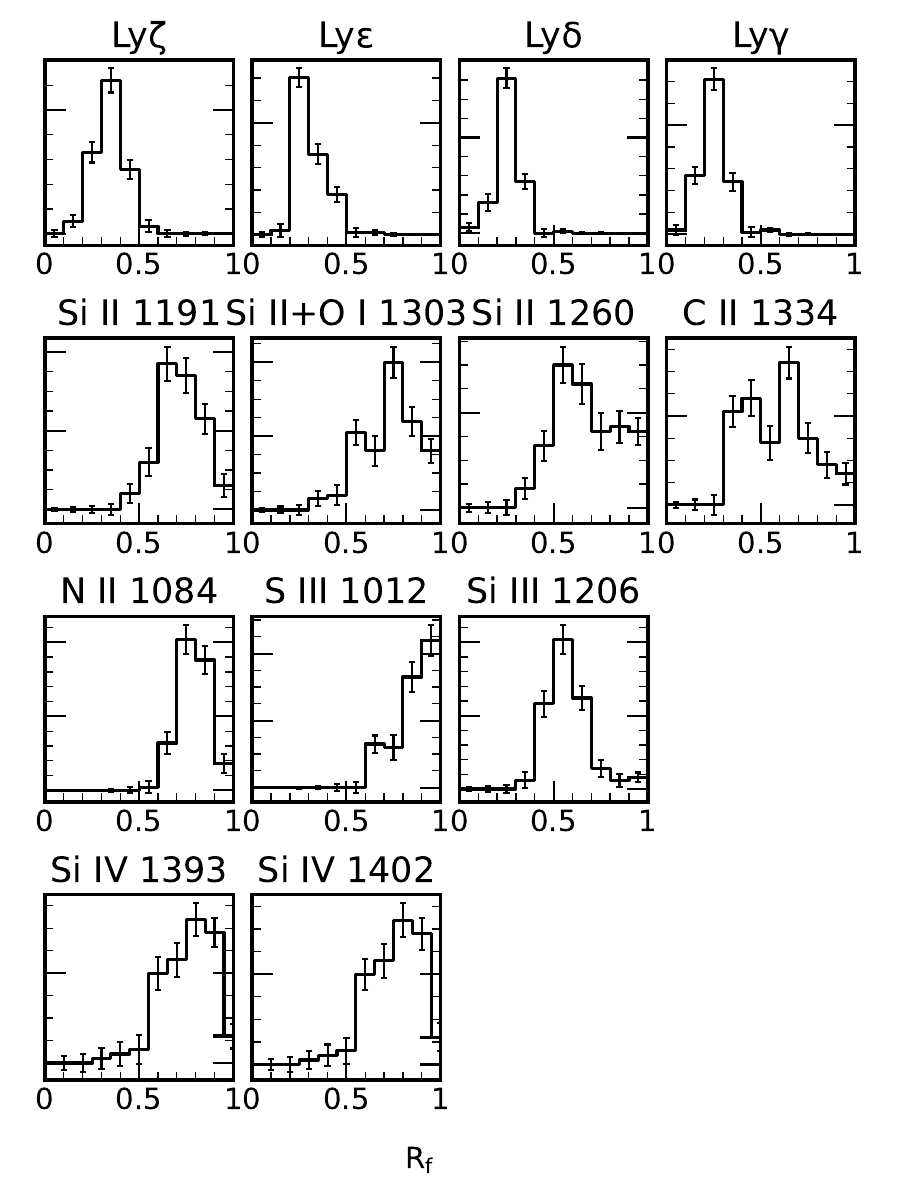}
    \caption{Same as Figure \ref{fig:Wl_hists} but for residual fluxes $R_f$.}
    \label{fig:Rf_hists}
\end{figure}

\subsection{The \cfh-\cfl\ Discrepancy}

\begin{figure}
    \centering
    \includegraphics[width=\columnwidth]{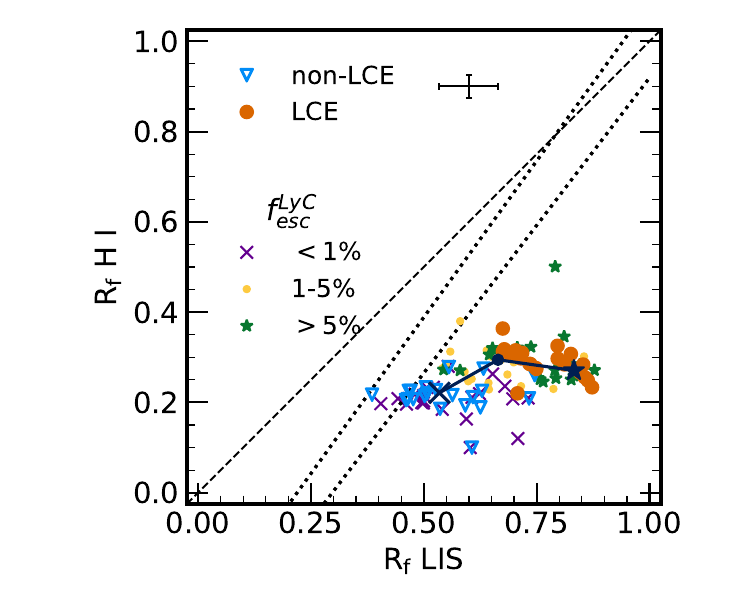}
    \caption{Comparison of residual fluxes $R_f=1-C_f$ for \ion{H}{1} and LIS species for stacked LzLCS+ subsets of non-detections (blue open triangles), LCEs (orange circles), non-LCEs (\fesc$<1$\%, violet crosses), weak LCEs (\fesc$=1$-5\%, yellow dots), and strong LCEs (\fesc$>5$\%, green stars). We illustrate the trend in \fesc\ with \emph{uncorrelated} stacks in \fesc\ (indigo line and symbols, with symbols matched to LCE strength). Black dashed line is the 1:1 line in the absence of observational and stacking effects. Black dotted lines indicate the two-sided 90\% confidence of simulated $R_f$ observations due to the COS G140L line spread function if $C_{f,\rm~H~I}=C_{f,\rm~LIS}$ implicitly; results between these two lines can be fully explained by combined effects of the LSF and stacking. See Appendix \ref{apx:abs} for details.}
    \label{fig:Rf_comp}
\end{figure}

In Figures \ref{fig:Wl_hists} and \ref{fig:Rf_hists}, we show $W_\lambda$ and $R_f$ results for individual lines measured using the linear approximation to the continuum. One notable result is that, as with individual galaxies \citep{2022A&A...663A..59S}, the \ion{H}{1} lines exhibit much lower $R_f$ values than other lines, which we highlight in Figure \ref{fig:Rf_comp}. Because the COS LSF is wavelength-dependent, it is possible that the COS optics can account for this discrepancy. We show the effects of the COS optics and stacking on the measured \rfh\ and \rfl\ for matched covering fractions in Figure \ref{fig:Rf_comp}. Simulations of the COS LSF demonstrate that the observed $C_f$ discrepancy does not arise from systematics such as stacking or optics effects at the 95\% confidence level for the majority of the measured $R_f$. This apparent discrepancy is therefore physical in nature.

\begin{figure}
\includegraphics[width=\columnwidth]{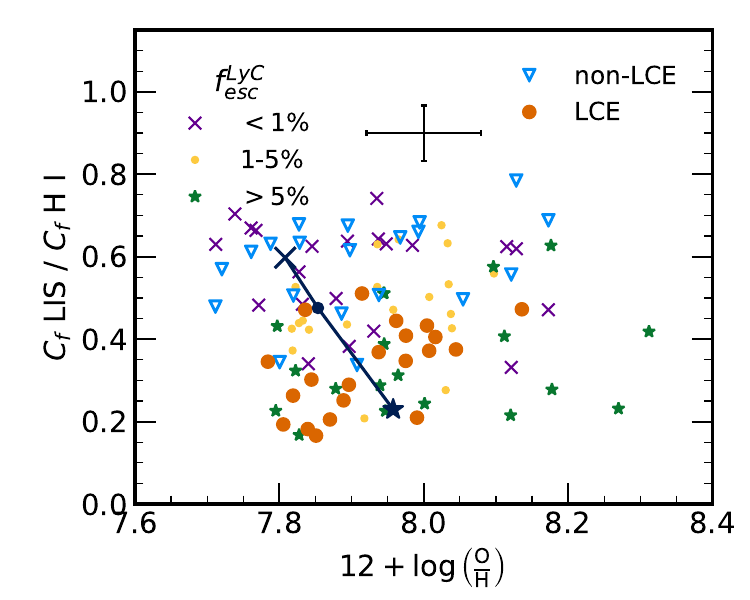}
\includegraphics[width=\columnwidth]{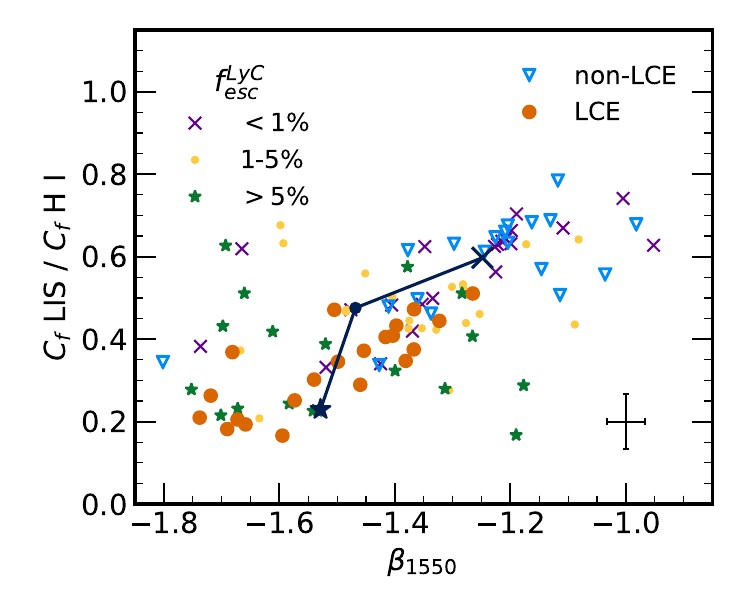}
\caption{Comparison of $C_{f}$ discrepancy \cfl$/$\cfh\ and gas phase metallicity \logoh\ ({\it top}) and dust-sensitive UV slope $\beta_{1550}$ ({\it bottom}). The dependence of the discrepancy on dust attenuation but \emph{not} on abundances indicates that differences in residual flux arise from the opacity and/or geometry of the intervening gas. Symbols as in Figure \ref{fig:Rf_comp}. Note that the trend in \fesc\ indicated by the indigo line in the top panel varies over little more than 0.1 dex and is not significant within the uncertainties or scatter in \logoh.}
\label{fig:zeta_expl}
\end{figure}

We quantify this discrepancy using the ratio of covering fractions \cfl/\cfh$=1-$\rfl$/1-$\rfh\ (henceforth the $C_f$ ratio) where covering fraction is given by $C_f=1-R_f$. As this discrepancy arises from a comparison of metal ions with \ion{H}{1}, one natural explanation is that metallicity differences lead to the observed spread in \rfl\ and, by extension, the differences in covering fraction.

To assess whether metallicity can account for the observed spread in \rfl\ and the $C_f$ ratio, we take two approaches: compare the \cfl\ and $C_f$ ratios of stacks of low metallicity and high metallicity galaxies in different subsets and calculate Kendall's $\tau$ relative to \logoh\ for both properties for the full set of stacks. Stacks of the lower metallicity galaxies with other subset criteria (\fesc\ value or LyC detection) do not have significantly different LIS covering fractions. In the case of stacks built solely on cuts in \logoh, the difference in \cfl\ is $0.177\pm0.101$, which is $<2\sigma$ significant; moreover, the covering fractions for the different \logoh\ stacks, 0.212 for \logoh$<8$ and 0.388 for \logoh$>8$, only account for a small amount of the total range of \cfl$\in[0.122,0.723]$ across all the stacks. The Kendall's $\tau$ correlation coefficient for \cfl\ and \logoh\ is weak ($\tau=0.026$) and insignificant ($p=0.657$), demonstrating that the spread in \cfl\ is not attributable to metallicity. Turning to the $C_f$ ratio, we find no significant differences in \cfl/\cfh\ between any of the \logoh\ stacks regardless of subset. { Moreover, these absorption lines are likely saturated in the low dispersion COS spectra \citep{2022A&A...663A..59S}, such that $W_\lambda$ ought to depend more on the velocity dispersion or the covering fraction of the gas than the column density of the metal ion species.} Thus, metallicity is not a favorable explanation for the observed discrepancy. Indeed, we compute Kendall's $\tau=0.009$ with $p=0.883$, indicating no correlation between the two.


However, compared to stacks with low $C_f$ ratio ($<0.3$), the stacks with high $C_f$ ratio also show higher \wfh\ by 0.5 \AA\ or more and higher \ebv\  by as much as 0.15. These differences suggest that the optical depths, both in \ion{H}{1} column density and in dust, are genuinely  distinct for different regimes of $C_f$ ratio. The mechanism responsible for the spread in \cfl\ and the $C_f$ ratio is therefore more likely due to the geometry or kinematic conditions of the ISM. Indeed, Kendall's $\tau$ corellation coefficient for the $C_f$ ratio is far higher for \wfh\ and $\beta_{1550}$\ ($\tau=0.295$ and $0.377$, respectively) and more significant ($p=6.922\times10^{-7}$ and $p=2.331\times10^{-10}$) than for stacked \logoh, which we illustrate in Figure \ref{fig:zeta_expl}. { As the lines are not resolved by the low resolution COS grating, the role of kinematics is observationally unconstrained. That said, a significant difference in gas kinematics between the \ion{H}{1} and LIS lines is unlikely to occur, particularly to the extent required to explain the observed discrepancies.} Therefore, we conclude that the apparent $C_f$ discrepancy must arise primarily from the effects of differences in gas distributions or the optical depth of \ion{H}{1} rather than from differences in chemical abundances. We discuss the implications of the range of $C_f$ ratios below in \S\ref{sec:geometry}.

\startlongtable
\begin{deluxetable*}{crcccccccccccc}
\tablecaption{Averaged properties of absorption lines measured from the stacks of \emph{HST}/COS G140L rest-UV spectra with direct LyC measurements. Columns indicate (i) the criterion/a for inclusion of a target in the stack, (ii) the \ion{H}{1} residual flux \rfh, (iii) the low-ionization species (LIS) residual flux \rfl, (iv) the \ion{H}{1} equivalent width \wfh, and (v) the LIS equivalent width \wfl.\label{tab:stack_lines}}
\tablewidth{\textwidth}
\renewcommand{\arraystretch}{0.75}
\tablehead{
\colhead{criterion} & \colhead{\rfh} & \colhead{\rfl} &
    \colhead{\wfh} & \colhead{\wfl} \\
     & & & \colhead{$[\AA]$} & \colhead{$[\AA]$} }
\startdata
\\[-8pt]
\multicolumn{14}{c}{\bf total} \\[2pt]
\hline
$\beta_{1550}>-2$                            & 0.228 $\pm$ 0.015 & 0.596 $\pm$ 0.042 & 1.933 $\pm$ 0.085 & 1.381 $\pm$ 0.098 \\
$\beta_{1550}<-2$                            & 0.278 $\pm$ 0.018 & 0.803 $\pm$ 0.053 & 1.466 $\pm$ 0.082 & 0.626 $\pm$ 0.135 \\
EW H$\beta$ $>100$ \AA                       & 0.264 $\pm$ 0.019 & 0.797 $\pm$ 0.072 & 1.337 $\pm$ 0.091 & 0.701 $\pm$ 0.165 \\
EW H$\beta$ $<100$ \AA                       & 0.276 $\pm$ 0.022 & 0.608 $\pm$ 0.031 & 1.694 $\pm$ 0.074 & 1.576 $\pm$ 0.084 \\
EW Ly$\alpha$ $>50$ \AA                      & 0.292 $\pm$ 0.028 & 0.727 $\pm$ 0.050 & 1.362 $\pm$ 0.078 & 1.166 $\pm$ 0.129 \\
EW Ly$\alpha$ $<50$ \AA                      & 0.251 $\pm$ 0.015 & 0.597 $\pm$ 0.034 & 1.821 $\pm$ 0.088 & 1.501 $\pm$ 0.096 \\
\fesclya$>0.2$                               & 0.294 $\pm$ 0.026 & 0.777 $\pm$ 0.048 & 1.304 $\pm$ 0.078 & 0.950 $\pm$ 0.113 \\
\fesclya$<0.2$                               & 0.227 $\pm$ 0.024 & 0.606 $\pm$ 0.035 & 1.766 $\pm$ 0.088 & 1.496 $\pm$ 0.100 \\
\muv$>-19.5$                                 & 0.270 $\pm$ 0.025 & 0.735 $\pm$ 0.065 & 1.513 $\pm$ 0.095 & 1.018 $\pm$ 0.178 \\
\muv$<-19.5$                                 & 0.228 $\pm$ 0.007 & 0.676 $\pm$ 0.032 & 1.822 $\pm$ 0.071 & 1.238 $\pm$ 0.084 \\
$\rm M_\star>10^9~M_\odot$                   & 0.268 $\pm$ 0.022 & 0.594 $\pm$ 0.043 & 1.957 $\pm$ 0.102 & 1.445 $\pm$ 0.106 \\
$\rm M_\star<10^9~M_\odot$                   & 0.289 $\pm$ 0.018 & 0.733 $\pm$ 0.045 & 1.398 $\pm$ 0.068 & 0.896 $\pm$ 0.115 \\
\orat$>5$                                    & 0.271 $\pm$ 0.024 & 0.809 $\pm$ 0.066 & 1.306 $\pm$ 0.090 & 0.811 $\pm$ 0.172 \\
\orat$<5$                                    & 0.273 $\pm$ 0.022 & 0.616 $\pm$ 0.033 & 1.679 $\pm$ 0.076 & 1.453 $\pm$ 0.086 \\
\logoh$>8$                                   & 0.293 $\pm$ 0.016 & 0.612 $\pm$ 0.034 & 1.741 $\pm$ 0.075 & 1.434 $\pm$ 0.091 \\
\logoh$<8$                                   & 0.264 $\pm$ 0.029 & 0.788 $\pm$ 0.067 & 1.095 $\pm$ 0.104 & 0.749 $\pm$ 0.165 \\
$r_{50}>0.6~\rm kpc$                         & 0.234 $\pm$ 0.019 & 0.550 $\pm$ 0.044 & 1.759 $\pm$ 0.089 & 1.489 $\pm$ 0.111 \\
$r_{50}<0.6~\rm kpc$                         & 0.300 $\pm$ 0.016 & 0.747 $\pm$ 0.038 & 1.359 $\pm$ 0.074 & 0.973 $\pm$ 0.107 \\
\sigsfr$\rm >10~\frac{M_{\sun}}{yr~kpc^{2}}$ & 0.317 $\pm$ 0.023 & 0.685 $\pm$ 0.034 & 1.533 $\pm$ 0.095 & 0.982 $\pm$ 0.124 \\
\sigsfr$\rm <10~\frac{M_{\sun}}{yr~kpc^{2}}$ & 0.231 $\pm$ 0.016 & 0.632 $\pm$ 0.045 & 1.855 $\pm$ 0.085 & 1.111 $\pm$ 0.107 \\
\hline
\\[-8pt]
\multicolumn{14}{c}{\bf LCEs} \\[2pt]
\hline
$\beta_{1550}>-2$                            & 0.364 $\pm$ 0.024 & 0.675 $\pm$ 0.050 & 1.480 $\pm$ 0.110 & 1.178 $\pm$ 0.133 \\
$\beta_{1550}<-2$                            & 0.266 $\pm$ 0.017 & 0.846 $\pm$ 0.052 & 1.339 $\pm$ 0.094 & 0.483 $\pm$ 0.147 \\
EW H$\beta$ $>100$ \AA                       & 0.284 $\pm$ 0.028 & 0.853 $\pm$ 0.070 & 0.959 $\pm$ 0.106 & 0.290 $\pm$ 0.176 \\
EW H$\beta$ $<100$ \AA                       & 0.315 $\pm$ 0.027 & 0.704 $\pm$ 0.041 & 1.505 $\pm$ 0.095 & 0.964 $\pm$ 0.125 \\
EW Ly$\alpha$ $>50$ \AA                      & 0.325 $\pm$ 0.013 & 0.796 $\pm$ 0.051 & 1.185 $\pm$ 0.085 & 0.627 $\pm$ 0.147 \\
EW Ly$\alpha$ $<50$ \AA                      & 0.285 $\pm$ 0.025 & 0.734 $\pm$ 0.040 & 1.698 $\pm$ 0.113 & 0.771 $\pm$ 0.138 \\
\fesclya$>0.2$                               & 0.307 $\pm$ 0.030 & 0.826 $\pm$ 0.060 & 1.338 $\pm$ 0.090 & 0.746 $\pm$ 0.148 \\
\fesclya$<0.2$                               & 0.306 $\pm$ 0.032 & 0.692 $\pm$ 0.058 & 1.483 $\pm$ 0.147 & 1.142 $\pm$ 0.148 \\
\muv$>-19.5$                                 & 0.318 $\pm$ 0.026 & 0.678 $\pm$ 0.083 & 1.154 $\pm$ 0.119 & 0.662 $\pm$ 0.220 \\
\muv$<-19.5$                                 & 0.310 $\pm$ 0.030 & 0.718 $\pm$ 0.044 & 1.479 $\pm$ 0.083 & 1.035 $\pm$ 0.115 \\
$\rm M_\star>10^9~M_\odot$                   & 0.220 $\pm$ 0.025 & 0.708 $\pm$ 0.044 & 1.812 $\pm$ 0.113 & 1.232 $\pm$ 0.133 \\
$\rm M_\star<10^9~M_\odot$                   & 0.249 $\pm$ 0.017 & 0.863 $\pm$ 0.061 & 1.258 $\pm$ 0.083 & 0.480 $\pm$ 0.153 \\
\orat$>5$                                    & 0.284 $\pm$ 0.028 & 0.812 $\pm$ 0.053 & 1.158 $\pm$ 0.097 & 0.528 $\pm$ 0.184 \\
\orat$<5$                                    & 0.297 $\pm$ 0.031 & 0.715 $\pm$ 0.045 & 1.551 $\pm$ 0.097 & 0.899 $\pm$ 0.128 \\
\logoh$>8$                                   & 0.276 $\pm$ 0.032 & 0.749 $\pm$ 0.045 & 1.709 $\pm$ 0.103 & 0.708 $\pm$ 0.130 \\
\logoh$<8$                                   & 0.256 $\pm$ 0.018 & 0.856 $\pm$ 0.068 & 0.937 $\pm$ 0.099 & 0.732 $\pm$ 0.171 \\
$r_{50}>0.6~\rm kpc$                         & 0.285 $\pm$ 0.025 & 0.736 $\pm$ 0.067 & 1.657 $\pm$ 0.111 & 0.778 $\pm$ 0.164 \\
$r_{50}<0.6~\rm kpc$                         & 0.297 $\pm$ 0.014 & 0.797 $\pm$ 0.049 & 1.164 $\pm$ 0.091 & 0.796 $\pm$ 0.130 \\
\sigsfr$\rm >10~\frac{M_{\sun}}{yr~kpc^{2}}$ & 0.313 $\pm$ 0.012 & 0.675 $\pm$ 0.043 & 1.364 $\pm$ 0.102 & 1.288 $\pm$ 0.157 \\
\sigsfr$\rm <10~\frac{M_{\sun}}{yr~kpc^{2}}$ & 0.234 $\pm$ 0.024 & 0.873 $\pm$ 0.068 & 1.578 $\pm$ 0.098 & 0.409 $\pm$ 0.176 \\
all                                          & 0.274 $\pm$ 0.010 & 0.749 $\pm$ 0.036 & 1.377 $\pm$ 0.065 & 0.669 $\pm$ 0.106 \\
\hline\\[-8pt]
\multicolumn{14}{c}{\bf non-detections} \\[2pt]
\hline
$\beta_{1550}>-2$                            & 0.225 $\pm$ 0.028 & 0.467 $\pm$ 0.052 & 2.077 $\pm$ 0.120 & 1.731 $\pm$ 0.142 \\
$\beta_{1550}<-2$                            & 0.209 $\pm$ 0.007 & 0.733 $\pm$ 0.101 & 2.025 $\pm$ 0.131 & 0.776 $\pm$ 0.327 \\
EW H$\beta$ $>100$ \AA                       & 0.275 $\pm$ 0.013 & 0.632 $\pm$ 0.122 & 1.418 $\pm$ 0.153 & 1.099 $\pm$ 0.384 \\
EW H$\beta$ $<100$ \AA                       & 0.207 $\pm$ 0.024 & 0.477 $\pm$ 0.046 & 2.075 $\pm$ 0.127 & 1.533 $\pm$ 0.127 \\
EW Ly$\alpha$ $>50$ \AA                      & 0.189 $\pm$ 0.045 & 0.625 $\pm$ 0.109 & 1.621 $\pm$ 0.137 & 0.695 $\pm$ 0.346 \\
EW Ly$\alpha$ $<50$ \AA                      & 0.225 $\pm$ 0.019 & 0.471 $\pm$ 0.047 & 2.012 $\pm$ 0.142 & 1.538 $\pm$ 0.138 \\
\fesclya$>0.2$                               & 0.260 $\pm$ 0.016 & 0.746 $\pm$ 0.086 & 2.024 $\pm$ 0.139 & 0.872 $\pm$ 0.248 \\
\fesclya$<0.2$                               & 0.205 $\pm$ 0.014 & 0.461 $\pm$ 0.048 & 2.048 $\pm$ 0.124 & 1.803 $\pm$ 0.141 \\
\muv$>-19.5$                                 & 0.207 $\pm$ 0.046 & 0.465 $\pm$ 0.084 & 1.266 $\pm$ 0.182 & 1.871 $\pm$ 0.293 \\
\muv$<-19.5$                                 & 0.186 $\pm$ 0.005 & 0.536 $\pm$ 0.052 & 2.218 $\pm$ 0.124 & 1.453 $\pm$ 0.133 \\
$\rm M_\star>10^9~M_\odot$                   & 0.217 $\pm$ 0.019 & 0.386 $\pm$ 0.050 & 1.636 $\pm$ 0.155 & 1.872 $\pm$ 0.163 \\
$\rm M_\star<10^9~M_\odot$                   & 0.226 $\pm$ 0.015 & 0.629 $\pm$ 0.088 & 1.915 $\pm$ 0.100 & 1.136 $\pm$ 0.205 \\
\orat$>5$                                    & 0.278 $\pm$ 0.025 & 0.556 $\pm$ 0.132 & 1.120 $\pm$ 0.185 & 0.876 $\pm$ 0.414 \\
\orat$<5$                                    & 0.214 $\pm$ 0.025 & 0.502 $\pm$ 0.048 & 2.170 $\pm$ 0.111 & 1.478 $\pm$ 0.129 \\
\logoh$>8$                                   & 0.227 $\pm$ 0.024 & 0.527 $\pm$ 0.043 & 1.919 $\pm$ 0.117 & 1.749 $\pm$ 0.129 \\
\logoh$<8$                                   & 0.100 $\pm$ 0.043 & 0.607 $\pm$ 0.174 & 1.426 $\pm$ 0.216 & 0.070 $\pm$ 0.410 \\
$r_{50}>0.6~\rm kpc$                         & 0.216 $\pm$ 0.020 & 0.564 $\pm$ 0.044 & 1.991 $\pm$ 0.129 & 1.841 $\pm$ 0.168 \\
$r_{50}<0.6~\rm kpc$                         & 0.212 $\pm$ 0.026 & 0.609 $\pm$ 0.072 & 1.958 $\pm$ 0.099 & 1.005 $\pm$ 0.205 \\
\sigsfr$\rm >10~\frac{M_{\sun}}{yr~kpc^{2}}$ & 0.193 $\pm$ 0.021 & 0.592 $\pm$ 0.093 & 1.259 $\pm$ 0.189 & 0.891 $\pm$ 0.361 \\
\sigsfr$\rm <10~\frac{M_{\sun}}{yr~kpc^{2}}$ & 0.220 $\pm$ 0.026 & 0.508 $\pm$ 0.054 & 2.021 $\pm$ 0.113 & 1.644 $\pm$ 0.132 \\
all                                          & 0.233 $\pm$ 0.023 & 0.504 $\pm$ 0.042 & 1.899 $\pm$ 0.096 & 1.644 $\pm$ 0.121 \\
\hline\\[-8pt]
\multicolumn{14}{c}{\bf non-LCEs: \fesc$<0.01$} \\[2pt]
\hline
$\beta_{1550}>-2$                            & 0.218 $\pm$ 0.021 & 0.481 $\pm$ 0.050 & 1.999 $\pm$ 0.118 & 1.578 $\pm$ 0.135 \\
$\beta_{1550}<-2$                            & 0.210 $\pm$ 0.007 & 0.731 $\pm$ 0.106 & 2.036 $\pm$ 0.132 & 0.747 $\pm$ 0.332 \\
EW H$\beta$ $>100$ \AA                       & 0.263 $\pm$ 0.014 & 0.653 $\pm$ 0.127 & 1.466 $\pm$ 0.141 & 1.103 $\pm$ 0.333 \\
EW H$\beta$ $<100$ \AA                       & 0.199 $\pm$ 0.005 & 0.499 $\pm$ 0.046 & 2.080 $\pm$ 0.116 & 1.531 $\pm$ 0.124 \\
EW Ly$\alpha$ $>50$ \AA                      & 0.236 $\pm$ 0.010 & 0.680 $\pm$ 0.125 & 1.625 $\pm$ 0.137 & 0.808 $\pm$ 0.336 \\
EW Ly$\alpha$ $<50$ \AA                      & 0.212 $\pm$ 0.008 & 0.503 $\pm$ 0.046 & 2.109 $\pm$ 0.129 & 1.531 $\pm$ 0.135 \\
\fesclya$>0.2$                               & 0.208 $\pm$ 0.038 & 0.697 $\pm$ 0.087 & 1.948 $\pm$ 0.122 & 0.833 $\pm$ 0.270 \\
\fesclya$<0.2$                               & 0.203 $\pm$ 0.012 & 0.500 $\pm$ 0.051 & 2.012 $\pm$ 0.116 & 1.617 $\pm$ 0.141 \\
\muv$>-19.5$                                 & 0.208 $\pm$ 0.045 & 0.443 $\pm$ 0.081 & 1.249 $\pm$ 0.182 & 2.039 $\pm$ 0.287 \\
\muv$<-19.5$                                 & 0.185 $\pm$ 0.007 & 0.541 $\pm$ 0.040 & 2.066 $\pm$ 0.112 & 1.265 $\pm$ 0.129 \\
$\rm M_\star>10^9~M_\odot$                   & 0.197 $\pm$ 0.007 & 0.405 $\pm$ 0.041 & 1.897 $\pm$ 0.151 & 1.675 $\pm$ 0.160 \\
$\rm M_\star<10^9~M_\odot$                   & 0.219 $\pm$ 0.014 & 0.623 $\pm$ 0.088 & 1.938 $\pm$ 0.100 & 1.204 $\pm$ 0.192 \\
\orat$>5$                                    & 0.280 $\pm$ 0.025 & 0.554 $\pm$ 0.132 & 1.123 $\pm$ 0.180 & 0.968 $\pm$ 0.406 \\
\orat$<5$                                    & 0.200 $\pm$ 0.015 & 0.496 $\pm$ 0.044 & 2.161 $\pm$ 0.111 & 1.439 $\pm$ 0.121 \\
\logoh$>8$                                   & 0.215 $\pm$ 0.017 & 0.500 $\pm$ 0.044 & 1.978 $\pm$ 0.117 & 1.598 $\pm$ 0.119 \\
\logoh$<8$                                   & 0.100 $\pm$ 0.043 & 0.603 $\pm$ 0.173 & 1.453 $\pm$ 0.224 & 0.004 $\pm$ 0.426 \\
$r_{50}>0.6~\rm kpc$                         & 0.196 $\pm$ 0.019 & 0.462 $\pm$ 0.052 & 2.072 $\pm$ 0.129 & 1.670 $\pm$ 0.148 \\
$r_{50}<0.6~\rm kpc$                         & 0.211 $\pm$ 0.026 & 0.607 $\pm$ 0.072 & 1.943 $\pm$ 0.100 & 1.032 $\pm$ 0.206 \\
\sigsfr$\rm >10~\frac{M_{\sun}}{yr~kpc^{2}}$ & 0.120 $\pm$ 0.027 & 0.708 $\pm$ 0.140 & 2.039 $\pm$ 0.720 & 0.000      $\pm$ 0.564 \\
\sigsfr$\rm <10~\frac{M_{\sun}}{yr~kpc^{2}}$ & 0.233 $\pm$ 0.023 & 0.507 $\pm$ 0.043 & 1.901 $\pm$ 0.096 & 1.644 $\pm$ 0.120 \\
detections                                   & 0.163 $\pm$ 0.020 & 0.595 $\pm$ 0.054 & 1.208 $\pm$ 0.187 & 0.216 $\pm$ 0.247 \\
non-detections                               & 0.233 $\pm$ 0.026 & 0.521 $\pm$ 0.055 & 2.030 $\pm$ 0.108 & 1.512 $\pm$ 0.134 \\
all                                          & 0.221 $\pm$ 0.020 & 0.535 $\pm$ 0.041 & 1.929 $\pm$ 0.094 & 1.489 $\pm$ 0.114 \\
\hline\\[-8pt]
\multicolumn{14}{c}{\bf weak LCEs: \fesc$\in[0.01,0.05]$} \\[2pt]
\hline
$\beta_{1550}>-2$                            & 0.315 $\pm$ 0.048 & 0.639 $\pm$ 0.055 & 1.424 $\pm$ 0.136 & 1.721 $\pm$ 0.162 \\
$\beta_{1550}<-2$                            & 0.302 $\pm$ 0.010 & 0.855 $\pm$ 0.102 & 1.100 $\pm$ 0.165 & 0.339 $\pm$ 0.195 \\
EW H$\beta$ $>100$ \AA                       & 0.247 $\pm$ 0.033 & 0.599 $\pm$ 0.084 & 0.957 $\pm$ 0.232 & 1.173 $\pm$ 0.324 \\
EW H$\beta$ $<100$ \AA                       & 0.300 $\pm$ 0.032 & 0.689 $\pm$ 0.067 & 1.731 $\pm$ 0.104 & 0.842 $\pm$ 0.154 \\
EW Ly$\alpha$ $>50$ \AA                      & 0.229 $\pm$ 0.027 & 0.644 $\pm$ 0.078 & 1.260 $\pm$ 0.140 & 1.032 $\pm$ 0.203 \\
EW Ly$\alpha$ $<50$ \AA                      & 0.289 $\pm$ 0.033 & 0.699 $\pm$ 0.099 & 1.589 $\pm$ 0.137 & 0.811 $\pm$ 0.230 \\
\fesclya$>0.2$                               & 0.243 $\pm$ 0.031 & 0.643 $\pm$ 0.098 & 1.597 $\pm$ 0.132 & 0.704 $\pm$ 0.241 \\
\fesclya$<0.2$                               & 0.252 $\pm$ 0.027 & 0.606 $\pm$ 0.069 & 1.447 $\pm$ 0.165 & 1.502 $\pm$ 0.180 \\
\muv$>-19.5$                                 & 0.313 $\pm$ 0.049 & 0.559 $\pm$ 0.128 & 1.486 $\pm$ 0.367 & 0.422 $\pm$ 0.436 \\
\muv$<-19.5$                                 & 0.315 $\pm$ 0.029 & 0.709 $\pm$ 0.043 & 1.475 $\pm$ 0.109 & 0.705 $\pm$ 0.148 \\
$\rm M_\star>10^9~M_\odot$                   & 0.262 $\pm$ 0.030 & 0.685 $\pm$ 0.066 & 1.788 $\pm$ 0.156 & 1.460 $\pm$ 0.180 \\
$\rm M_\star<10^9~M_\odot$                   & 0.276 $\pm$ 0.025 & 0.542 $\pm$ 0.059 & 1.340 $\pm$ 0.167 & 0.812 $\pm$ 0.199 \\
\orat$>5$                                    & 0.297 $\pm$ 0.033 & 0.694 $\pm$ 0.094 & 1.019 $\pm$ 0.166 & 1.218 $\pm$ 0.245 \\
\orat$<5$                                    & 0.300 $\pm$ 0.023 & 0.693 $\pm$ 0.068 & 1.459 $\pm$ 0.121 & 0.852 $\pm$ 0.174 \\
\logoh$>8$                                   & 0.380 $\pm$ 0.024 & 0.581 $\pm$ 0.055 & 1.378 $\pm$ 0.137 & 0.645 $\pm$ 0.189 \\
\logoh$<8$                                   & 0.268 $\pm$ 0.030 & 0.591 $\pm$ 0.095 & 1.071 $\pm$ 0.144 & 0.767 $\pm$ 0.235 \\
$r_{50}>0.6~\rm kpc$                         & 0.236 $\pm$ 0.019 & 0.716 $\pm$ 0.082 & 1.505 $\pm$ 0.116 & 0.853 $\pm$ 0.194 \\
$r_{50}<0.6~\rm kpc$                         & 0.302 $\pm$ 0.001 & 0.649 $\pm$ 0.077 & 1.481 $\pm$ 0.168 & 0.976 $\pm$ 0.203 \\
\sigsfr$\rm >10~\frac{M_{\sun}}{yr~kpc^{2}}$ & 0.284 $\pm$ 0.026 & 0.549 $\pm$ 0.053 & 1.520 $\pm$ 0.179 & 1.177 $\pm$ 0.208 \\
\sigsfr$\rm <10~\frac{M_{\sun}}{yr~kpc^{2}}$ & 0.230 $\pm$ 0.028 & 0.787 $\pm$ 0.074 & 1.603 $\pm$ 0.128 & 0.773 $\pm$ 0.205 \\
all                                          & 0.295 $\pm$ 0.029 & 0.664 $\pm$ 0.040 & 1.673 $\pm$ 0.115 & 1.108 $\pm$ 0.133 \\
\hline\\[-8pt]
\multicolumn{14}{c}{\bf strong LCEs: \fesc$>0.05$} \\[2pt]
\hline
$\beta_{1550}>-2$                            &        ---        &        ---        &        ---        &        ---        \\
$\beta_{1550}<-2$                            & 0.251 $\pm$ 0.017 & 0.827 $\pm$ 0.063 & 1.222 $\pm$ 0.119 & 0.460 $\pm$ 0.170 \\
EW H$\beta$ $>100$ \AA                       & 0.306 $\pm$ 0.032 & 0.645 $\pm$ 0.088 & 0.883 $\pm$ 0.124 & 0.463 $\pm$ 0.274 \\
EW H$\beta$ $<100$ \AA                       & 0.289 $\pm$ 0.050 & 0.801 $\pm$ 0.064 & 1.471 $\pm$ 0.194 & 0.755 $\pm$ 0.222 \\
EW Ly$\alpha$ $>50$ \AA                      & 0.284 $\pm$ 0.025 & 0.846 $\pm$ 0.077 & 1.029 $\pm$ 0.127 & 0.092 $\pm$ 0.210 \\
EW Ly$\alpha$ $<50$ \AA                      & 0.271 $\pm$ 0.045 & 0.878 $\pm$ 0.108 & 1.620 $\pm$ 0.156 & 0.152 $\pm$ 0.278 \\
\fesclya$>0.2$                               & 0.273 $\pm$ 0.017 & 0.544 $\pm$ 0.067 & 1.081 $\pm$ 0.112 & 0.113 $\pm$ 0.200 \\
\fesclya$<0.2$                               & 0.272 $\pm$ 0.013 & 0.790 $\pm$ 0.122 & 0.881 $\pm$ 0.243 & 0.429 $\pm$ 0.301 \\
\muv$>-19.5$                                 & 0.312 $\pm$ 0.028 & 0.720 $\pm$ 0.099 & 0.862 $\pm$ 0.139 & 0.000 $\pm$ 0.295 \\
\muv$<-19.5$                                 & 0.251 $\pm$ 0.045 & 0.758 $\pm$ 0.071 & 1.477 $\pm$ 0.146 & 0.564 $\pm$ 0.284 \\
$\rm M_\star>10^9~M_\odot$                   & 0.346 $\pm$ 0.117 & 0.811 $\pm$ 0.070 & 1.181 $\pm$ 0.404 & 0.754 $\pm$ 0.280 \\
$\rm M_\star<10^9~M_\odot$                   & 0.254 $\pm$ 0.018 & 0.793 $\pm$ 0.083 & 1.212 $\pm$ 0.108 & 0.404 $\pm$ 0.219 \\
\orat$>5$                                    & 0.321 $\pm$ 0.033 & 0.707 $\pm$ 0.098 & 0.763 $\pm$ 0.127 & 0.080 $\pm$ 0.273 \\
\orat$<5$                                    & 0.276 $\pm$ 0.049 & 0.836 $\pm$ 0.055 & 1.231 $\pm$ 0.224 & 0.358 $\pm$ 0.183 \\
\logoh$>8$                                   & 0.271 $\pm$ 0.032 & 0.580 $\pm$ 0.112 & 1.585 $\pm$ 0.150 & 0.594 $\pm$ 0.249 \\
\logoh$<8$                                   & 0.501 $\pm$ 0.028 & 0.791 $\pm$ 0.095 & 0.683 $\pm$ 0.140 & 0.325 $\pm$ 0.258 \\
$r_{50}>0.6~\rm kpc$                         &        ---        &        ---        &        ---        &        ---        \\
$r_{50}<0.6~\rm kpc$                         & 0.283 $\pm$ 0.016 & 0.825 $\pm$ 0.060 & 1.080 $\pm$ 0.107 & 0.666 $\pm$ 0.162 \\
\sigsfr$\rm >10~\frac{M_{\sun}}{yr~kpc^{2}}$ & 0.323 $\pm$ 0.030 & 0.737 $\pm$ 0.061 & 1.172 $\pm$ 0.125 & 1.166 $\pm$ 0.215 \\
\sigsfr$\rm <10~\frac{M_{\sun}}{yr~kpc^{2}}$ & 0.246 $\pm$ 0.027 & 0.765 $\pm$ 0.105 & 1.710 $\pm$ 0.127 & 0.085 $\pm$ 0.308 \\
all                                          & 0.271 $\pm$ 0.015 & 0.833 $\pm$ 0.053 & 1.201 $\pm$ 0.094 & 0.583 $\pm$ 0.148
\enddata
\end{deluxetable*}

\section{Feedback and ISM Geometry}\label{sec:discuss}

\subsection{Ionizing Feedback\label{sec:ionfdbk}}

Because the earliest type stars produce the bulk of the intrinsic LyC budget, the youngest stellar populations will drive any feedback that is primarily in the form of ionizing photons
. In principle, these LyC photons could provide feedback by ``brute force'' ionization, creating wide swaths of optically-thin, density bounded regions simply by ionizing all of the neutral gas along a sight line. LyC photons can also drive winds via radiation pressure on dust grains or even on neutral \ion{H}{1} \citep[e.g., ][2024 in prep]{2021ApJ...920L..46K}, accelerating ionized gas up to several $10^3$ \kms. Such radiatively driven winds might provide the momentum necessary to move gas out from the line of sight or, at the very least, increase the optically thin openings between dense gas clumps so that LyC photons can escape. { Extreme ionization environments like those of the Green Peas can alternatively exhibit suppressed mechanical feedback \citep[e.g.,][]{2017ApJ...851L...9J} especially via weak feedback and catastrophic cooling \citep[e.g.][]{2023ApJ...958..149J,2023ApJ...958L..10O}, suggesting that ionizing photons escape when gas conditions are conducive to dense clumping of the ISM. Whether by direct photoionization or by the winds driven by momentum from the LyC, ionization remains the primary feedback mechanism of the youngest stellar populations at low metallicity.}

\begin{figure}
    \centering
    \includegraphics[width=\columnwidth]{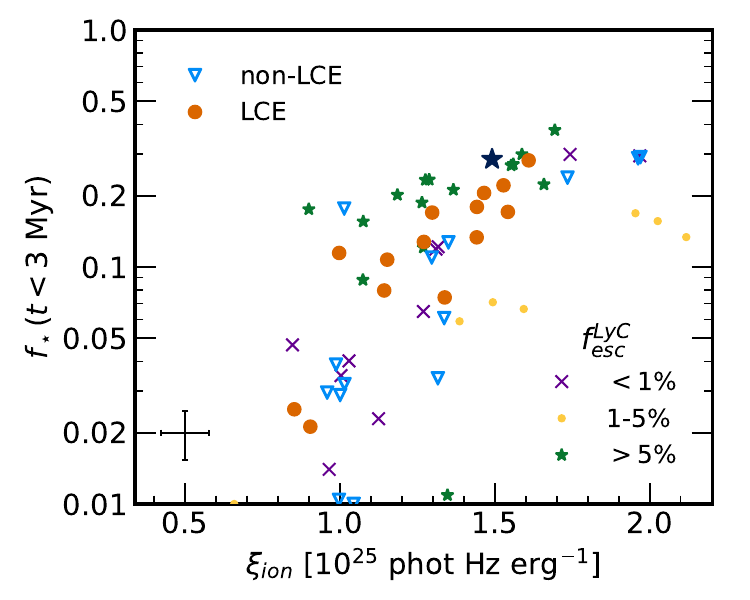}
    \caption{Prevalence of very young stellar populations \fyng\ in the stacked SEDs compared to the relative ionizing budget \xiion$=Q/L_{1500}$, a proxy for the number of LyC photons per unit stellar mass. Symbols as in Figure \ref{fig:Rf_comp}. 
    }
    \label{fig:fyng_xiion}
\end{figure}

We assess the significance of ionizing feedback using the \fyng\ parameter, which quantifies the prevalence of very young stellar populations in the stacked SEDs. { As an additional metric of ionizing feedback, we consider the instrinsic $\xi_{ion}$, the number of ionizing photons $Q$ per unit UV luminosity $L_{1500}$, both quantities measured from the best-fit stellar populations--see \S\ref{sec:specfit}, as a more direct measurement of the relative ionizing budget and effectively a proxy for ``LyC photons per stellar mass''.} We compare these two metrics in Figure \ref{fig:fyng_xiion}, finding a strong correlation between the two metrics (Kendall's $\tau=0.517$ with $p\sim0$). Later stellar populations could still contribute to the total LyC budget. Correlation coefficients indicate that neither \fwr\ nor \fold\ correlate with \xiion\ ($\tau=0.044$ and 0.051, respectively); however, \fwr\ does correlate with the total number of ionizing photons ($\tau=0.244$ with $p=4\times10^{-5}$). Thus, while the youngest stellar populations drive the ionizing feedback, stellar populations in the Wolf-Rayet stage of 3-6 Myr likely augment the total LyC budget.

\begin{figure}
    \centering
    \includegraphics[width=\columnwidth]{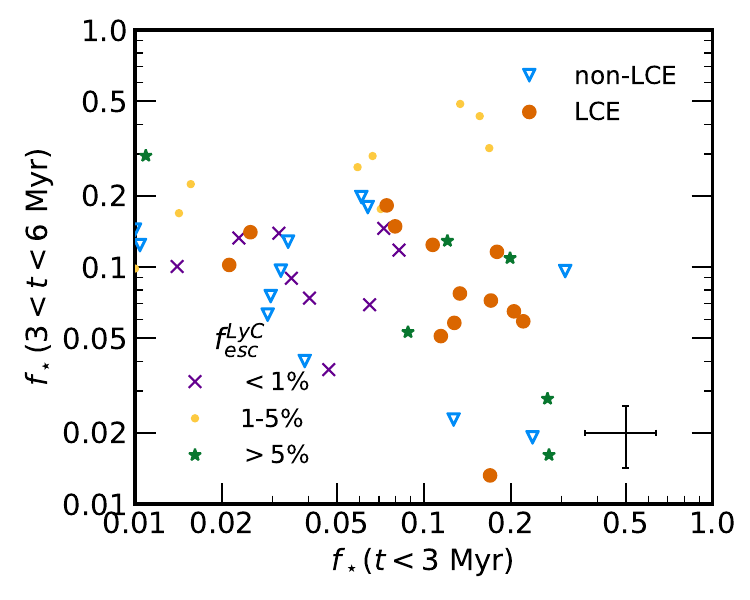}
    \caption{Comparison of \fyng\ and \fwr. Symbols as in Figure \ref{fig:Rf_comp}. We do not show the trend in uncorrelated \fesc\ as the stack of all strong LCEs have \fyng$>0$ and \fwr$=0$ while the stacks of all weak and all non-LCEs have \fyng$=0$ and \fwr$>0$.}
    \label{fig:fyng_fwr}
\end{figure}

This 3-6 Myr population contribution to the LyC (and the UV SED in general) is most pronounced in stacks of LCEs with \fesc$=$1-5\% and, to a lesser degree, stacks of galaxies with \fesc$<1$\%. In contrast, the stacks with the highest \fesc ($> 5$\%) tend to have high fractions of young ($<3$ Myr) populations.  { A comparison of \fyng\ and \fwr\ in Figure \ref{fig:fyng_fwr} indicates a relationship between the prevalence of each population age ($\tau=-0.375$, $p=2.840\times10^{-10}$). One possible interpretation of the anticorrelation between \fwr\ and \fyng\ is an evolution or duty cycle in LyC escape: (i) the youngest populations have the highest \fesc; then (ii) \fesc\ drops as \fwr\ increases and \fyng\ decreases, even though some of these galaxies may have high total \xiion\ due to low levels of ongoing star formation \citep[see similar interpretations by][for LAE stacks at $z=2$ and Carr et al. submitted for \ion{Mg}{2} profiles of LCEs]{2022MNRAS.510.4582N}.} In such a duty cycle, geometry may also change from a clumpy medium permitting LyC escape to a swept-up superbubble \citep[][]{2019ApJ...885...96J}. { Such extreme changes in ISM geometry are not implausible in compact dwarf star-forming galaxies. A $10^3\rm~km~s^{-1}$ radiatively-driven wind, the likes of which are observed in Green Peas \citep[e.g.,][]{2021ApJ...920L..46K}, is capable of reaching kiloparsec scales on the order of megayears.}

In lieu of the duty cycle interpretation, 3-6 Myr stellar populations may be entirely disconnected from the very young populations and augment the intrinsic LyC but undergo no LyC escape; i.e., the \fesc\ is only associated with the youngest OB cluster and only the \xiion\ of these young clusters is relevant for the LyC emergent into the IGM. This disconnect may be especially important in weak (\fesc$=1-5$ \%) LCEs with extended star formation. By mapping high ionization channels and identifying locations of young stellar populations, high spatial resolution of key spectroscopic features in the optical and UV (achievable with the planned Habitable Worlds Observatory) may help to establish a LyC escape duty cycle in greater detail. 

While the role of 3-6 Myr stellar populations is not entirely clear, we have demonstrated a relationship between \fyng\ and the efficiency with which galaxies produce ionizing photons. Moreover, \fesc\ is linked to the youngest stellar populations. We discuss the implications that this relationship has for LyC escape in \S\ref{sec:lycnear}.

\subsection{Mechanical Feedback\label{sec:mechfdbk}}

Feedback from supernovae (SNe) is often invoked to explain or otherwise account for LyC escape \citep[e.g.,][]{2001ApJ...558...56H,2011ApJ...730....5H}.
Due to the metallicity dependence of line-driven stellar winds, SNe are typically an order of magnitude more powerful in mechanical luminosity than winds at the subsolar metallicities common among LCEs, as indicated by predictions from both {\tt Starburst99} and {\tt BPASS}. As a result, stellar winds have a limited time window in which to dominate the mechanical feedback for a single stellar population, with SNe dominating for several tens of Myr afterward \citep[e.g.,][]{1999ApJS..123....3L,2005ARA&A..43..769V,Leitherer_2014,2023ApJ...958..149J,2023MNRAS.tmp.1148S}.
However, the upper mass limit for SNe is not well constrained. \citet{2023ApJ...958..149J} demonstrate that a delay in this intense mechanical feedback is possible, with onsets as late as 10 Myr and  precipitous decline in stellar winds in the intervening 3-10 Myr period as the most massive stars undergo direct black hole collapse \citep[e.g.,][]{2011ApJ...730...70O,2020MNRAS.499.2803P} and/or dynamical ejection \citep[e.g.,][]{2016A&A...590A.107O}. Direct collapse can be particularly significant at low metallicities, with possible reductions in mechanical feedback by as much as $\sim$80\% \citep[e.g.,][]{2023ApJ...958..149J}. Since LCEs tend to have lower metallicities \citep{2022ApJ...930..126F}, we anticipate such delays and reductions in SNe feedback. 

\begin{figure}
    \centering
    \includegraphics[width=\columnwidth]{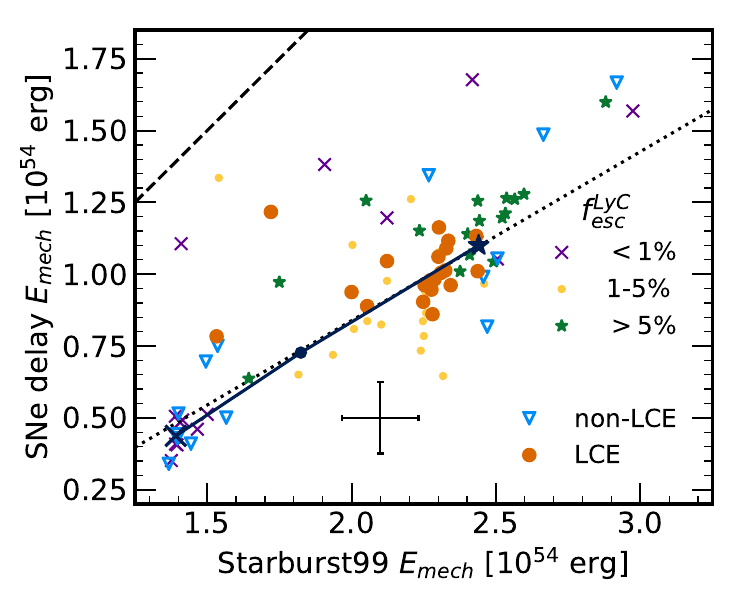}
    \caption{Comparison of standard {\tt Starburst99} cumulative mechanical energy $E_{mech}$ with supernovae starting after 3 Myr (abscissa) and $E_{mech}$ with a time delay in supernovae until after 8 Myr (ordinate). Dashed line indicates 1:1 agreement. Dotted line indicates a 60\% reduction in mechanical feedback due to time delay of supernovae. Symbols as in Figure \ref{fig:Rf_comp}.}
    \label{fig:Emech_comp}
\end{figure}

To determine possible effects of postponed mechanical feedback, in Figure \ref{fig:Emech_comp} we compare the cumulative mechanical energy $E_{mech}$ injected into the ISM by stellar populations with standard SNe onset and with onset delayed until after 8 Myr. The cumulative injected energy $E_{mech}$ for a population of age $t$ is given by integrating over model mechanical luminosities $L_{mech}$ (stellar and SNe mass loss rates scaled by wind terminal velocities $v_\infty$) from the birth of the population until its present age such that
\begin{equation}
    E_{mech}(t) = \int\limits_0^t L_{wind}(t^\prime)+L_{SNe}(t^\prime){\rm d}t^\prime
\end{equation}
\citep[see discussion in][]{2023MNRAS.519L..26H,2023MNRAS.520.5903H}. We calculate $E_{mech}$ from the mechanical luminosities for wind and SNe in each of the 40 {\tt Starburst99} models used in the SED fits, once using the default SNe onset and a second time with SNe onset delayed until after 8 Myr. The effect of delaying mechanical feedback is clear in Figure \ref{fig:Emech_comp} where the delay reduces the mechanical feedback by a factor of $\sim0.6$ in the stacks. While this effect appears not to vary substantially among the stacks, a reduction in \emech\ due to a metallicity-dependent time delay may affect how we interpret \fold\ and \fwr, namely that the latter age will not be associated with strong mechanical feedback. { Lacking further constraints on the SN time delay, in part due to the problematic stellar metallicities discussed in \S\ref{sec:specfit} and in part due to the large scatter in \logoh\ demonstrated in Figure \ref{fig:zeta_expl}, we proceed with the default {\tt Starburst99} predictions for the remainder of our analysis.}

\begin{figure}
    \centering
    \includegraphics[width=\columnwidth]{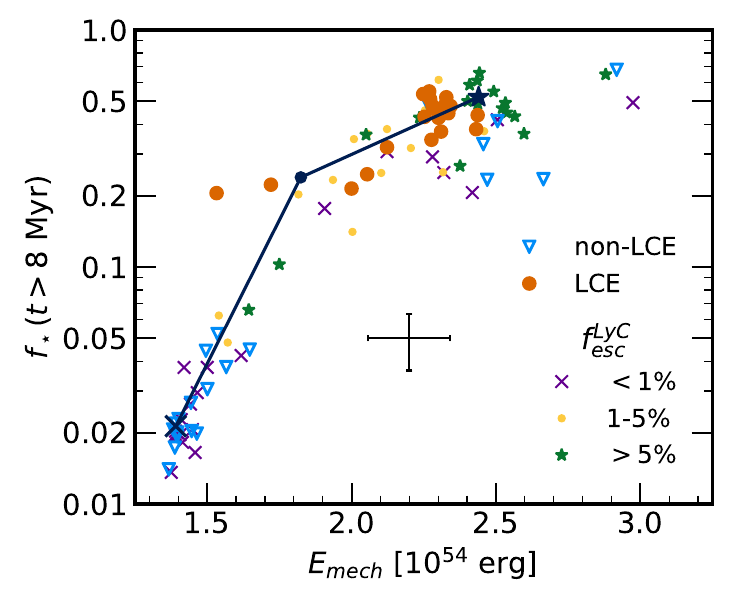}
    \caption{Comparison of \fold\ and cumulative mechanical energy $E_{mech}$ injected into the ISM by stellar populations. Symbols as in Figure \ref{fig:Rf_comp}.}
    \label{fig:fold_Emech}
\end{figure}

As seen with the \fyng-\xiion\ relation, \fold\ and $E_{mech}$ are strongly correlated ($\tau=0.513$). From Figure \ref{fig:fold_Emech}, both \fold\ and \emech\ are far higher in LCEs than in non-LCEs. { Notable exceptions (blue open triangles and purple crosses with \emech$>2\times10^{54}\rm~erg$) include stacks of non-detections and non-LCEs with \orat$>5$ or \logoh$<8$ and non-LCEs with high \sigsfr$>\rm10~M_\odot~yr^{-1}~kpc^{-2}$, all likely genuine LCEs given their properties as discussed in \S\ref{sec:nondets}. Non-detections with $\rm M_\star>\rm10^9~M_\odot$ also exhibit enhanced \emech\ and \fold. Such galaxies should intrinsically have a higher SNe rate assuming a constant \ssfr.} While metallicity introduces caveats regarding SNe time delay, in general this result suggests a substantial role of SN feedback in LyC escape. Regardless, older stellar populations are certainly far more prevalent in LCEs than in non-LCEs, { and these results do not depend on assumptions about SNe onset}.

The prominence of both $<$3 Myr and $>$8 Myr stellar populations in the LCE stacks points to either (i) two modes of LyC escape, as evidenced by UV and optical properties of individual LCEs \citep{2022ApJ...930..126F} or (ii) bursty star formation where subsequent generations of star formation provide the one-two punch of mechanical energy and ionizing radiation, which is consistent with the radio continua of individual LzLCS+ galaxies \citep{2024A&A...688A.198B}. One or both interpretations must be characteristic of LCEs. What we can conclude about the implications for \emech\ is less clear as the cumulative mechanical feedback requires an a priori knowledge of the SNe onset. However, the significance of older stellar populations likely suggests that mechanical feedback, and SN feedback in particular, does indeed play a key role in LyC escape.

\begin{figure}
    \centering
    \includegraphics[width=\columnwidth]{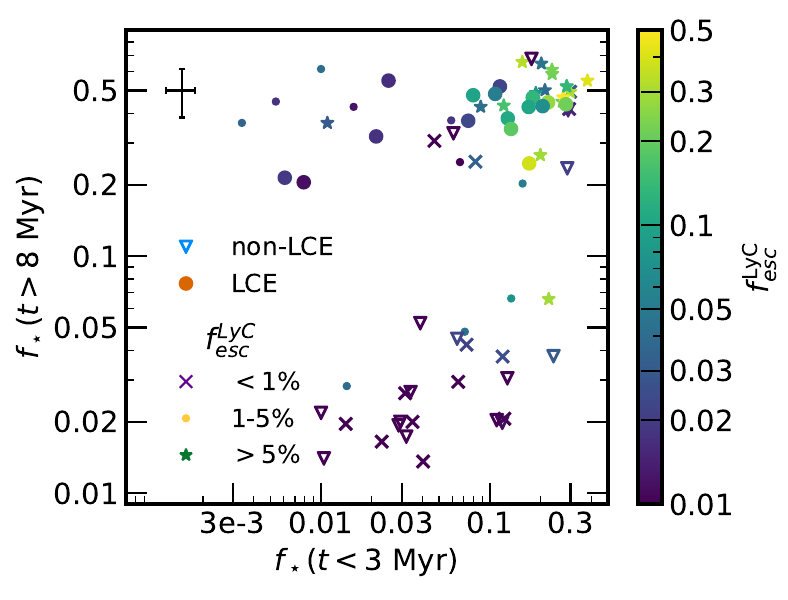}
    \caption{Comparison of \fold\ and \fyng, color-coded by \fesc. Symbols as in Figure \ref{fig:Rf_comp}.}
    \label{fig:enter-label}
\end{figure}

\subsection{Geometry}\label{sec:geometry}

Inferring the geometry of the ISM from absorption features is nuanced given the degeneracies of various parameters and the accompanying assumptions of the covering fraction model. Moreover, the COS G140L LSF is markedly more winged than typical Gaussian profiles and can dramatically affect measurements of gas geometry via the residual flux (see \S\ref{sec:absn} and Appendices \ref{apx:conv} and \ref{apx:abs}). { As reflected in our simulations, a single LSF correction cannot be applied to the measured $R_f$ and $W_\lambda$ values -- as such, we can only make approximate corrections and cannot derive precise physical values such as column density for any individual stack.}

\begin{figure}
    \centering
    \includegraphics[width=\columnwidth]{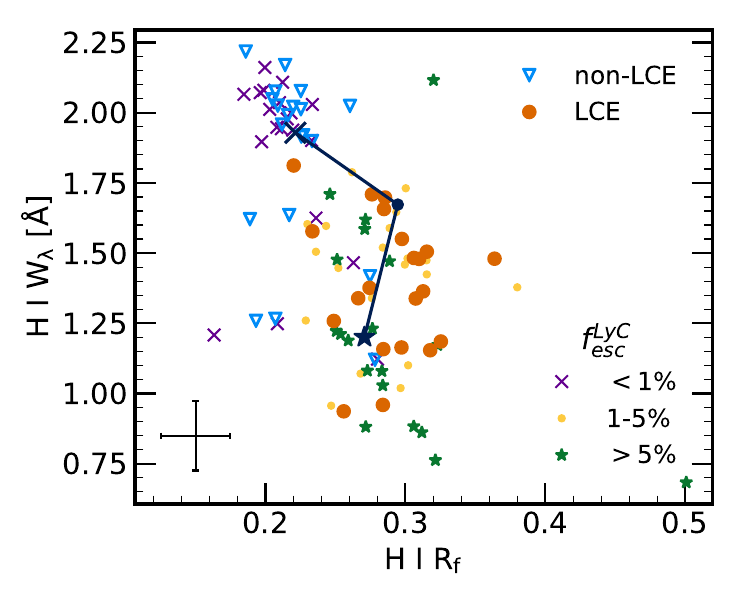}
    \includegraphics[width=\columnwidth]{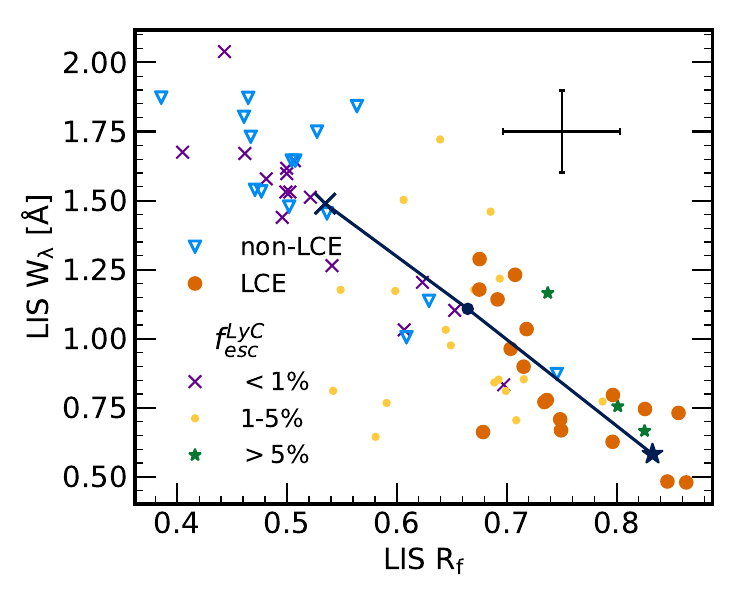}
    \caption{Comparison of $W_\lambda$ and $R_f$ for \ion{H}{1} (\emph{top}) and LIS (\emph{bottom}). Symbols as in Figure \ref{fig:Rf_comp}.}
    \label{fig:wl_rf}
\end{figure}

\begin{figure}
    \centering
    \includegraphics[width=\columnwidth]{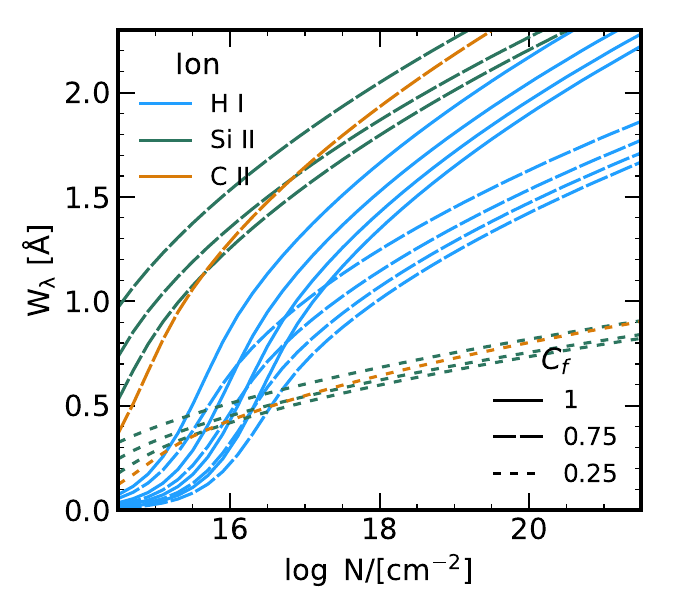}
    \caption{Curves of growth for LIS ion transitions \ion{Si}{2}$\lambda1190,93,1260$ (green) and \ion{C}{2}$\lambda1334$ and \ion{H}{1} transitions Ly$\gamma,\delta,\varepsilon,\zeta$ (blue) for $C_f=1$ (solid lines), 0.75 (long dashes), and 0.25 (short dashes) assuming $b=100\rm~km~s^{-1}$ and accounting for the COS G140L LSF (see Appendices \ref{apx:conv} and \ref{apx:abs}). $C_f=0.75$ to 1 is characteristic of \ion{H}{1} in the stacks while LIS span $C_f=0.25$ to 0.7, in both cases assuming a $\approx30$\% reduction in $R_f$ due to the LSF.}
    \label{fig:cog}
\end{figure}

Despite these caveats, our measurements of the absorption lines still provide meaningful insight into the geometry of LCEs and their non-leaking counterparts. To begin, we compare $W_\lambda$ and $R_f$ in Figure \ref{fig:wl_rf} for \ion{H}{1} and LIS lines to assess whether clumpiness, column density, or a combination of the two accounts for variations in $W_\lambda$. While \wfh\ is negatively correlated with \rfh\ ($\tau=-0.397$, $p<10^{-7}$, which bootstrapping suggests is robust to outliers evident in the top panel of Figure \ref{fig:wl_rf}), its dependence on \rfh\ appears predominantly determined by the separation of LCEs and non-LCEs. Indeed, from Figure \ref{fig:wl_rf} (top panel), \rfh\ is roughly constant at 0.3 for LCEs as \wfh\ decreases, indicating that any decrease in \wfh\ must be due to a decline in $N_{\rm H~I}$; for non-LCEs, the same is true at slightly higher covering fraction (\rfh$\sim0.2$). The Lyman series curves of growth shown in Figure \ref{fig:cog} suggest the low values of \wfh=1 \AA\ and \cfh$\approx0.75$ (from correcting \rfh$=0.3$ for the LSF) correspond to column densities as low as $\rm10^{17}~cm^{-2}$, sufficient to allow isotropic LyC escape as high as \fesc$\sim25$\%. A similar interpretation holds for non-LCEs. LyC optical depth along the line of sight is decreasing with \wfh, but the distribution of \ion{H}{1} remains more or less uniform.

On the other hand, the strong dependence of \wfl\ on \rfl\ ($\tau=-0.508$ with $p=0$, Figure \ref{fig:wl_rf} bottom panel) indicates that decreasing covering fraction is responsible, at least in part, for the differences in \wfl. The LSF-adjusted curve of growth for LIS lines, as depicted in Figure \ref{fig:cog}, suggests the range of \cfl\ (0.25 to 0.75 from a 30\% increase in $R_f$ due to LSF effects, see Appendix \ref{apx:abs}) can account for much of the observed range in \wfl. Assuming covering fraction is the primary driver of variations in \wfl, the observed \wfl\ would consistently correspond to column densities in excess of $N\sim10^{16}\rm~cm^{-2}$ and, in cases when \wfl$\ga2$ \AA, as high as $10^{18}\rm~cm^{-2}$. Under such conditions, the LIS lines trace especially dense gas, as the \ion{H}{1} column will be even higher in regions occupied by the LIS gas. Combined with the strong $C_f$ dependence of $W_\lambda$, the LIS lines likely arise from clumps or clouds of dense gas embedded in the ambient neutral ISM.

Since the \ion{H}{1} lines trace the ambient neutral ISM and the LIS lines trace clouds of dense gas, we can anticipate some discrepancy in the line properties such as that observed in the LzLCS+ stacks and presented in \S\ref{sec:absn}. As demonstrated both in \S\ref{sec:absn} and Appendix \ref{apx:abs}, observational effects cannot account for the observed differences between the \ion{H}{1} and LIS residual fluxes. Neither does metallicity, which should primarily affect the LIS column densities, correlate with the LIS $R_f$, LIS $W_\lambda$, or the $C_f$ ratio (see \S\ref{sec:absn}). Thus, any variations in LIS $R_f$ or the $C_f$ ratio may be attributable instead to differences in the geometry of the gas. Given the $W_\lambda$-$R_f$ trends discussed above and shown in Figure \ref{fig:wl_rf}, geometric differences traced by the $C_f$ ratio would correspond to the distribution of dense gas clouds in the ambient ISM. We explore the geometry in greater detail here by comparison of this discrepancy with \ion{H}{1} LyC optical depth (as measured by \wfh) and dustiness (as measured by $\beta_{1550}$).

\begin{figure}
    \centering
    \includegraphics[width=\columnwidth]{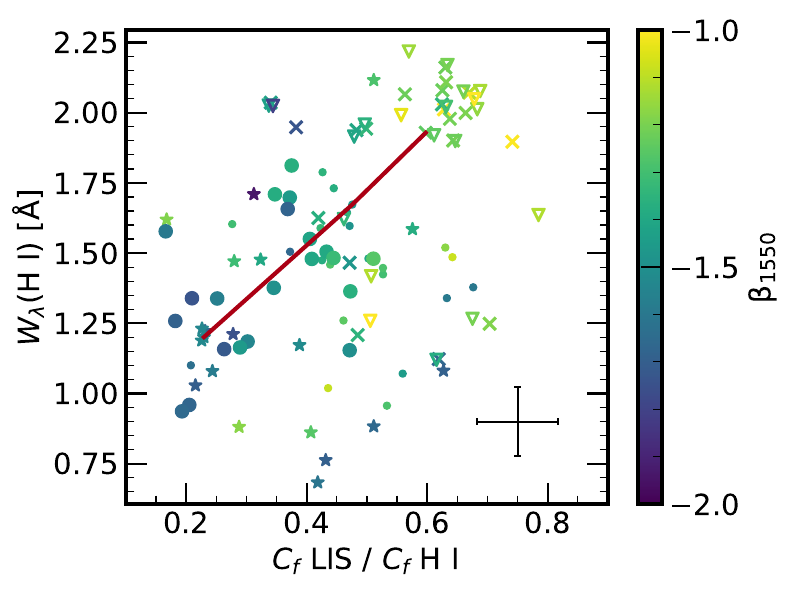}
    \caption{Comparison of the \ion{H}{1} $W_\lambda$ with the residual flux discrepancy expressed as a ratio of covering fractions $C_f = 1-R_f$. Symbols as in Figure \ref{fig:Rf_comp} and color-coded by $\beta_{1550}$, the slope of the UV continuum which is primarily sensitive to dust attenuation. Red line indicates the trend in uncorrelated stacks of all non-LCEs (\fesc$<1$\%, cross), all weak LCEs (\fesc$=1$-5\%, circle), and all strong LCEs (\fesc$>5$\%).}
    \label{fig:wh1_zeta_ebv}
\end{figure}

We compare the ratio of covering fractions, \wfh, and $\beta_{1550}$ in Figure \ref{fig:wh1_zeta_ebv}. The \ion{H}{1} column appears to decline as the dense LIS gas becomes increasingly sparse, indicating a decline in ambient neutral gas coincident with a decline in dense gas clouds. Indeed, the $C_f$ ratio and \wfh\ are strongly correlated ($\tau=0.295$, $p=6.922\times10^{-7}$) with relatively few outliers. $\beta_{1550}$ also correlates strongly with the $C_f$ ratio ($\tau=0.462$, $p<10^{-7}$), indicating a decrease in ISM opacity due to dust grains along the line of sight as fewer clouds of dense gas populate the ISM. As shown in Figure \ref{fig:wh1_zeta_ebv}, dust attenuation also declines significantly with \wfh, with $\tau=0.311$ and $p<10^{-7}$. These relationships indicate a lockstep decline in dust attenuation, ambient neutral gas column density, and the prevalence dense gas clouds along the line of sight. Our finding is consistent with results from individual galaxies \citep[e.g.,][]{2020A&A...639A..85G,2022A&A...663A..59S} and from stacks of galaxy spectra at higher redshifts \citep[$z=3$][]{2016ApJ...828..108R}.

\begin{figure*}
    \centering
    \includegraphics[width=\linewidth,trim={0.5in 0.35in 0.5in 2.15in},clip=True]{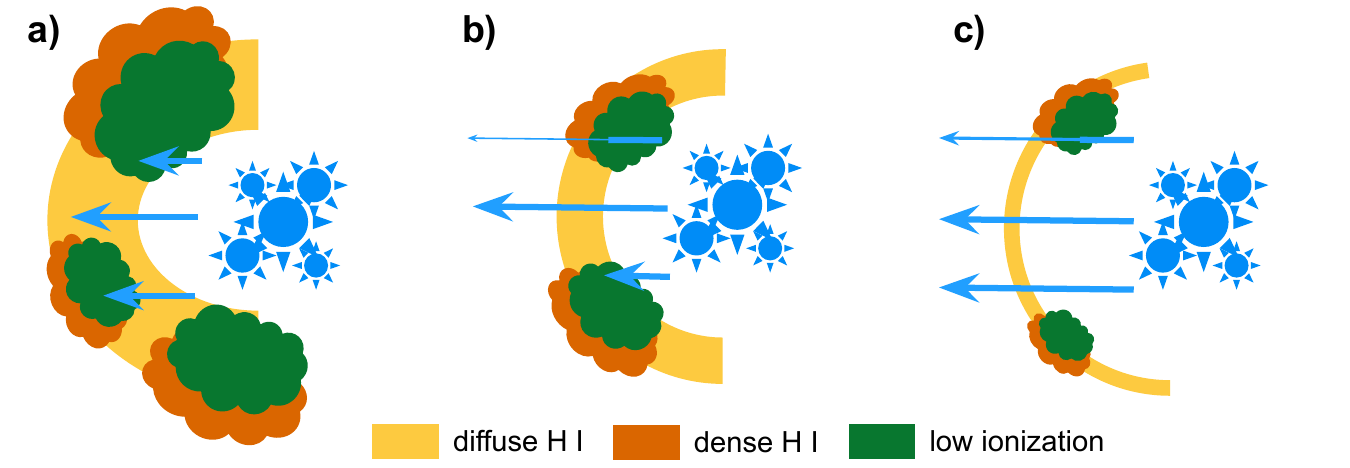}
    \caption{Cartoon depiction of different ISM geometries determined from LIS (green) and \ion{H}{1} (orange, yellow) absorption lines. The three panels depict different $C_f$ ratio regimes with {\bf a)} some \ion{H}{1}-LIS discrepancy, many dense clouds (orange and green), and abundant diffuse \ion{H}{1} (yellow), {\bf b)} substantial \ion{H}{1}-LIS discrepancy with few dense clouds but abundant diffuse \ion{H}{1}, and {\bf c)} extreme ionization with few dense clouds and loss of most ambient diffuse \ion{H}{1}.}
    \label{fig:tau_cartoon}
\end{figure*}

To illustrate a holistic interpretation of our geometry results, in Figure \ref{fig:tau_cartoon} we present as a visual aid a cartoon model of the relationship between $C_f$ discrepancies and ISM geometry. Here, a foreground screen through which the intrinsic stellar SED is transmitted consists of ambient diffuse \ion{H}{1} populated by clouds of dense gas containing \ion{H}{1} and LIS ions. When the $C_f$ is high, both \ion{H}{1} and LIS absorption are ubiquitous. When \cfl\ decreases (\rfl\ increases), there are fewer dense clouds along the line of sight; however, diffuse \ion{H}{1} persists, resulting in a high \cfh\ (low \rfh). In a more extreme scenario, even the diffuse \ion{H}{1} becomes ionized, causing the \wfh\ to decrease even if dense clouds are present.

\subsection{Feedback and Geometry}\label{sec:fdbkgeom}

\begin{figure}
    \centering
    \includegraphics[width=\columnwidth]{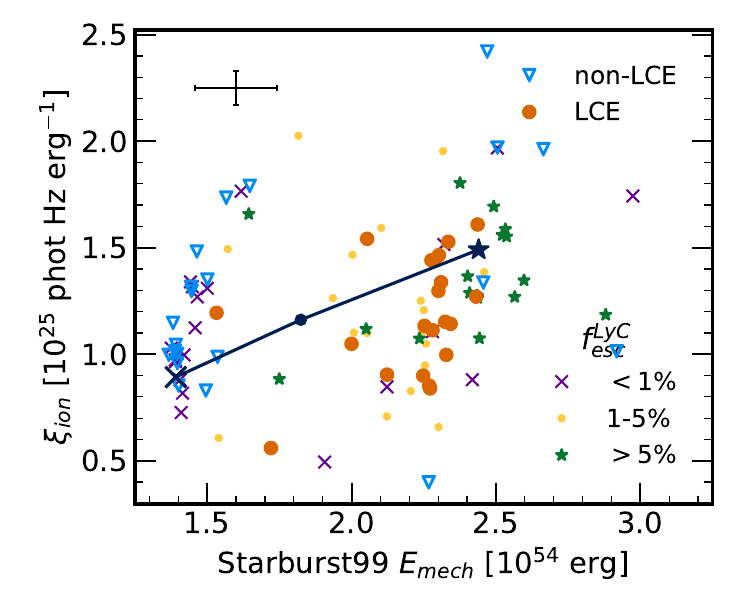}
    \caption{Comparison of the relative ionizing photon production rate \xiion\ and the cumulative mechanical feedback \emech\ without any SN delay. Symbols as in Figure \ref{fig:Rf_comp}. While \xiion\ may be mildly enhanced in LCEs, \emech\ is markedly higher among LCEs than non-LCEs.}
    \label{fig:IonMech}
\end{figure}

Perhaps the most critical question at this juncture is whether the stellar feedback and ISM geometry are in any capacity related to one another. In turn, such insight can inform \emph{how} the unique conditions for LyC escape arise. To begin, we compare the different types of feedback in Figure \ref{fig:IonMech} to determine whether the two are related in some fashion. While \xiion\ and $E_{mech}$ do not appear particularly correlated, the stacks of LCEs and high \fesc\ stacks exhibit $E_{mech}$ higher than most of the non-leaker and non-detection stacks by as much as 50\%.

The few cases where non-detection stacks do exhibit higher $E_{mech}$ include stacks on high \orat\ or low \logoh, which accounts for the high \xiion\ associated with these stacks. However, other non-detection stacks on high mass, high \sigsfr, and high UV $r_{50}$ also demonstrate enhanced $E_{mech}$, as one might anticipate given the increased likelihood of SNe in these high UV luminosity systems. In some cases, these non-detection stacks are true LCEs in disguise, such as the high \orat\ stacks which have non-negligible \fesc.

For each type of feedback, we assess $W_\lambda$, $R_f$, the $C_f$ ratio, and $\beta_{1550}$ in concert both with \xiion\ and with \emech. We quantify the relationships between each type of feedback and ISM metric using Kendall's $\tau$ and list the results in Table \ref{tab:kendall1}. Following the relations between stellar populations and feedback explored in \S\ref{sec:ionfdbk}-\ref{sec:mechfdbk}, we also list Kendall's $\tau$ for correlations between feedback and ISM metrics with \fyng, \fwr, and \fold\ in Table \ref{tab:kendall2}. \emech\ strongly and significantly correlates with all of the ISM metrics. However, \xiion\ only correlates strongly and significantly with the absorption line $W_\lambda$. As discussed in \S\ref{sec:geometry}, \wfh\ traces ambient diffuse neutral gas while \wfl\ corresponds to the prevalence of clouds of dense gas. The $W_\lambda$ correlation with \xiion\ therefore suggests that that ionizing feedback primarily affects the column densities of gas, both the diffuse \ion{H}{1} and dense clouds, and could lead to increasingly isotropic LyC escape with \fesc\ higher along channels of diffuse gas. Meanwhile, mechanical feedback likely plays the pivotal role in shaping the anisotropic distribution of dense, dusty gas, as evidenced by the strong correlation of $\beta_{1550}$, \rfh, \rfl, and \cfl/\cfh\ with \emech\ but \emph{not} with \xiion. While both ionizing and mechanical feedback affect the optical depth of gas, these results suggest that SNe and stellar winds are the critical mechanisms for shaping the anisotropic geometry of the ISM.


\begin{deluxetable*}{l l l l l}
\tablecaption{Kendall's $\tau$ correlation coefficients (columns 2 and 4) with null hypothesis probabilities $p$ (columns 3 and 5) for various stellar population and ISM metrics with \xiion\ (columns 2-3) and \emech\ (columns 4-5).\label{tab:kendall1}}
\tablewidth{\textwidth}
\tablehead{
\colhead{Metric} & \colhead{$\tau$ \xiion} & \colhead{$p$ \xiion} & \colhead{$\tau$ \emech} & \colhead{$p$ \emech}}
\startdata
\multicolumn{5}{c}{stellar populations}\\
\hline
\fyng          &  ~0.500$\pm$0.027  &  $<2\times 10^{-52}$     &  ~0.396$\pm$0.034  &  $2.945\times 10^{-11}$  \\
\fwr           &  ~0.008$\pm$0.041  &  $8.931\times 10^{ -1}$  &  -0.176$\pm$0.027  &  $3.115\times 10^{ -3}$  \\
\fold          &  ~0.090$\pm$0.038  &  $1.283\times 10^{ -1}$  &  ~0.687$\pm$0.019  &  $<2\times 10^{-52}$     \\
\hline
\multicolumn{5}{c}{ISM}\\
\hline
$\beta_{1550}$ &  -0.090$\pm$0.037  &  $1.309\times 10^{ -1}$  &  -0.337$\pm$0.015  &  $1.445\times 10^{ -8}$  \\
\wfh           &  -0.375$\pm$0.021  &  $2.803\times 10^{-10}$  &  -0.400$\pm$0.056  &  $1.862\times 10^{-11}$  \\
\wfl           &  -0.243$\pm$0.052  &  $4.504\times 10^{ -5}$  &  -0.370$\pm$0.021  &  $5.179\times 10^{-10}$  \\
\rfh           &  ~0.093$\pm$0.039  &  $1.165\times 10^{ -1}$  &  ~0.248$\pm$0.059  &  $3.079\times 10^{ -5}$  \\
\rfl           &  ~0.150$\pm$0.037  &  $1.179\times 10^{ -2}$  &  ~0.345$\pm$0.031  &  $6.461\times 10^{ -9}$  \\
\cfl/\cfh      &  -0.161$\pm$0.035  &  $6.650\times 10^{ -3}$  &  -0.334$\pm$0.049  &  $1.942\times 10^{ -8}$
\enddata
\end{deluxetable*}

\begin{deluxetable*}{l l l l l l l}
\tablecaption{Kendall's $\tau$ correlation coefficients (columns 2, 4, and 6) with null hypothesis probabilities $p$ (columns 3, 5, and 7) for various feedback and ISM metrics with \fyng\ (columns 2-3), \fwr\ (columns 4-5), and \fold (columns 6-7).\label{tab:kendall2}}
\tablewidth{\textwidth}
\tablehead{
\colhead{Metric} & \colhead{$\tau$ \fyng} & \colhead{$p$ \fyng} & \colhead{$\tau$ \fwr} & \colhead{$p$ \fwr} & \colhead{$\tau$ \fold} & \colhead{$p$ \fold}}
\startdata
\xiion         &   0.500$\pm$0.027  &  $<2\times 10^{-52}$  &   0.008$\pm$0.031  &  $8.931\times 10^{ -1}$  &   0.090$\pm$0.053  &  $1.283\times 10^{ -1}$  \\
\emech         &   0.396$\pm$0.040  &  $2.945\times 10^{-11}$  &  -0.176$\pm$0.038  &  $3.115\times 10^{ -3}$  &   0.687$\pm$0.024  &  $<2\times 10^{-52}$  \\
$\beta_{1550}$ &  -0.181$\pm$0.024  &  $2.401\times 10^{ -3}$  &   0.203$\pm$0.051  &  $6.302\times 10^{ -4}$  &  -0.394$\pm$0.033  &  $3.524\times 10^{-11}$  \\
\wfh           &  -0.376$\pm$0.052  &  $2.660\times 10^{-10}$  &   0.130$\pm$0.055  &  $2.907\times 10^{ -2}$  &  -0.341$\pm$0.045  &  $1.034\times 10^{ -8}$  \\
\wfl           &  -0.276$\pm$0.033  &  $3.591\times 10^{ -6}$  &   0.153$\pm$0.031  &  $9.894\times 10^{ -3}$  &  -0.394$\pm$0.031  &  $3.428\times 10^{-11}$  \\
\rfh           &   0.112$\pm$0.031  &  $6.050\times 10^{ -2}$  &   0.077$\pm$0.031  &  $1.982\times 10^{ -1}$  &   0.320$\pm$0.050  &  $7.667\times 10^{ -8}$  \\
\rfl           &   0.180$\pm$0.036  &  $2.417\times 10^{ -3}$  &  -0.036$\pm$0.040  &  $5.413\times 10^{ -1}$  &   0.433$\pm$0.063  &  $3.328\times 10^{-13}$  \\
\cfl/\cfh      &  -0.170$\pm$0.024  &  $4.258\times 10^{ -3}$  &   0.044$\pm$0.034  &  $4.611\times 10^{ -1}$  &  -0.405$\pm$0.046  &  $9.615\times 10^{-12}$  \\
\enddata
\end{deluxetable*}

\section{Implications for LyC Escape}\label{sec:lycescape}

Above, we have considered three different properties related to LyC escape: stellar populations (production of the LyC), feedback (clearing the path for LyC escape), and ISM geometry (the LyC escape route). We now assess individual and multivariate correlations related to these different properties to determine which are most fundamentally important for the production and escape of the LyC. In order to obtain a holistic interpretation of our results, we employ canonical correlation analysis (CCA) to determine how multiple variables correlate together with \fesc\ and elaborate on this technique in Appendix \ref{apx:cca}.

\begin{deluxetable*}{l c c c c c c}
\tablecaption{Correlation of stack properties with \fesc\ as determined with single- and multivariate analysis. For individual variables, we report Kendall's $\tau$ (column 2) with null hypothesis probability $p$ (column 3). For a multivariate assessment, we report the canonical correlation coefficients $\mathbf{a}$ (columns 4-7). The final row indicates the total multivariate correlation between the combined variables and \fesc\ determined using CCA.\label{tab:cca}}
\tablewidth{\textwidth}
\tablehead{
\colhead{Property} & \colhead{$\tau$} & \colhead{$p$} & \colhead{$\rm\mathbf{a}_{stars}$} & \colhead{$\rm\mathbf{a}_{fdbk}$} & \colhead{$\rm\mathbf{a}_{ISM}$} & \colhead{$\rm\mathbf{a}_{total}$}
}
\startdata
\fyng          &  0.400$\pm$0.025  &  $1.688\times 10^{-11}$ & 8.756 & --    & --     &  9.210 \\
\fwr           & -0.106$\pm$0.048  &  $7.518\times 10^{ -2}$ & 0.103 & --    & --     &  2.567 \\
\fold          &  0.456$\pm$0.017  &  $1.754\times 10^{-14}$ & 1.100 & --    & --     & -1.297 \\
\xiion         &  0.344$\pm$0.078  &  $7.033\times 10^{ -9}$ & --    & 1.927 & --     & -0.947 \\
\emech         &  0.456$\pm$0.000  &  $1.876\times 10^{-14}$ & --    & 5.443 & --     &  0.302 \\
$\beta_{1550}$ & -0.553$\pm$0.012  &  $<2\times 10^{-52}$    & --    & --    & -2.348 & -2.295 \\
\wfh           & -0.550$\pm$0.032  &  $<2\times 10^{-52}$    & --    & --    & -1.706 & -0.727 \\
\wfl           & -0.456$\pm$0.025  &  $1.898\times 10^{-14}$ & --    & --    & -0.422 & -0.224 \\
\rfh           &  0.341$\pm$0.040  &  $9.860\times 10^{ -9}$ & --    & --    & -1.974 & -0.769 \\
\rfl           &  0.478$\pm$0.034  &  $9.992\times 10^{-16}$ & --    & --    & ~4.751 &  4.684 \\
\cfh/\cfl      & -0.440$\pm$0.031  &  $1.336\times 10^{-13}$ & --    & --    & ~5.719 &  4.777 \\
\hline
$\rho$ & & & 0.640 & 0.477 & 0.737 & 0.804
\enddata
\end{deluxetable*}

\subsection{Optimal LyC Escape Conditions}\label{sec:lycnear}

While the ISM conditions consistently have the strongest correlations with \fesc, multivariate analysis indicates that the prevalence of very young stellar populations (as measured by \fyng) is the most significant term for maximizing multivariate correlations with \fesc, both among the stellar populations and among all the variables together. To some degree, this result is intuitive because O stars will inherently dominate the intrinsic LyC budget and indeed are necessary to achieve non-negligible emergent LyC flux from a galaxy. However, the significance of these young populations over both those with WR stars generating substantial stellar winds and those after SNe onset is less clear given that older populations produce the mechanical feedback which our models indicate is necessary for clearing out the ISM. Our results thus differ from previous studies which suggest that WR-stage stellar populations are key to LyC escape as a result of, e.g., stellar winds, LyC production, and the onset of SNe \citep[e.g.,][]{2013ApJ...766...91J,2013ApJ...779...76Z,2016MNRAS.459.3614M}.

Perhaps the significance of \fyng\ over \fwr\ and \fold\ indicates that, for the youngest stellar populations, high \fesc\ could be inevitable among LzLCS+ galaxies due to extreme ionization, which decreases \ion{H}{1} and promotes increasingly isotropic escape. Indeed, the strongest LCEs in the LzLCS+ are Green Peas \citep{2022ApJ...930..126F}. These galaxies, known for concentrated star formation, very young stellar populations, and extreme ionization, comprise a significant portion of the galaxies contributing to LCE stacks and, especially, high \fesc\ stacks. The Green Pea LCEs may even be dominating the apparent correlations due to their characteristically young ages and high \fesc. With that caveat in mind, on average, the ionization mechanisms at play in LyC escape in Green Peas are likely the key for the most prodigious LCEs. I.e., radiative feedback dominates LyC emission, particularly when mechanical feedback could be delayed in low metallicity stellar populations.

{ Although Green Peas can exhibit strong WR features \citep{2012ApJ...754L..22A,2022MNRAS.511.2515F}, the weak correlation of \fwr\ with \fesc\ indicates that the presence of WR stars is insufficient to account for the escape of LyC photons. Moreover, higher \fwr\ appears consistent with the weaker LCEs (\fesc$=1-5$\%) as in Figure \ref{fig:fyng_fwr}, suggesting that Green Peas with significant WR contributions may be in transition between LCE modes or evolving in an LCE duty cycle. Indeed, most of the Green Peas with WR features exhibit \orat$\la5$, indicating lower values of \fesc\ \citep{2018MNRAS.478.4851I,2022ApJ...930..126F}. LCEs with WR features may instead contain very massive stars (VMS) or young stellar populations on the verge of transition, as with the Sunburst arc (\citealt{2023ApJ...945...53V,2024arXiv240408884R}, see \S\ref{sec:lycfar}).}

That being said, of the types of feedback quantified by \xiion\ and \emech, the latter is more important for maximizing the correlation with \fesc\ both individually as quantified by $\tau$ and holistically when quantified by CCA. Thus, while very young stellar populations are more strongly associated with LyC escape than older ones, mechanical feedback is also more important than ionizing feedback for clearing out the ISM. The significance of both \rfl\ and \emech\ in the CCA illustrates the key role mechanical feedback likely plays in producing this geometry (see also Table \ref{tab:kendall2}). Moreover, the paucity of dense, dusty gas clouds relative to ambient \ion{H}{1}, as quantified by the \cfl/\cfh\ ratio, is one of the most important among the ISM tracers for maximizing correlation with \fesc, reinforcing the idea that patchy dense gas distributions produced by mechanical feedback lead to more efficient LyC escape. A sign change in the correlation coefficient for the $C_f$ ratio when accounting for additional variables, though, indicates that \cfl/\cfh\ traces something more than a combination of optical depth via \ion{H}{1} and prevalence of dense clouds via LIS. In other words, given additional constraints, a higher $C_f$ ratio can imply higher \fesc. As discussed in \S\ref{sec:geometry}, the $C_f$ ratio could be indicating extreme ionization a fundamental to high \fesc.

Incorporating all properties simultaneously points to a possible combination of ionization and mechanical feedback to achieve substantial \fesc. However, while the Kendall's $\tau$ coefficients indicate that feedback is clearly necessary for LyC escape, neither \xiion\ nor \emech\ appear as relevant for \fesc\ as other properties. Instead, the youngest stellar populations, coupled with \cfl/\cfh, $\beta_{1550}$, and $R_f$, collectively drive the correlation with \fesc\ with \fyng\ dominating. One interpretation of this result is that \xiion\ and \emech\ are implicitly expressed by the ISM terms since feedback correlates with the geometry. However, feedback produces a relatively weak CCA correlation coefficient of $\rho=0.477$, indicating that the LyC per unit mass and mechanical feedback are not nearly as important as which stellar populations are present ($\rho=0.640$) or the ISM geometry ($\rho=0.737$). Perhaps instead SN feedback is helpful but unnecessary for LyC escape--favoring two different LCE mechanisms rather than a two-stage burst.

The ISM measurements are the most strongly correlated with \fesc. This result is somewhat unsurprising given that ISM conditions are intrinsically coupled to \fesc\ both in observations \citep[e.g.,][]{2018ApJ...869..123S,2020A&A...639A..85G,2016ApJ...826L..24S} and in simulations \citep[e.g.,][]{2021A&A...646A..80M,2024ApJ...969...50G}. Indeed, all of the properties alone correlate strongly and significantly with \fesc, most of all \wfh\ and $\beta_{1550}$. Tracing neutral gas and dust, respectively, \wfh\ and $\beta_{1550}$ indicate the total LyC optical depth along the line of sight. However, CCA predicts that the paucity of density clouds, as measured by \rfl, and extreme ionization, as measured by \cfl/\cfh, are relatively more correlated with \fesc\ among both the ISM and the full set of properties when accounting for other variables. Thus, a confluence of extreme ionization, traced by low $C_f$ ratios, and mechanically-widened optically-thin channels traced by high \rfl\ likely optimize for high \fesc. { Such an interpretation is not without precedent: earlier analysis by \citet{2020A&A...644A..21R} of optical emission lines implies a similar scenario.}

\begin{figure}
    \centering
    \includegraphics[width=\columnwidth,clip=True,trim={0in 0.5in 0in 0in}]{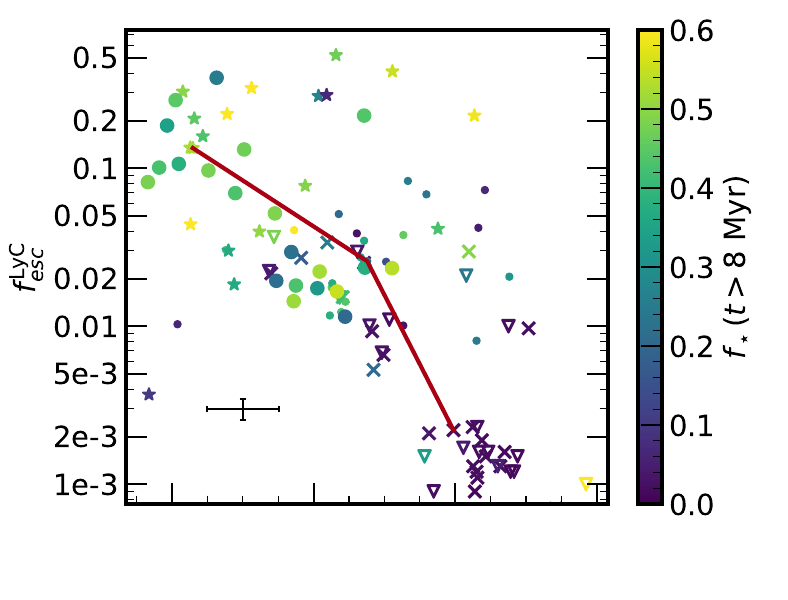}
    \includegraphics[width=\columnwidth]{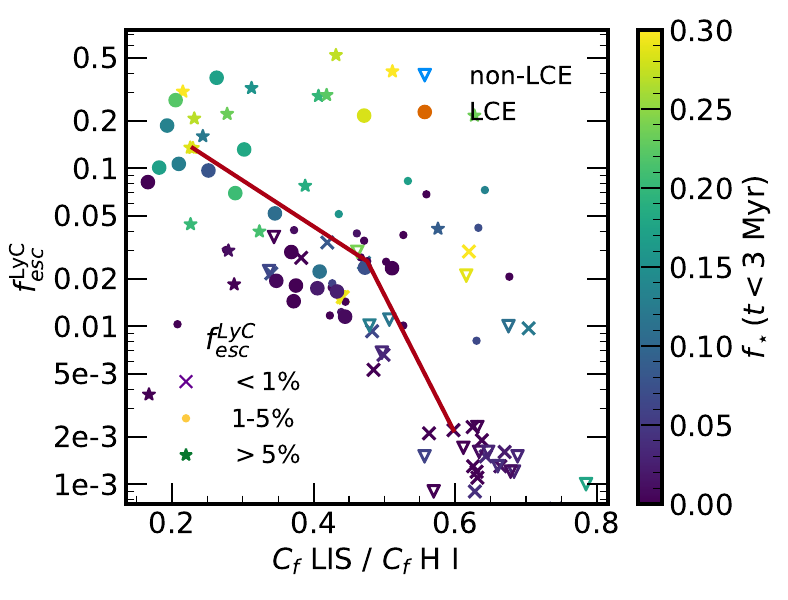}
    \caption{Comparison of \fesc\ with \cfl/\cfh\ colored by ({\it top}) \fold\ and ({\it bottom}) \fyng. We highlight the trend in uncorrelated stacks in red. Trends among \fesc, \cfl/\cfh, and population prevalence are apparent in both cases.}
    \label{fig:fesc_tau_fyng}
\end{figure}

\begin{figure*}
    \centering
    \includegraphics[width=\textwidth,clip=True,trim={0in 0.25in 0in 0.5in}]{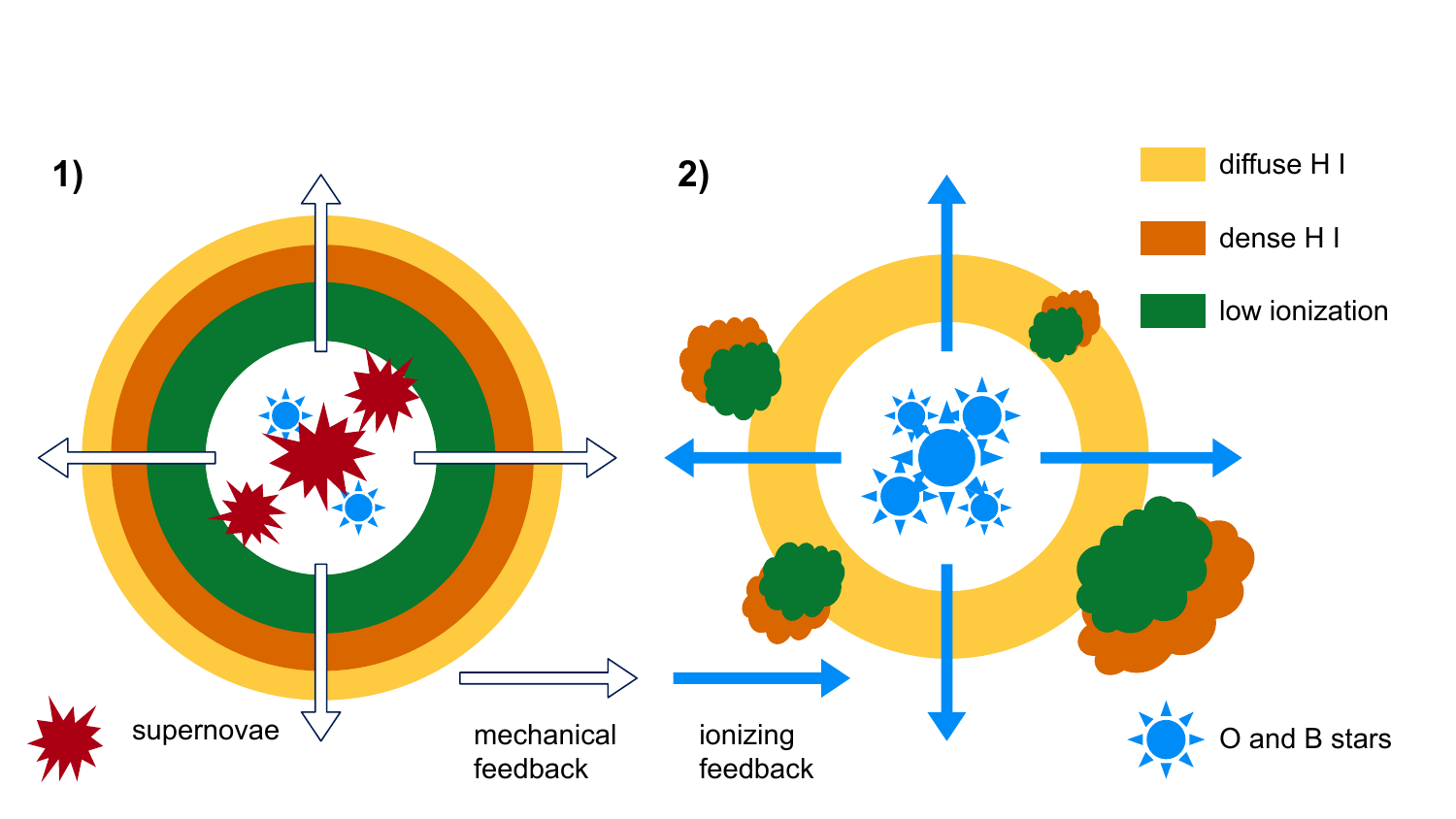}
    \caption{Toy model cartoon illustrating the optimal LyC escape scenario of a two-stage burst. In stage 1) (\emph{left}), supernovae (red) anisotropically inject mechanical energy (white arrows) into the ISM, which will clear optically-thin channels in the dense ISM (orange, green). In stage 2), ionizing photons ionize these channels (yellow) and much of the dense gas (orange, green), allowing the LyC from very young stellar populations (blue) to escape over large opening angles occupied only by diffuse \ion{H}{1} (yellow).}
    \label{fig:fesc_cartoon}
\end{figure*}

Regarding stellar populations, while both \fyng\ and \fold\ correlate strongly with \fesc, \fyng\ dominates in significance for every CCA scenario in which it is considered, including the total set of properties. Clearly, while mechanical feedback and bursty star formation are fundamental to shaping the ISM geometry, the LyC budget provided by the youngest stars serves as the foundation of LyC escape. Given that the most prevalent ISM indicators require a combination of two-stage bursts of star formation and extreme ionization, it is not surprising that the youngest stars achieve the highest \fesc. Thus, a complete picture must include both very young stellar populations (a key ingredient even when considering the ISM conditions) and a patchy geometry even when LyC escape is roughly isotropic. We illustrate this scenario with results in Figure \ref{fig:fesc_tau_fyng} and a cartoon in Figure \ref{fig:fesc_cartoon}.


Figure \ref{fig:fesc_tau_fyng} compares \fesc, the $C_f$ ratio, and \fyng\ or \fold. In Figures \ref{fig:fyng_xiion}-\ref{fig:fold_Emech}, we established that strong LyC leakers (\fesc$>5$\%) tend to have prominent young and old stellar populations. Here, in Figure \ref{fig:fesc_tau_fyng}, we illustrate that the \emph{degree} of \fesc\ corresponds directly to the prevalence of the very young and older stellar populations. Given the relationship between \fyng\ and \xiion\ and between \fold\ and \emech, Figure \ref{fig:fesc_tau_fyng} could imply that two-stage bursts of star formation provide optimal LyC escape conditions by maximizing the radiative and mechanical feedback simultaneously or that two LCE modes exist \cite[e.g.,][]{2022ApJ...930..126F}. Moreover, \fesc\ appears increase significantly with \fyng. Figure \ref{fig:fesc_cartoon} illustrates the effects of a two-stage burst where, in panel 1), SNe provide mechanical feedback to open up channels in the ISM and then, in panel 2), a subsequent generation of stars further clears the path for LyC escape via radiative feedback.

Two outliers of interest in Figure \ref{fig:fesc_tau_fyng} are stacks of non-LCEs which exhibit high \fyng\ but high \cfl/\cfh\ and weak LyC escape (\fesc$\sim0.02$). These two stacks are built from non-detections and non-leakers, both with \orat$>5$. In each case, the galaxies composing these stacks are likely leaking LyC substantially but have only weak \fesc\ along the line of sight. Such cases are within the regime of isotropic LyC escape but spatially varying \fesc, as discussed in \ref{sec:lycescape}. In Figure \ref{fig:fesc_cartoon}, the two outliers' lines of sight would most likely be obstructed by one of the dense clouds in panel 2) even though most sightlines are not obscured as in Knot A of Haro 11 \citep{2024ApJ...967..117K}. The size and/or optical depth of the cloud may allow for marginal LyC escape which falls below the \emph{HST}/COS detection threshold for the stacks' constituent galaxies. 

\subsection{LyC Escape at High Redshift}\label{sec:lycfar}

Cosmological simulations have been a frequent tool for predicting the escape of LyC from galaxies at $z>4$ and have been an excellent baseline for comparisons of nearby $z\sim0.3$ analogs with our expectations for galaxies at the epoch of reionization \citep[see, e.g., comparisons of the LzLCS+ with simulation results in][]{2022ApJ...930..126F,2024arXiv240610179J}. Simulations predict that instantaneous bursts of star formation precede high \fesc\ \citep[e.g.,][]{2014ApJ...788..121K,2017MNRAS.470..224T,2019MNRAS.486.2215K,2020ApJ...902L..39B,2020MNRAS.498.2001M,2022ApJS..259...21K}. Given arguments for bursty star formation in galaxies at redshifts $>6$ \citep{2023MNRAS.524.2312E,2024MNRAS.527.6139S,2024arXiv240505111M}, the case for burstiness as a path toward LyC escape seems plausible at the epoch of reionization, particularly given LyC photons might self-regulate subsequent star formation.

However, simulations indicate a lag of anywhere from 3.5 Myr \citep[e.g.,][]{2023arXiv230408526C} to 10 Myr \citep[e.g.,][]{2014ApJ...788..121K} or more \citep[e.g.,][]{2020ApJ...902L..39B} between a burst of star formation and intense LyC escape. As such, these simulations indicate slightly older stellar populations are more closely associated with high \fesc, owing largely to the mechanical feedback injected by SNe to clear out the ISM. { These \fesc\ lagging predictions arise in part from poor sampling of the SN delay time distribution, which can over-estimate the population age at which SNe occur. Proper treatment of SN onset times may bring simulation predictions into better agreement with the observations. Additionally, the spatial resolution of simulations may not properly treat radiative feedback and \ion{H}{2} region geometry on small scales, which may allow for earlier LyC escape \citep[e.g.,][]{2019MNRAS.486.2215K,2021ApJ...908...30K}.}

While we do find older stellar populations and high \emech\ to be important for LyC escape, our stacking analysis indicates that populations younger than $3$ Myrs are the signpost for \fesc\ in excess of 5\%. Stellar populations aged 3-6 Myr are primarily associated with LyC escape below 1\%, exhibiting the weakest individual correlation with \fesc\ of all the properties considered (see Table \ref{tab:cca}). Populations older than 8 Myr are ubiquitous among LCEs and exhibit a strong correlation with \fesc\ on their own, relative to \fyng. On the whole, these older populations are most significantly correlated with LyC escape of all population age groups (see $\tau$ in Table \ref{tab:cca}) and indicate whether a galaxy is a LyC leaker. Distinguishing between a weak (\fesc$=1$-5\%) and strong (\fesc$>5$\%) LCE, though, requires prevalent young stellar populations (see Figure \ref{fig:fesc_tau_fyng} and \S\ref{sec:fdbkgeom}). 

\citet{2020MNRAS.498.2001M} argue SNe-triggered bursts of star formation may be uniquely optimal for LyC escape. Such burstiness is consistent with our results; however, the role of radiative feedback in bursty models is still understated.
{\it Perhaps one of the most surprising and significant results of this study is the importance of young stellar populations and dominance of ionizing feedback in LCEs.}
As such, prescriptions for feedback and/or stellar populations and radiative transfer in cosmological simulations may need revision in order to properly predict the LyC escape mechanisms and the critical contributors to cosmic reionization. Indeed, incorporating robust ionizing feedback into simulations reproduces our result that stellar populations younger than 3 Myr are key to the most prodigious LyC escape. \citet{2017MNRAS.466.4826K} found ionizing feedback to be critical to \fesc\ when incorporated into high resolution simulations with stochastic star formation. 

{ Results from higher redshift galaxies also emphasize the role of ionizing feedback. The Sunburst arc, a confirmed LCE at $z=2.37$ \citep{2019Sci...366..738R}, exhibits extreme outflows capable of evacuating gas from star-forming sites \citep{2022ApJ...940..160M,2022A&A...659A...2V}. The LyC is emergent from a massive clump with extremely young ($<4\rm~Myr$) stellar populations and concentrated star formation (\sigsfr$=5\times10^3\rm~M_\odot~yr^{-1}~kpc^{-2}$) indicative of bursty star formation \citep{2022A&A...659A...2V,2024arXiv240408884R}.  
Surrounding the young stellar population are dense clouds of gas with openings allowing LyC escape and suggestive of catastrophic cooling \citep{2023ApJ...957...77P}. Local analog Mrk 71, the nearest Green Pea and a LCE candidate, exhibits similar extreme conditions: a massive young star cluster with dense gas undergoing catastrophic cooling \citep{2017ApJ...849L...1O,2023ApJ...958L..10O} as well as LyC-driven outflows presenting as broad nebular lines \citep{2021ApJ...920L..46K}.

Massive star clusters ($M\sim10^6\rm~M_\odot$) with concentrated star formation and young stellar populations akin to Mrk 71 and the LyC-leaking clump in the Sunburst have been identified at the epoch of reionization ($z=6$, \citealt{2023ApJ...945...53V}, and $z=10$, \citealt{2024Natur.632..513A}), featuring paucities of dust \citep{2024Natur.632..513A} and high ionizing photon budgets (\xiion$\ga5\times10^{25}\rm~phot~Hz~erg^{-1}$) capable of clearing out LyC escape channels \citep{2023ApJ...945...53V}. Similarly, faint (\muv$=-15$ to $-17$) galaxies at $z=6-8$ exhibit \xiion$\sim6\times10^{25}\rm~phot~Hz~erg^{-1}$, a dearth of dust ($\beta<-2$), and bursty star formation \citep{2024Natur.626..975A}. Extreme ionizing feedback implied by the high \xiion\ values may even indicate the presence of VMS, as in the case of Mrk 71 \citep{2023ApJ...958..194S}. Such VMS may be necessary to provide both the high \xiion\ and the intense ionizing feedback necessary for LyC escape at $z>6$. Whatever the case, these results form higher redshifts are consistent with the picture of prodigious LCEs our stacks have brought into focus at $z\sim0.3$: concentrated star formation with significant amounts of ionizing feedback.}

Whether the association of 3-6 Myr starbursts with non-LCEs indicates a separate class of galaxies or a transition in LyC escape remains to be seen. Comparisons between \fwr\ and geometry indicate mild positive correlations between the presence of WR populations and \wfh, which may point to the redistribution of ISM via stellar winds \citep{2019ApJ...885...96J}. Given the bursty star formation exhibited by the LzLCS stacks, it is possible that contributors to cosmic reionization ``turn on'' as LCEs at the start of a burst of star formation preceded by another burst by at least 8 Myrs. These LCEs would start with very high \fesc\ which subsequently declines once the most massive O stars begin to evolve off the main sequence, in turn reducing the ionizing feedback needed for the most extreme LyC escape (cf. Carr et al. submitted). Ionizing feedback can regulate star formation, which may contribute to burstiness \citep[][]{2017MNRAS.466.4826K,2023MNRAS.519.1425P}, thereby contributing to a two-stage burst scenario. 
Additional constraints from inferred bursty star formation history at $z>6$, combined with observables tracing LyC escape, would subsequently provide an empirical assessment of future simulation results.

\section{Conclusion}\label{sec:concl}

Through stacking analysis of \emph{HST}/COS G140L spectra from the LzLCS+ observations of 89 galaxies in the rest FUV, we investigate (i) robustness of non-detections of the LyC, (ii) the stellar populations, feedback, and ISM geometry of LCEs vs non-LCEs, and (iii) the nature of LyC escape, including its optimal conditions. We find the following results:

\begin{itemize}
    \item The shape of the LyC is too flat to contain significant emergent nebular continuum and is instead consistent with photoionization of \ion{H}{1}. Thus, nebular continuum does not enhance \fesc\ in the LzLCS+. Furthermore, the lack of nebular continuum points to a clumpy ISM geometry with channels that are poor in \ion{H}{1}.
    \item In most cases, non-detections correspond to true non-leakers. Stacking allows us to push below the COS sensitivity limits by boosting the LyC signal with averaged spectra of non-detections. Only stacks of non-detections with Green Pea properties (e.g., high \orat, high H$\beta$ EW, low \logoh) exhibit \fesc$>1$\%. The detection of missing LyC flux in the stacks of Green Peas points to isotropic LyC escape with line-of-sight variations in \fesc.
    \item Stellar populations older than 8 Myr are ubiquitous among LCEs (\fesc$>1$\%). 
    However, the youngest stellar populations are most prominent in the most prodigious LCEs (\fesc$\ga5$\%).
    \item Mechanical feedback from SNe accompanies older stellar populations and is enhanced in LCEs compared to non-LCEs.
    \item Ionizing feedback, associated with the youngest stellar populations, characterizes the most extreme LCEs.
    \item LCEs exhibit a paucity of dense gas clouds which correlates strongly with declining \ion{H}{1} column densities and dust content. These correlations indicate that, as openings in the dense gas widen, the column density of the diffuse gas decreases.
    \item Mechanical feedback reduces the global covering fraction of dense gas clouds.
    \item Ionizing feedback reduces the neutral gas and dust in the low density gas channels. In extreme cases, ionizing feedback can also reduce the LyC opacity of dense gas clouds.
    \item While mechanical feedback appears to be key for all LCEs, only a combination of mechanical feedback and extreme ionizing feedback can achieve high \fesc.
\end{itemize}

From these results, we can paint a holistic picture of LyC escape. SNe open channels in dense gas in the ISM via mechanical feedback, clearing the path for LyC escape. For high \fesc, subsequent young stellar populations provide ionizing feedback to further dissipate the dense gas and reduce the optical depth of diffuse gas in the ISM. A two-stage burst of star formation is therefore necessary to obtain these optimal LyC escape conditions by providing both mechanical and ionizing feedback to move or eliminate LyC-obstructing gas and dust.

Given the argument for bursty star formation at high redshift, galaxies at $z>6$ might readily exhibit this two-stage burst scenario to achieve the optimal conditions for LyC escape and thus drive cosmic reionization. Future work to investigate this possibility should involve more detailed examinations of star formation histories at cosmic dawn as well as confirmation via simulations of these feedback mechanisms at play. Such results could distinguish between the possibilities of a two-stage starburst LCE and two modes of LyC escape.

\begin{acknowledgments}

Support for this work was provided by NASA through grant number \emph{HST}-GO-15626 from the Space Telescope Science Institute. Additional work was based on observations made with the NASA/ESA Hubble Space Telescope, obtained from the data archive at the Space Telescope Science Institute from \emph{HST} proposals 13744, 14635, 15341, and 15639. STScI is operated by the Association of Universities for Research in Astronomy, Inc. under NASA contract NAS 5-26555.

Funding for the Sloan Digital Sky Survey IV has been provided by the Alfred P. Sloan Foundation, the U.S.Department of Energy Office of Science, and the Participating Institutions. SDSS-IV acknowledges support and resources from the Center for High Performance Computing  at the University of Utah. The SDSS website is \url{www.sdss.org}. SDSS-IV is managed by the Astrophysical Research Consortium for the Participating Institutions of the SDSS Collaboration.

SRF acknowledges support from NASA/FINESST (grant number 80NSSC23K1433) and the MSGC. S. R. Flury thanks N. Choustikov, R. Endsley, S. Gazagnes, and A. Saxena for discussions regarding feedback and ISM conditions and S. Gorsky for discussions regarding multivariate statistics. ASL acknowledges support from Knut and Alice Wallenberg Foundation.

\end{acknowledgments}

\software{{\tt\ astropy\ } \citep{astropy:2013,astropy:2018}, {\tt\ CCA\ }\citep{CCAsoft},{\tt\ Cloudy\ } \citep{2013RMxAA..49..137F},
{\tt FiCUS} \citep{2023MNRAS.522.6295S}, {\tt\ KaplanMeier\ } \citep{Flury_KaplanMeier},
{\tt lmfit\ } \citep{lmfit}, {\tt\ matplotlib\ } \citep{matplotlib}, {\tt\ numpy\ } \citep{numpy}, {\tt\ scipy\ } \citep{scipy}, {\tt Starburst99\ } \citep{1999ApJS..123....3L}}


\bibliographystyle{aasjournal}
\bibliography{biblio}

\appendix

\section{Approaches to Stacking}

{
Below, we assess various approaches to stacking the LzLCS+ \emph{HST}/COS spectra, addressing caveats to each choice and its effect (if any) on the science results.

\subsection{Normalization}

Perhaps one of the most straightforward yet complicated choices is the normalization of the galaxies' spectra before undertaking stacking. Doing so is key to preventing luminosity bias: without normalization, the brightest sources will be over-represented in the stacks. Due to Galactic and telluric contamination (absorption and emission lines, respectively), which occur at different rest-frame wavelengths due to differences in cosmological redshift, certain portions of the spectrum are not sampled for every source. Moreover, features from galaxies themselves cause substantial variations due to the underlying stellar populations (photospheric and P Cygni wind lines) and gas content (e.g., Lyman series and low ionization absorption lines). In an effort to preserve these features and better detect them in the stacked spectra, we need to select relatively featureless portion continuum which is well-detected in every source.

Blueward of 1050 \AA\ in the rest frame, the spectra suffer from telluric contamination from geocoronal Ly$\alpha$ and \ion{O}{1} $\lambda1305$ emission, plus \ion{H}{1} Lyman series absorption, and many prominent stellar features including \ion{C}{3} $\lambda997$, \ion{N}{3} $\lambda991$, \ion{O}{6} $\lambda1038$, and \ion{S}{4} $\lambda1060$. The stellar \ion{C}{3} $\lambda1175$ multiplet and \ion{N}{5} $\lambda1238,42$ P Cygni lines, ISM \ion{Si}{2} $\lambda\lambda1191,1193,1260,1304$, \ion{O}{1} $\lambda1302$, \ion{C}{2} $\lambda\lambda1334,5$, and of course Ly$\alpha$ together prevent reliably obtaining featureless continuum from 1170 to 1350 \AA. The COS FUV A sensitivity redward of observed frame 1600 \AA\ is a factor of four lower than at its peak around 1350-1400 \AA, meaning that continuum at wavelengths longer than 1300 \AA\ in the rest frame are subject to significantly more noise and therefore substantial uncertainty in any continuum measured there. Given the peak sensitivity of COS at 1050-1100 \AA\ in the LzLCS+ rest-frame and the relative lack of features from 1080-1110 \AA\ in the underlying continuum, we opt to normalize the spectra at 1100 \AA. The choice of 1100 \AA\ is also in keeping with previous LzLCS studies' use of the 1100 \AA\ flux as representative of non-ionizing stellar continuum \citep{2022ApJS..260....1F,2022ApJ...930..126F}.

We note that any choice of normalization wavelength is subject to bias due to stellar population age and dust attenuation; however, the choice of flux estimator in each wavelength can be similarly (or more) important to determining the representative spectral shape for the stacked spectrum.

\subsection{Resolution}

Observed-frame grating resolution with \emph{HST}/COS G140L is roughly seven pixels which, with a plate scale of 0.0803 \AA\ px$^{-1}$, corresponds to 0.5621 \AA. The LzLCS+ characteristic $z\sim0.3$ thus implies a rest-frame resolution of 0.4324 \AA. Given the unbinned LzLCS+ spectra, we can effectively choose the rest-frame wavelength binning down to a characteristic pixel. We consider (i) a sampling of 7 pixels as listed in the COS documentation (rest-frame resolution of 0.43 \AA), (ii) a conservative 10 pixels (rest-frame resolution of 0.62 \AA) to further boost the signal, and (iii) a sampling of 8 pixels to reflect the non-Gaussian wings of the COS LSF and provide relatively even spacing in the stack wavelengths (rest-frame resolution of 0.50 \AA). We find the resolution of 0.43 \AA\ provides lower S/N than desired in many of the stacks. While 0.62 \AA\ does an excellent job of boosting the S/N, some spectroscopic features (mainly photospheric and ISM absorption lines) are increasingly ``washed out'' due to the decrease in resolution and artificial ``infilling'' by inclusion of starlight from adjacent pixels. The third case of 0.50 \AA\ resolution optimizes between the two scenarios.

As a test, both of the effects of resolution on our results and on the possible low resolution of more extended galaxies in the LzLCS+, we build all our stacks using both 0.5 \AA\ and 0.62 \AA\ resolution. We then measure absorption line properties from all stacks at each resolution. We find no significant difference in either LIS or \ion{H}{1} lines due to changes in resolution

\subsection{Flux Estimators}

In each wavelength bin, we obtain a distribution of fluxes from some (losses due to Galactic or telluric contamination) or all of the galaxies used to build the stack. We must choose an estimator to characterize this distribution and thus represent it in the stack. We consider three different estimators for the first and second moments (the stack flux and uncertainty, respectively): (i) mean, (ii) variance-weighted mean, and (iii) median. We calculate the second moment by bootstrapping subsets of the distribution of fluxes in the wavelength bin and recalculating the first moment $10^4$ times, meaning that the reported uncertainty is comparable to a standard error of the mean. The first estimator (mean) is biased towards outliers in each bin while the second estimator (variance-weighted mean) is biased towards the highest signal to noise sources in each bin (thereby undermining the effort of normalization, and our impetus for stacking in general). The third approach (the median), while sometimes subject to larger uncertainties, provides the least biased estimate of the stacked flux in each bin. We confirm the bias of the first two methods towards brighter sources relative to the median by comparing mean, variance-weighted, and median estimators for the stacked flux for test cases with 50 contributing sources, making these tests more statistically robust than less populated stacks.

While it is entirely possible that the median estimator could select a single COS spectrum as representative of the entire set, this result is highly unlikely: a simple frequentist calculation suggests a chance of only one part in ten thousand of selecting just one galaxy spectrum for the least populated stacks (just 5 galaxies) assuming no Galactic or telluric contamination. Accounting for this contamination would reduce the probability of obtaining this result even further by eliminating at least one stack from a subset of wavelength bins.

\subsection{Considerations for Absorption Lines}

Of particular import is the effect of stacking on the ISM absorption lines. While ideally, absorption lines could be normalized in the observed source spectra and stacked solely as line profiles, various effects prevent us from accurately undertaking this approach. The caveats discussed above regarding normalization apply here.

For LIS lines like \ion{Si}{2} and \ion{C}{2}, the continuum itself is not well-detected due to the drop off of COS sensitivity, making any attempt at proper normalization by a fit to the continuum tenuous at best. In our efforts to test this procedure for these important ISM lines, we found that the median ``normalized'' continuum in the resulting stack varied anywhere from 0.5 to 1.5 when it should be roughly unity.

Further investigation with \ion{H}{1} Lyman series lines yielded even less promising results: due to contamination by stellar features, other ISM lines, and steep changes in the continuum due to the intrinsic stellar SED (the peak of a 30 000 K blackbody falls at roughly 966 \AA) and the effects of dust, the continuum is difficult to estimate accurately even when S/N is high and telluric contamination is not present. In our investigation, the ``normalized'' continuum for many of the stacked \ion{H}{1} lines was highly non-linear when it should not vary at all with wavelength, instead containing spurious bumps and wiggles.

These combined effects indicate that stacks of the individual ISM lines are not reliable. Therefore, we do not include such an approach in our stacking analysis of the LzLCS+.
}

\section{Kaplan-Meyer Estimator and Survival Functions}\label{apx:kaplanmeier}

{
For galaxies used to build a given stack, the distribution of \fesc\ often contains non-detections, especially when assessing non-leakers. To appropriately handle these cases in our comparison of galaxy and stack \fesc\ measurements, we use the non-parametric Kaplan-Meier estimator \citep[KM,][]{KaplanMeier1958}\footnote{Our implementation \citep{Flury_KaplanMeier} of the KM estimator is available at \url{https://github.com/sflury/KaplanMeier}}, which accounts for upper limits in the cumulative distribution function, to determine the probability \pfesc\ of obtaining \fesc\ less than the escape fraction \fst\ measured in the stack given the escape fractions \fg\ of the galaxies used to build that stack. The KM estimator is effectively a survival function accounting for censoring and is given by
\begin{equation}
    P\left(<f_{\rm s}|f_{\rm g}\right) = \prod\limits_{i=1}^{f_{\rm s}<f_{{\rm g},i}}\left(\frac{N-m_i}{N-m_{i-1}}\right) = 1 - \frac{m_i}{N}
\end{equation}
where $i$ is the rank order of $1-f_{{\rm g}}$ corresponding to the lowest \fg\ which satisfies \fg$>$\fst, $m_i$ is the number of galaxies which satisfy $f_{\rm g}>f_{{\rm g},i}$ if the LyC is detected or $f_{\rm s}>f_{{\rm g},i-1}$ if the LyC is censored, and $n_i$ is the number of galaxies with $f_{\rm g}>f_{{\rm g},i-1}$. The ordering by $1-\mathfesc$ here arises from the left-censoring KM formalism, meaning than we must treat the fraction of LyC \emph{absorbed} rather than the fraction which has escaped. Thus, \pfesc$=S(1-\mathfesc)$ where $S$ is the ``survival'' of LyC. We show examples of the KM survival function in Figure \ref{fig:km_examps}

\begin{figure}
    \centering
    \includegraphics[width=0.49\columnwidth]{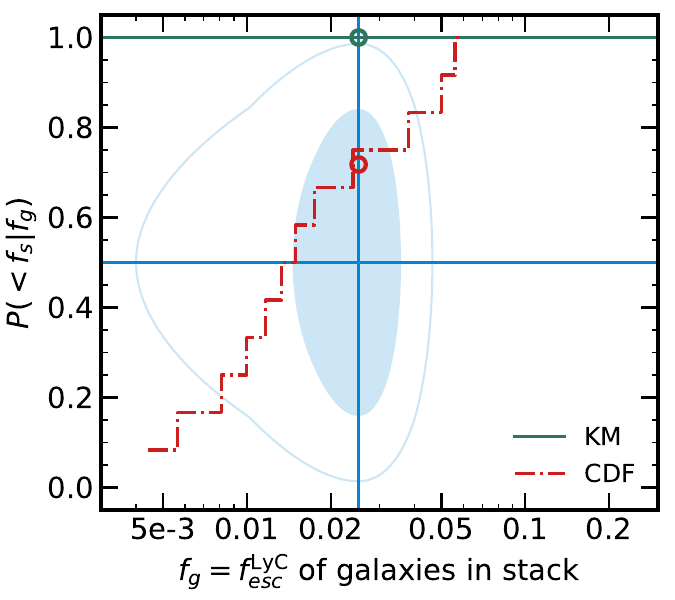}
    \includegraphics[width=0.49\columnwidth]{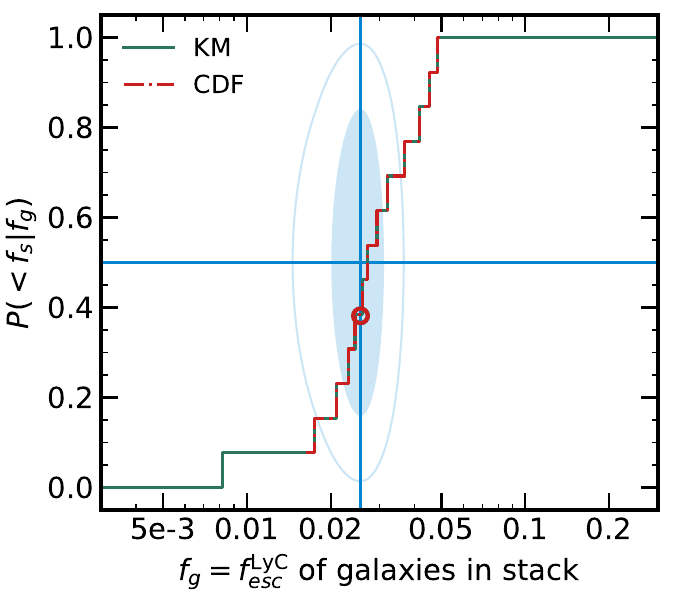}
    \caption{ Examples of the Kaplan-Meier survival function (KM, solid green line) compared to a traditional cumulative distribution function (CDF, dashed red line) in cases where upper limits significantly affect the distribution ({\it left}) and where they do not ({\it right}). Green open circle indicates the KM estimator for \pfesc\ while the red open circle indicates that for the CDF. Blue horizontal line indicates the 50th percentile of the KM survival function. Blue vertical line indicates the \fesc\ value measured from the stack (\fst). The filled (open) ellipsoid indicates the 1$\sigma$ (2$\sigma$) confidence on \fst\ and corresponding distribution functions. 
    }
    \label{fig:km_examps}
\end{figure}

Values of \pfesc$\in[0.1587,0.8413]$ indicate agreement between \fst\ and \fg\ at the 1$\sigma$ level while $P\in[0.025,0.975]$ indicates only tentative rejection of the null hypothesis. 
To assist in the interpretation of \pfesc, we can express \pfesc\ as the confidence $\sigma_0$ with which we reject the null hypothesis that \fst\ is representative of \fg. This confidence $\sigma_0$ corresponds to \pfesc\ via the probit function such that
\begin{equation}
\sigma_0 = {\rm probit}(P) = \sqrt{2}{\rm erf}^{-1}(2P-1).
\end{equation}
Put simply, $\sigma_0$ is the significance of the difference between \fst\ and the median \fg.
}

\section{Convolution with COS templates\label{apx:conv}}

The COS line-spread function (LSF) is highly non-Gaussian with prominent wings which lead to contamination of spectral features by extent continuum starlight. Furthermore, the COS LSF is wavelength dependent, which prevents straightforward convolution of model spectra with a single kernel to account for transmission through the \emph{HST}/COS optics. Here, we outline our use of tabulated wavelength-dependent COS LSFs to obtain convolved model spectra for assessments such as detectability and systematic uncertainties\footnote{Tabulated LSFs sampled at regular 5 \AA\ intervals are available at \url{https://www.stsci.edu/hst/instrumentation/cos/performance/spectral-resolution} with relevant documentation at \url{https://github.com/spacetelescope/notebooks/blob/master/notebooks/COS/LSF/LSF.ipynb}.}. The majority of the observations considered in this study utilized the COS G140L grating at lifetime position 4. We proceed using the tabulated LSFs for this instrument setting. It is worth noting that these LSFs are only valid for point sources (applicable for most of the LzLCS+ sources: 83/89 have COS NUV halflight radii $<0.3\arcsec$, placing them squarely in the point-source regime for the $2.5\arcsec$ COS aperture, \citealt{2022ApJS..260....1F}): extended sources will suffer from a broader LSF, exacerbating the effects illustrated and discussed below.

First, all model spectra with fluxes $f_{true}$ are evaluated or resampled to the 0.0803 \AA\ COS detector resolution. We interpolate the LSF at each wavelength $\lambda$. Finally, we compute the entire convolution integral at each wavelength
\begin{equation}
    f_{conv}(\lambda) = \int\limits_{-\Delta\lambda}^{+\Delta\lambda} {\rm LSF}(\delta\lambda,\lambda)f_{true}(\lambda+\delta\lambda){\rm d}\delta\lambda
\end{equation}
where $\Delta\lambda=12.5268\rm~\AA$ is the half-width of the LSF kernel and k$\delta\lambda$ is an arbitrary position within the LSF relative to the central wavelength. As the COS LSFs are tabulated, we approximate the integral using the trapezoid method. Typically, evaluating all integrals over the range of model wavelengths ($\lambda\in[900,1500]$) is rapid, taking only a few seconds on a conventional laptop to process a single spectrum.

\begin{figure}
    \centering
    \includegraphics[width=0.49\columnwidth]{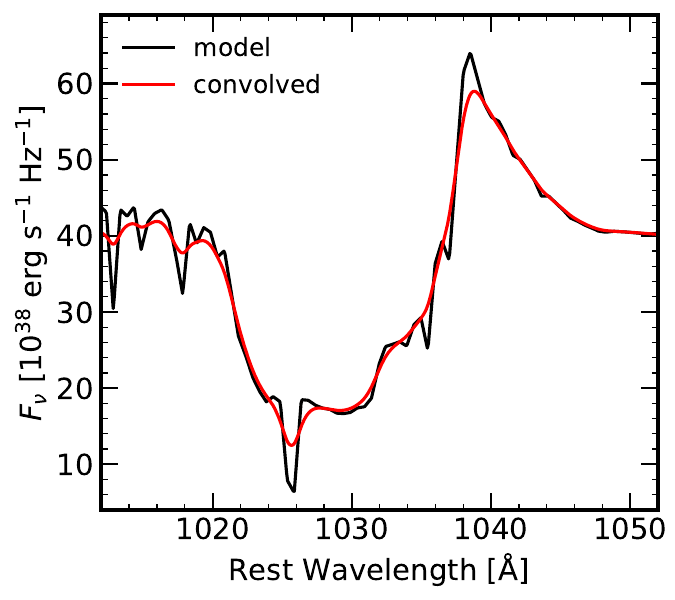}
    \includegraphics[width=0.49\columnwidth]{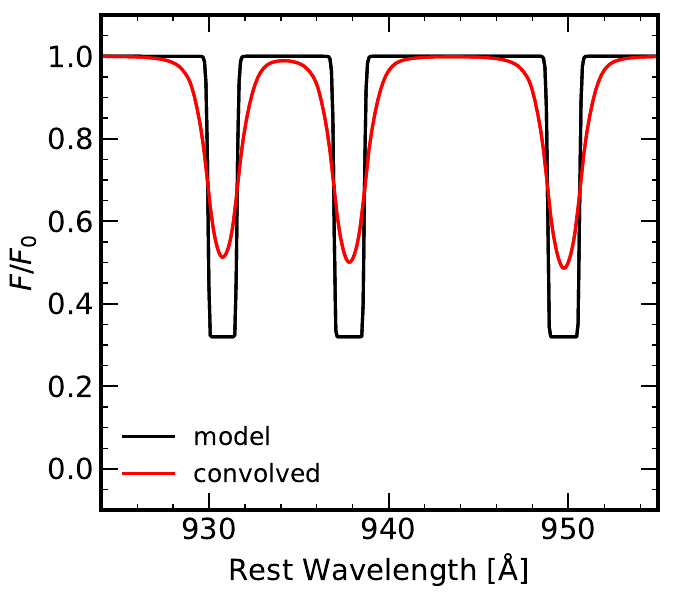}
    \caption{Examples of model stellar (left) and absorption line (right) spectra before (black) and after (red) convolution with the COS wavelength-dependent line-spread function (LSF). The effect of the COS optics is relatively muted for the continuum but quite pronounced where features are concerned. The \ion{O}{6} P Cygni feature on the left exhibits a weaker emission component due to the LSF, which affects the use of wind lines like \ion{O}{6} and \ion{N}{5} as age diagnostics. The COS LSF also causes absorption lines to appear as having substantially higher residual flux than reality, as shown by the saturated Lyman series \ion{H}{1} lines on the right.}
    \label{fig:mod_conv}
\end{figure}

Examples of convolved stellar and absorption line models are shown in Figure \ref{fig:mod_conv}. The effects of \emph{HST}/COS optics on photospheric and ISM absorption lines is particularly pronounced as the features become much more shallow with respect to the continuum.

\section{Assessment of Stacked Absorption Lines\label{apx:abs}}

We produce theoretical absorption lines for each object in the combined LzLCS+ assuming
\begin{equation}
    \frac{F}{F_0} = C_f\exp\left[-\frac{\sigma f\lambda N}{b}\phi(\lambda,b)\right]+1-C_f
\end{equation}
where $C_f$ is the gas covering fraction, $\lambda$ is the line center wavelength in \AA, $b$ is the Doppler-broadening parameter in km s$^{-1}$, $N$ is the gas column density (assumed to be $10^{19}\rm~cm^{-2}$ for \ion{H}{1} and $10^{15}\rm~cm^{-2}$ otherwise, as these lines are likely saturated, e.g., \citealt{2022A&A...663A..59S}), $\sigma=\sqrt{\pi}e^2/m_e c\approx10^{-14.8247}\rm~cm^{2}$ is the absorption cross-section (absorbing units from $\lambda$ and $b$), and $\phi$ is the line profile. We assume a Voigt line profile as approximated by the real part of the Fadeeva function as implemented by {\tt scipy}. Atomic data for $f$ and $\lambda$ are taken from \citet{2003ApJS..149..205M}. For purposes of this assessment, we consider the following lines which were consistently used throughout our analysis:
Ly$\zeta$, Ly$\epsilon$, Ly$\delta$, Ly$\gamma$, \ion{Si}{2} 1190,93, \ion{Si}{2} 1260, \ion{O}{1} 1302, \ion{Si}{2} 1304, \ion{C}{2} 1334.

For each line in each object, we generate two sets of line profiles, each over a grid of $b=50,100,150,200\rm~km~s^{-1}$ and $C_f=0.1,0.2,0.3,0.5,0.75$. The first set of profiles assume $C_{f,\rm LIS}=C_{f,\rm H I}$ while the second imposes a grid of $C_{f,\rm LIS}$ while $C_{f,\rm H I}$ is held fixed (similar to invoking the $C_f$ ratio demonstrated in \S\ref{sec:absn}). Line profiles for a given $C_f$ and $b$ are combined multiplicatively as dictated by radiative transfer to form a single simulated absorption spectrum. We convolve each absorption spectrum for each of the 89 possible redshifts with the LSF as discussed in Appendix \ref{apx:conv} for a total of 3 560 simulated line spectra.

To simulate stacking, we randomly select subsets of 10, 15, 20, 30, 40, and 50 simulated spectra of a given $C_f$ and $b$, with and without a discrepancy between the \ion{H}{1} and LIS $C_f$. For each spectrum in the subset, we introduce noise corresponding to a fixed S/N. We coadd all the noisy convolved spectra in the rest-frame in wavelength bins of 0.5 \AA\ in keeping with the procedure outlined in \S\ref{sec:UVStacks}. We repeat this stacking exercise for S/N values from 0.05 to 5 in 0.05 increments (recall that stacking of S/N$<1$ signals can still result in a ``detection''). For reference, we also build stacks of the model and convolved clean spectra to assess recovery of assumed model parameters after (i) processing by the \emph{HST}/COS optics and (ii) noise introduction from the detector.

\begin{figure}
    \centering
    \includegraphics[width=0.49\linewidth]{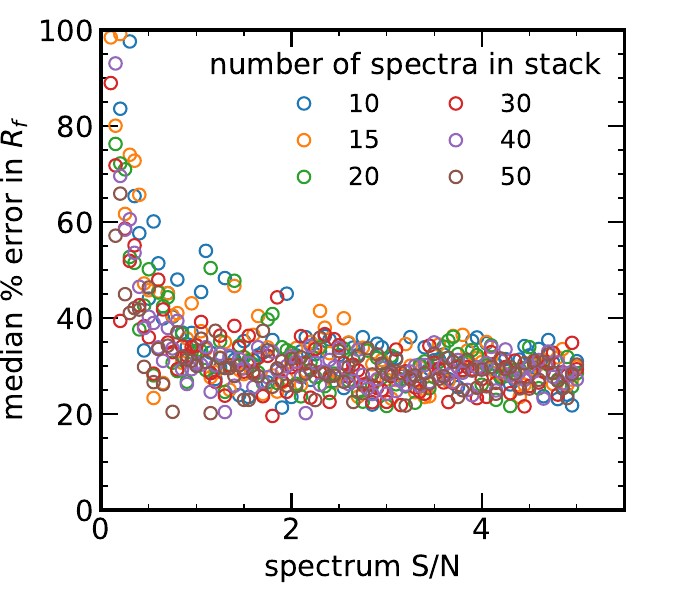}
    \includegraphics[width=0.49\linewidth]{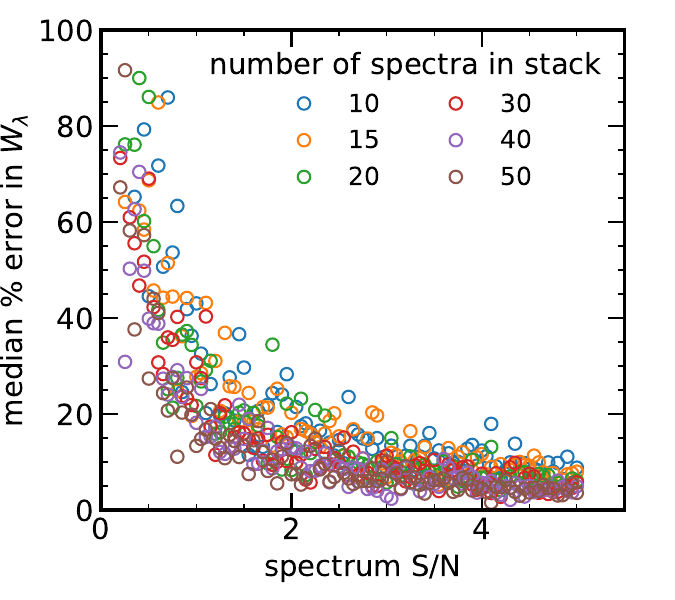}
    \caption{Recovery rate (i.e., systematic uncertainty) in \ion{Si}{2} 1260 $R_f$ and $W_\lambda$ as determined from simulated stacks of LSF-convolved model absorption line spectra. Symbols color-coded by number of spectra included in a given model stack. Simulations cover a grid of $b$ and $C_f$ with toggling of the residual flux discrepancy.}
    \label{fig:abs_sim}
\end{figure}

In each stack, we measure model, convolved, and noisy $W_\lambda$ and $R_f$ for each of the simulated lines following \ref{sec:absn}. To illustrate the accuracy of measuring $W_\lambda$ and $R_f$, we show full results for the \ion{Si}{2} $\lambda1260$ line in Figure \ref{fig:abs_sim} as a function of S/N in the individual spectra. From these results, it is clear that, as a result of the effects of convolution redistributing the emergent flux, the measured $R_f$ is always systematically higher than the model values. In other words, the true $R_f$ cannot be recovered. E.g., for $C_f=1$ in \ion{Si}{2} $\lambda1260$, one will measure a minimum of $R_f\sim0.25$ ($C_f\sim0.75$) from the stack, not $R_f=0$. Recovery in $W_\lambda$ continues to improve with increasing S/N, likely due to the fact that area of the absorption feature is more or less conserved during the convolution process even as the emergent flux is redistributed.

\begin{figure}
    \centering
    \includegraphics[width=0.495\linewidth]{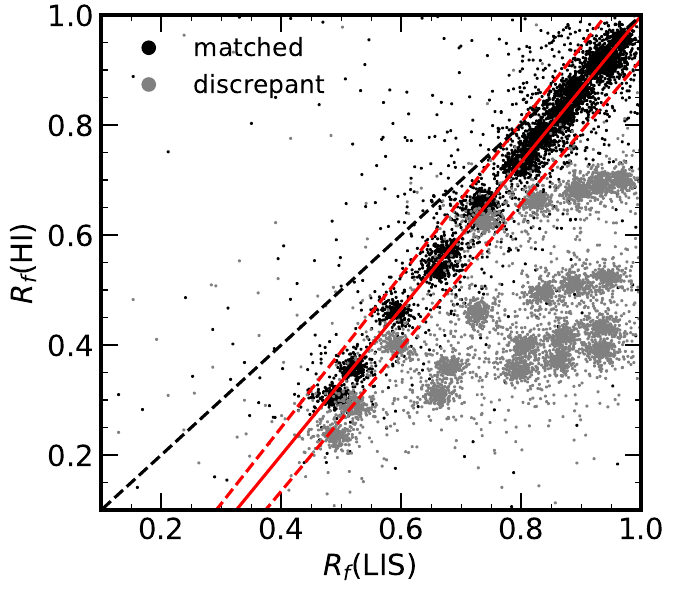}
    \includegraphics[width=0.495\linewidth]{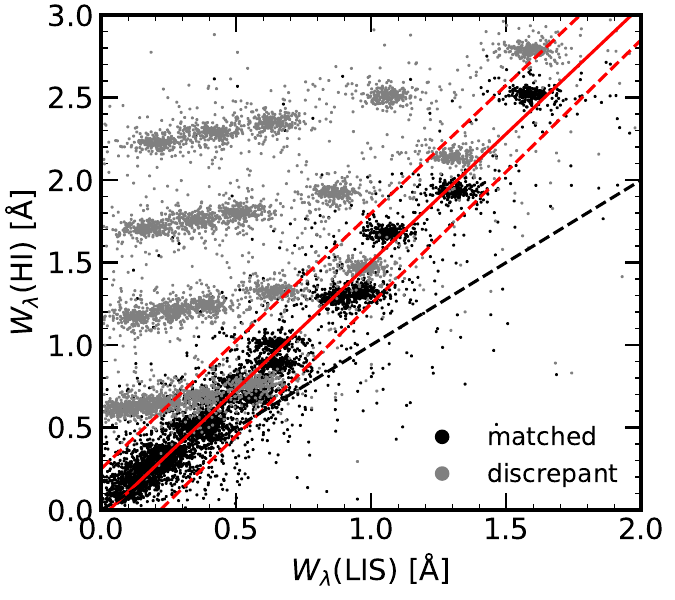}
    \caption{Simulated absorption line residual fluxes $R_f$ (\textit{left}) and equivalent widths $W_\lambda$ (\textit{right}) with a discrepancy imposed by assuming \rfh=0 as \rfl\ varies (grey circles) or with no discrepancy by assuming $R_{f,\rm {H}{I}}=R_{f,\rm LIS}$ (black squares). Black dashed lines are the intrinsic lines of agreement \rfh=\rfl\ and \wfh=\wfl. Red solid lines indicate the \rfh=\rfl\ and \wfh=\wfl\ agreement after accounting for effects of the COS LSF and stacking. Red dashed lines indicate the two-sided 90\% confidence intervals between which measured discrepancy can be attributed solely to LSF effects. The wavelength dependence of the LSF is increasingly apparent as $R_f$ decreases and $W_\lambda$ increases: the \ion{H}{1} lines are measured blueward of Ly$\alpha$ while the LIS lines are redward of Ly$\alpha$. The discrepancy in $R_f$ measured in the LzLCS+ stacks cannot be reproduced solely by instrumental effects and can only be recovered if a genuine discrepancy exists.}
    \label{fig:tau_sim}
\end{figure}

To assess whether the $C_f$ discrepancy is purely an instrumental artefact rather than a genuine attribute of the stacks, we calculate the variance-weighted average LIS and \ion{H}{1} line properties $W_{\lambda}$ and $R_f$ for all the simulated stacks using the methods outlined in \S\ref{sec:absn}. From the averaged LIS and \ion{H}{1} $R_f$ without any discrepancy imposed (i.e., $R_{f,\rm {H}{I}}=R_{f,\rm LIS}$ intrinsically), we determine the two-sided 0.9 quantile for the intrinsic 1:1 agreement and describe the distribution with the linear expressions
\begin{equation}
    R_{f,\rm {H}{I}}\in[\ 1.308R_{f,\rm LIS}-0.389,\ 1.385R_{f,\rm LIS}-0.304\ ]
\end{equation}
and verify that $>90$\% of the simulated $R_{f,\rm {H}{I}}=R_{f,\rm LIS}$ results fall between these. Thus, we can with confidence say that results with $R_{f,\rm {H}{I}}$ between these two lines are likely the result of COS optics alone. Above and below this region, results cannot be explained solely by the LSF, meaning the observed discrepancy is physical (i.e., ``real") with $p<0.05$ confidence. We compare this limit to the simulation results in Figure \ref{fig:tau_sim}. We perform a similar assessment for the 1:1 agreement between \wfh\ and \wfl\, finding 90th percentile limits of
\begin{equation}
    W_{\lambda,\rm {H}{I}}\in[\ 1.60W_{\lambda,\rm {H}{I}}-0.35,\ 1.55W_{\lambda,\rm {H}{I}}+0.25\ ]
\end{equation}
which we show in Figure \ref{fig:tau_sim}.


\section{Canonical Correlation Analysis}\label{apx:cca}

{
To undertake our assessment of results measured from the stacks, we employ two separate tools: the non-parametric Kendall's $\tau$ for univariate correlations (introduced in \S\ref{sec:absn}) and canonical correlation analysis (CCA) for multivariate correlations. CCA is a generalization of principal component analysis (PCA) and computes the linear combination of predictor variables $\mathbf{X}$ which maximizes their correlation with response variables $\mathbf{Y}$.
The canonical coefficients $\mathbf{a}$ are the so-called ``loadings'' of $\mathbf{X}$ and represent the relative significance of each parameter to maximizing the multivariate correlation. In other words, a relatively large value of some element $a_k\in\mathbf{a}$ indicates that variable $X_k\in\mathbf{X}$ contributes significantly to the correlation between $\mathbf{X}$ and $\mathbf{Y}$. Using the R package {\tt CCA} \citep{JSSv023i12,CCAsoft}, we solve for the canonical correlation coefficients $\mathbf{a}$ for $\mathbf{Y}^\prime=$\fesc\ for different subsets of properties associated with stellar populations (\fyng, \fwr, \fold), feedback (\xiion, \emech), ISM ($\beta_{1550}$, \wfh, \rfl, \cfl/\cfh), and all three subsets together. We list the results for both Kendall's $\tau$ and the loading factors $\mathbf{a}$ from CCA in Table \ref{tab:cca}.
}

\end{document}